\documentclass[a4paper,11pt]{article}
\pdfoutput=1
\usepackage{jcappub}
\usepackage[T1]{fontenc}
\usepackage{graphicx}
\usepackage{amsmath,amssymb}
\usepackage{hyperref}
\usepackage{url}
\usepackage{epstopdf}
\allowdisplaybreaks
\numberwithin{equation}{section}
\usepackage{enumitem}
\usepackage{bm}
\usepackage[utf8]{inputenc}
\usepackage{color}
\usepackage{bm}
\usepackage{multirow}
\usepackage{hyperref}

\usepackage{subfigure}
\usepackage{array}
\usepackage{todonotes}
\usepackage[english]{babel}
\usepackage{tensor}
\usepackage{comment}
\usepackage[normalem]{ulem}

\hypersetup{
    colorlinks=true,
    linkcolor=blue,
   citecolor=blue,
   urlcolor=blue
}
\graphicspath{{Figures/}}

\definecolor{abpurple}{rgb}{0.75, 0.0, 1.0}

\definecolor{mordantred19}{rgb}{0.68, 0.05, 0.0}

\def\lcdm{$\Lambda$CDM}

\usepackage{orcidlink}
\title{Prospects of testing late-time cosmology with weak lensing of gravitational waves and galaxy surveys}
\author[a,b]{Anna Balaudo\orcidlink{0000-0003-4109-8094},}
\emailAdd{balaudo@strw.leidenuniv.nl}
\author[b]{Alice Garoffolo\orcidlink{0000-0002-0125-4577},}
\emailAdd{garoffolo@lorentz.leidenuniv.nl}
\author[c,d]{Matteo Martinelli\orcidlink{0000-0002-6943-7732},}
\emailAdd{matteo.martinelli@inaf.it}
\author[e]{Suvodip Mukherjee\orcidlink{0000-0002-3373-5236},}
\emailAdd{suvodip@tifr.res.in}
\author[b]{Alessandra Silvestri\orcidlink{0000-0001-6904-5061}}
\emailAdd{silvestri@lorentz.leidenuniv.nl}

\affiliation[a]{Leiden Observatory, Leiden University, PO Box 9513, Leiden 2300 RA, The Netherlands}
\affiliation[b]{Institute Lorentz, Leiden University, PO Box 9506, Leiden 2300 RA, The Netherlands}
\affiliation[c]{INAF-Osservatorio Astronomico di Roma, Via Frascati 33, I-00078 Monteporzio Catone, Italy}
\affiliation[d]{INFN-Sezione di Roma, Piazzale Aldo Moro, 2 - c/o Dipartimento di Fisica, Edificio G. Marconi, I-00185 Roma, Italy}
\affiliation[e]{Department of Astronomy \& Astrophysics, Tata Institute of Fundamental Research, 1, Homi Bhabha Road, Colaba, Mumbai 400005, India}

\abstract{We investigate the synergy of upcoming galaxy surveys and gravitational wave (GW) experiments in constraining late-time cosmology, examining the cross-correlations between the weak lensing of gravitational waves (GW-WL) and the galaxy fields. Without focusing on any specific GW detector configuration, we benchmark the requirements for the high precision measurement of cosmological parameters by considering several scenarios, varying the number of detected GW events and the uncertainty on the inference of the source luminosity distance and redshift.
We focus on $\Lambda$CDM and scalar-tensor cosmologies, using the Effective Field Theory formalism as a unifying language. We find that, in some of the explored setups, GW-WL contributes to the galaxy signal by doubling the accuracy on non-$\Lambda$CDM parameters, allowing in the most favourable scenarios to reach even percent and sub-percent level bounds.
Though the most extreme cases presented here are likely beyond the observational capabilities of currently planned individual GW detectors, we show nonetheless that – provided that enough statistics of events can be accumulated – GW-WL offers the potential to become a cosmological probe complementary to LSS surveys, particularly for those parameters that cannot be constrained by other GW probes such as standard sirens.}

\begin{document}

\maketitle
\flushbottom


\section{Introduction}
\label{sec:introduction}
The historical direct detection of a gravitational wave (GW) by the LIGO-Virgo collaboration in 2015~\cite{LIGOScientific:2016aoc}, marked the beginning of a series of observing campaigns that led to the detection of around one hundred sources~\cite{LIGOScientific:2018mvr,LIGOScientific:2020ibl,LIGOScientific:2021djp}, one of which (GW170817) with a confirmed electromagnetic counterpart~\cite{LIGOScientific:2017vwq,LIGOScientific:2017zic,LIGOScientific:2017ync}. In the near future, the KAGRA interferometer~\cite{KAGRA:2020cvd} will join LIGO-Virgo in their observing runs, while the space interferometer LISA~\cite{LISA:2017} is expected to launch in the late '30s. Further down the lane, the third generation of ground-based GW detectors will see light with the network of Einstein Telescope (ET)~\cite{Maggiore:2019uih} and Cosmic Explorer (CE)~\cite{Reitze:2019iox}, which can drastically improve the sensitivity to GW signals and measure hundreds of thousands of events over 10 years of observations. While GW170817 led the way, allowing the first GW measurement of the Hubble constant~\cite{LIGOScientific:2017adf}, multi-band GW observations will effectively open a new window in observational cosmology, targeting several classes of sources among which stellar black holes binaries (BBH), neutron star-black hole systems (NS-BH), binary neutron stars (BNS), massive black hole binaries (MBHB), extreme mass-ratio inspirals and a stochastic gravitational wave background.

Meanwhile, the intense worldwide effort towards mapping the Universe via wide galaxy and weak lensing surveys has started to deliver large scale structure (LSS) data of unprecedented precision and extension, such as those provided by KiDS ~\cite{KiDS:2013,vanUitert:2017ieu} and DES ~\cite{DES:2016jjg,DES:2021wwk} collaborations. With Stage IV missions (such as {\it Euclid} ~\cite{Laureijs2011}, the Vera C. Rubin Observatory ~\cite{LSST:2009,LSSTDarkEnergyScience:2018jkl,Ivezic:2008fe} and Nancy Grace Roman Space Telescope ~\cite{2015arXiv150303757S}) we will see a paradigm shift in the volume of data available. This will lead to a new level of scrutiny of the standard model of cosmology, Lambda Cold Dark Matter (\lcdm{}).

This model successfully describes the Universe in terms of few parameters, yet the rise of precision cosmology  has seen the emergence of some tensions between datasets when interpreted within it~\cite{Abdalla:2022yfr,Perivolaropoulos:2021jda}, which could simply signal  the first cracks in \lcdm{} as we are achieving a new level of precision in the measurement of its parameters~\cite{DiValentino:2021izs}. 

With the ongoing/upcoming cosmological surveys, we have the opportunity to probe gravity on cosmological scales, and shed light on the nature of dark energy, sourcing cosmic acceleration, and dark matter, supporting gravitational clustering.  
In this context, many alternatives to the \lcdm{} model have been proposed, trying to fill the gaps in the interpretation of the available datasets. We focus on extensions of \lcdm{} that address the phenomenon of cosmic acceleration by means of a dynamical dark energy (DE) component or modifying the laws of gravity on large scales~\cite{Silvestri:2009hh,Clifton:2011jh}. Effectively adding one scalar degree of freedom to the gravitational theory, we adopt the formalism of effective field theory of dark energy (EFT of DE)~\cite{Cheung:2007st,Creminelli:2008wc,Park:2010cw,Bloomfield:2011np,Gubitosi:2012hu,Piazza:2013coa} to  explore the  class of  Horndeski gravity~\cite{Horndeski:1974wa}, which will undergo a high degree of scrutiny with the upcoming cosmological missions. We refer to these models with the common designation of \emph{scalar-tensor theories} (ST), or simply \emph{modified gravity} (MG); popular examples are f(R) gravity, generalized Brans-Dicke models and Generalized Galileons~\cite{Clifton:2011jh}.

Wide surveys that combine galaxy clustering (GC) and weak lensing (WL), such as the ESA-led {\it Euclid} mission~\cite{Euclid} and the Vera Rubin Observatory~\cite{LSST}, are expected to test gravity on cosmological scales at unprecedented precision, thus providing important constraints on ST theories, see e.g.~\cite{Amendola:2016saw}.  In order to further tighten such constraints, and better understand the nature of DE and the theory of gravity, it will be crucial to bring in new probes able to provide independent measurements and also break degeneracies with the cosmological parameters. Multi-band GW observations are very interesting to this extent, as they offer complementary probes that are characterized by a different set of systematic errors with respect to galaxy surveys.

The dynamics of GWs depends on the extended parameters on two level: explicitly, as they enter directly their propagation equation, and implicitly via the different expansion rate of the Universe and growth of  gravitational potentials.
In this work, we consider gravitational theories in which photons are not directly coupled to the additional scalar field, which is a typical assumption also in the EFT of DE description. As a result, the dynamics of photons is affected only implicitly, while gravitational waves can display an explicitly modified behavior due to the direct couplings between the metric and the scalar field, which results in a modification of the equations for the propagation of tensors. This opens the possibility of gaining constraining power on DE/MG through probes that combine electromagnetic and GW sources.
 One example of such breaking of degeneracies lies in the cosmological friction term which enters the equation for propagation of photons and GWs on cosmological background; in ST theories, this term is modified by a factor depending on the conformal coupling only for GWs, while it remains unaffected for photons.
This results in a difference between the luminosity distance inferred from electromagnetic sources and the one inferred from the amplitude of GWs (see Sec.~\ref{sec:theory})~\cite{Belgacem:2017ihm,Amendola:2017ovw,Nishizawa:2017nef} and can be used to test theories of gravity as shown in~\cite{Belgacem:2017ihm, Belgacem:2018lbp,LISACosmologyWorkingGroup:2019mwx,Mukherjee:2020mha,Baker:2020apq,Leyde:2022orh, Finke:2021aom, Finke:2021eio, Finke:2021znb}.

In this paper, we focus on the potentialities of weak lensing of GWs (GW-WL). For high enough redshift sources ($z \gtrsim 0.5$), GW-WL can cause up to $\sim 5-10 \%$ distortion in the GW strain, posing a serious limitation to the precision with which the luminosity distance of the sources is measured~\cite{Holz:2005df, Hirata:2010PhRvD}. However, it can also be exploited as an interesting signal~\cite{Camera:2013xfa}, especially when cross-correlated with galaxies~\cite{Mukherjee:2019wcg,Libanore:2021jqv,LISACosmologyWorkingGroup:2022jok}, CMB lensing~\cite{Mukherjee:2019wfw}, HI intensity mapping~\cite{Scelfo:2021fqe, Scelfo:2022lsx} and other GW probes~\cite{Congedo:2018wfn,Mpetha:2022xqo}. Cross-correlation of the GW sources with galaxies is shown to be a powerful technique for several other science topics ranging from cosmology and astrophysics~\cite{Oguri:2016dgk, Calore:2020bpd,Mukherjee:2020hyn, Scelfo:2020jyw,Scelfo:2021fqe,Diaz:2021pem} to its recent application to GWTC-3~\cite{Mukherjee:2022afz}.
Building on the relativistic corrections to the luminosity distance of GWs in MG,  we introduce a lensing convergence estimator for GWs which contains explicitly  a contribution from the conformal coupling characteristic of ST theories, thus maximizing the constraining power on cosmological parameters when cross correlating GW-WL with the galaxy fields (GC,WL). 

As we will discuss in our analysis, an accurate measurement of the GW-WL signal will require the detection of numerous GW events, as well as high precision in the determination of the GW source luminosity distance and in the inference of its redshift. We choose to focus this work on bright GW sources, i.e. sources for which the redshift is known with high accuracy, and we consider separately the cases in which the source redshift is observed photometrically or spectroscopically.
Additionally, we decide not to specialize to a given GW detector with its noise and sensitivity. Rather, we investigate the requirements on both the number of GW sources and the tolerance in the luminosity distance measurement uncertainty for GW-WL to contribute significantly to the GC+WL signal, improving on the constraints on gravity that will become available with upcoming galaxy surveys. We then compare our benchmarks with the current observational expectations for planned GW detectors.

We show that the information contained in GW-WL and it cross-correlations with the galaxies can push cosmological bounds significantly beyond the capabilities of galaxy surveys, but only if high precision measurement of distances are possible for bright sirens, in addition to the detection of a consistent statistics of events. Though we have limited our work to bright GW events and its synergies with galaxies, combination with other EM probes can also be explored, and can further improve the measurement of beyond-GR effects (see e.g.~\cite{Mukherjee:2019wfw}).

The paper is organized as follows: in Section~\ref{sec:gravitationa_waves} we shortly review the principal features of GW observations;
in Section~\ref{sec:theory} we setup our theoretical framework and define the estimator that can be used to extract the lensing convergence from GW data; in Section~\ref{sec:observables} we present the observables used to probe the cosmological models, while in Section~\ref{sec:analysis} we report the details of our analysis, and collect our results in Section~\ref{sec:results}; finally in Section~\ref{sec:conclusions} we draw our conclusions.

\section{Gravitational waves observation}
\label{sec:gravitationa_waves}

General relativity predicts two polarization modes, $h_+$ and $h_\times$, for the strains of GW emitted from coalescing compact objects. The strain for a binary system of masses $m_1$ and $m_2$ can be modelled in terms of the redshifted chirp mass $\mathcal{M}_z= (1+z)\mathcal{M} \equiv (m_1m_2)^{3/5}/(m_1+m_2)^{1/5}$. Assuming GR, and under the approximation of circular orbits, at leading order in frequency this reads
\begin{eqnarray}\label{eq:gw_strain}
    h_{+}(f) &\propto& \, \frac{1+\cos^2\iota}{2d_L^{\rm gw}} \, \big(\mathcal{M}_z^{5/6}\big) \big(f^{-7/6}\big) \, e^{i\psi_+(\mathcal{M}_zf)},\\
    h_{\times}(f) &\propto& \, \frac{\cos \iota}{d_L^{\rm gw}} \, \big(\mathcal{M}_z^{5/6}\big) \big(f^{-7/6}\big) \, e^{i\psi_{\times}(\mathcal{M}_zf)},
\end{eqnarray}

where $\iota$ is the inclination angle and $\psi$ denotes the phase of the GW strain, with $\psi_{\times} = \psi_+ + \pi/2$. The observed GW signal $h_{\rm obs} (t)= F_+(t) h_+(t) + F_\times(t) h_\times (t)$ is a linear composition of the two polarization strain of the GW modulated by the antenna pattern $F_{+,\times}$. Thus, the luminosity distance $d_L$ to the GW sources can be measured directly from the GW signal, as it is inversely proportional to the amplitude of the strain~\cite{Schutz:1986gp}. Its accuracy can be increased by combining the two polarizations to mitigate the degeneracy between $d_L$ and the inclination angle. This same degeneracy can also be reduced through the measurement of higher modes of the GW signal~\cite{Varma:2016dnf,Varma:2014jxa} or exploiting external information such as, in case the binary merger is accompanied by  the emission of a relativistic jet, the angle of the jet with respect to the line of sight~\cite{Guidorzi:2017ogy,Mooley:2018qfh,Hotokezaka:2018dfi,Wang:2020vgr}

As the GW strain can be parameterized in terms of the redshifted masses of the binary objects, GW sources cannot probe independently the redshift of the source unless there is a known mass-scale or physical scale which can be used to break the degeneracy between mass and redshift ~\cite{Taylor:2012db, Farr:2019twy, 2021PhRvD.104f2009M, Mukherjee:2021rtw, Ezquiaga:2022zkx}. For some GW sources, a detectable electromagnetic counterpart can potentially be observed and used to measure its redshift directly with a spectroscopic (or photometric) follow-up of the host galaxy. Among these sources are BNS like GW170817, whose coincident jet emission has been observed as a gamma-ray burst~\cite{LIGOScientific:2017vwq, LIGOScientific:2017ync, LIGOScientific:2017zic} and later in X-rays~\cite{Troja:2017nqp} and radio~\cite{Hallinan:2017woc, Mooley:2018qfh, Ghirlanda:2018uyx}, while the kilonova emission resulting from the synthesis of heavy elements in the merger ejecta has been followed up in optical and near-infrared~\cite{LIGOScientific:2017ync, Pian:2017gtc, Smartt:2017fuw, Kasen:2017sxr}. NS-BH are expected to produce EM emission through similar mechanisms~\cite{Janka:1999qu}, while stellar origin and intermediate mass BBH that merge in a gas-rich environment, such as an AGN disk, could emit X-rays in the late inspiral phases as a consequence of super-Eddington accretion~\cite{Miller:2004bu, Caputo:2020irr}. Finally, MBHB embedded in accretion disks can potentially emit X-rays and gamma rays immediately prior to or after the merger through two main channels: the perturbations induced on the disk by the recoil kick imparted on the merger product by the GW emission~\cite{Schnittman:2008ez, Rossi2010} and the possible formation of relativistic jets~\cite{Palenzuela:2010nf, Moesta:2011bn, Gold:2014dta, Khan:2018ejm}.
These sources are typically dubbed 'bright' sirens.
Beside the hunt for EM signatures accompanying the GW emission, if the direction of the incoming GW is sufficiently well localized it is in principle possible to identify the galaxy hosting the merging binary using the GW signal alone. In practice, this is not going to be possible with third-generation detectors, as they are expected to localize GW event in a sky area of several squared degrees, encompassing thousands of galaxies~\cite{Maggiore:2019uih, Ronchini:2022gwk, Pieroni:2022bbh}. However, fourth-generation detectors like DECIGO~\cite{Kawamura:2020pcg}, BBO~\cite{Crowder:2005nr} and ALIA~\cite{Baker:2019ync} are expected to provide localization areas of the order of squared arcminutes, even arcseconds in the best scenarios~\cite{Baker:2019ync, Crowder:2005nr}, which will largely facilitate the identification of individual host galaxies within the catalogues of surveys like Euclid or LSST.

For the purposes of this study, we will focus on bright standard sirens, broadly defined as GW detections for which a photometric or spectroscopic redshift can be measured, including the case of host identification when possible.

However, if the GW sources are such that a direct EM counterpart can be expected, then the total number and redshift distribution of bright detections will depend on various astrophysical assumptions. These assumptions concern the EM emission mechanism, the intensity and frequency of the emitted EM radiation and the availability and characteristics of the EM detectors tasked with the follow-up observation of the GW signal. Similarly, if the redshifts are determined through the identification of the host galaxy, the redshift distribution of bright (in the broad sense that we intend here) sources will depend on the completeness and depth of the galaxy catalogues used to identify the hosts.

Realistically, we can envision a future scenario in which the final catalogue of GW events usable for analyses similar to the one presented in this paper will be made up of sources whose redshifts have been measured in different ways. These include both host identification and direct EM counterparts that are emitted through different astrophysical channels. Thus, we do not make any restrictive assumptions regarding a specific type of counterpart observation. Rather, we directly make an assumption on the redshift distribution of the final catalogue of GW sources. We detail such assumption in Sec.~\ref{subsec:analysis_surveys}, and investigate its impact on the cosmological constraints in App.~\ref{app:redshifts}.

Since the weak lensing phenomenon will affect GW events generated by binary systems of all masses, a higher statistics of events can be stacked by combining observations from space-based detectors such as LISA~\cite{LISA:2017}, Taiji~\cite{Ruan:2018tsw}, TianQin~\cite{TianQin:2020hid} or DECIGO, and ground-based detectors like LVK~\cite{KAGRA:2013rdx}, CE~\cite{Reitze:2019iox, Hall:2019} and the ET~\cite{Punturo:2010zz}. These observatories can detect numerous bright GW sources, among which BNS, NS-BH and massive BBH. Moreover, a limited number of these sources might be observed in different frequency bands by both space and ground-based detectors, whose observations can then be combined to reduce uncertainties in the inference of the GW source parameters~\cite{Klein:2022rbf}. This combined analysis is expected to provide better determination of the luminosity distance measurement, though it might be applicable only to a small number of events~\cite{Seto:2022xmh, Moore:2019pke, Gerosa:2019dbe}.


\section{Theoretical framework}
\label{sec:theory}

With the new generation of LSS and GWs data, we will have the opportunity to test gravity on cosmological scales with unprecedented precision. Alternatives to \lcdm{} introducing a dynamical dark energy or modifications of GR, will be explored and constrained.    
In this work, we focus on dynamical models of dark energy and modifications of GR that could be relevant at late times, on large cosmological scales. While the theory landscape is vast, most of the models in this category introduce a propagating scalar degree of freedom, either directly or modifying GR equations from constraints to dynamical ones~\cite{Clifton:2011jh}. In order to satisfy some minimal stability requirements, they must lead to second order equations of motion for the propagating degrees of freedom: one massless tensor field (with the two standard polarization modes) and one scalar. Models that meet these characteristics are broadly referred to as \emph{scalar-tensor} theories, or Horndeski gravity~\cite{Horndeski:1974wa}.

We restrict our study to Horndeski models in which tensor modes propagate luminally, $i.e.$ $c_T^2=1$, at all redshifts, satisfying the bound from GW170817~\cite{LIGOScientific:2017zic} as studied in~\cite{Ezquiaga:2017ekz,Creminelli:2017sry, Baker:2017hug}. In fact, with the advent of space based GW detectors, constraints on $c_T$ could be further narrowed through multi-wavelength observations, see e.g.~\cite{Baker:2022eiz}.
Rather than choosing a specific Horndeski model, we opt for a more agnostic exploration based on the effective field theory of dark energy (EFT of DE)~\cite{Cheung:2007st,Creminelli:2008wc,Park:2010cw,Bloomfield:2011np,Gubitosi:2012hu,Piazza:2013coa} which spans a vast range of beyond-\lcdm{} models, while containing \lcdm{} as a subcase. For a thorough review of the EFT of DE formalism and its applications to cosmological tests of gravity, we refer the reader to~\cite{Frusciante:2019xia}.  EFT of DE offers a powerful unifying framework in terms of a handful of free functions of time, commonly referred to as \emph{ EFT functions}. 
A peculiar feature of EFT of DE is that it allows to formulate a set of conditions that the EFT functions need to satisfy in order for the resulting theory to be viable, $i.e.$ for instance to not develop instabilities~\cite{Gubitosi:2012hu,Gleyzes:2013ooa,Frusciante:2016xoj,Frusciante:2018vht,Frusciante:2019xia}. These can be quite powerful in restricting a priori the parameter space for which it makes sense to explore the phenomenology~\cite{Raveri:2014cka,Peirone:2017lgi}.

The EFT of DE action for Horndeski models can be written in unitary gauge, upon identifying the slices of constant time with the hypersurfaces of a uniform scalar field. Imposing second-order equations of motion, and additionally a luminal speed of propagation for tensors,  the resulting quadratic action reads
\begin{align}\label{eq:EFTaction}
  S = \int d^4 x \sqrt{-g} \bigg\{ &\frac{m^2_0}{2} \left[1+ \Omega (\tau) \right]\, R + \Lambda(\tau) - c(\tau) a^2 \delta g^{00} +  \nonumber\\
    &+\gamma_1(\tau) \frac{m^2_0 H^2_0}{2} (a^2 \delta g^{00})^2 - \gamma_2(\tau) \frac{m^2_0 H_0}{2} (a^2 \delta g^{00})\delta K^\mu_\mu  \bigg\}\, +  S_m [g_{\mu\nu}, \chi_i].
\end{align}
where $m_0^2=(8\pi G)^{-2}$ and $\delta g^{00}, \delta K^\mu_\mu$ are, respectively, the reduced Planck mass, the perturbations of the time-time component of the metric, and the trace of the perturbations to the extrinsic curvature of constant-time hypersurfaces; $S_m$ is the action for matter fields $\chi_i$, minimally coupled to the metric (for more details on the construction of this action see~\cite{Bloomfield:2011np,Gubitosi:2012hu}).  
The free functions of time $\Omega(a), \Lambda(a), c(a)$ and $\gamma_1(a),\gamma_2(a)$ are the EFT functions; the first three affect the dynamics both of the background and linear cosmological perturbations, while the latter two affect only perturbations. \lcdm{} is included in this framework, and it corresponds to the choice $\Lambda(a) = const$, with the rest of EFT functions being zero.

Different EFT functions correspond to different characteristics of the theory: the non-minimal coupling $\Omega(a)$, leads to a running Planck mass;  the kineticity $\gamma_1(a)$, quantifies the independent dynamics of the scalar field; the braiding $\gamma_2(a)$, broadly signals a coupling between the metric and the scalar degree of freedom. Note that we adopt the convention of~\cite{Hu:2013twa, Hu:2014oga} for the EFT functions. Alternative conventions are found in the literature, most commonly the so-called $\alpha_i$ parametrizations~\cite{Bellini:2014fua} in terms of $\{\alpha_M,\alpha_K, \alpha_B\}$ (while $\alpha_T=c_T^2 -1$ is zero for our case) for which there is a simple direct correspondence with  $\{\Omega, \gamma_1, \gamma_2\}$ (see e.g.~\cite{Frusciante:2019xia} and references therein). For reference, non-minimally coupled quintessence, f(R) gravity and Brans-Dicke theories would be characterized by the equation of state parameter of DE, $w_{\rm DE}$, and $\Omega$, while having $\gamma_1=\gamma_2=0$; k-essence would further have $\gamma_1\neq 0$ and k-mouflage correspond to all functions being non-zero. 

\subsection{Luminosity distance in Scalar-Tensor theories.}
\label{subsec:theory_gw}

In scalar-tensor scenarios, the non-minimal coupling,  parameterized by $\Omega(z)$,  induces an additional dissipation term in the equation for propagation of GWs on the cosmological background ~\cite{LISACosmologyWorkingGroup:2019mwx,Bellini:2014fua,Belgacem:2017ihm}. This is not the case for photons; the dependence of their luminosity distance on the expansion rate remains unaltered because photons do not couple directly to the scalar field~\cite{Belgacem:2017ihm,Amendola:2017ovw,LISACosmologyWorkingGroup:2019mwx}. It follows that the luminosity distance to a given redshift differs if determined via an electromagnetic or GW wave, according to:
\begin{equation}\label{eq:DlGW}
     {\bar d}^{\rm gw}_L (z) = \sqrt{1 + \Omega(z)} \: {\bar d}^{\rm \,em}_L(z)\,.
\end{equation}
where
\begin{equation}\label{eq:distance-to-z}
  \bar d^{\rm \,em}_L(z) = (1+z) \chi(z) = \frac{(1+z)}{H_0} \int_{0}^z \, \frac{d z'}{E(z')}\, ,
\end{equation}
with $E(z)\equiv H(z)/H_0$.
The non-minimal coupling $\Omega(z)$ in Eq.~\eqref{eq:DlGW}, is more commonly written in terms of the running Planck mass as $ M_P(z) = M_P(z=0) \sqrt{1+ \Omega(z)}$.

On their journey through the expanding Universe, photons and GWs encounter structure which induces direction-dependent corrections to their amplitude, and correspondingly inferred luminosity distance. These include  weak lensing, Doppler, Sachs-Wolfe, Integrated Sachs-Wolfe, Shapiro time delays and volume effects on the propagating waves, which translate into corrections to the luminosity distance to the corresponding source~\cite{Bertacca:2018,Laguna:2009re}. The  luminosity distance becomes then a function not only of redshift but also of angular direction, i.e.
\begin{equation}
    d_L^{\rm \,gw} (z, \hat n) = \bar d_L^{\rm \,gw} (z) + \Delta \, d_L^{\rm \,gw} (z, \hat n)\,,
\end{equation}
where we have assumed that the effects due to the inhomogeneities are small compared to the background luminosity distance so that
$\Delta \, d_L^{\rm \,gw} (z, \hat n)$ is the correction due to large scale structures.

For a detailed derivation of relativistic corrections to $d_L^{\rm \,gw}$ in scalar-tensor theories we refer the reader to~\cite{Garoffolo:2019mna}. These have been computed only in the subclass of Horndeski theories with $c^2_T = 1$, and this is the reason we consider the same set of theories. They  were derived, using the covariant formulation of Horndeski gravity, and have been subsequently mapped into the EFT of DE language, as well as implemented in \texttt{EFTCAMB} in~\cite{Garoffolo:2020vtd}.  

Depending on the angular scale and redshift, some of the relativistic effects can be dominant. For instance, in~\cite{Bertacca:2018,Dalal:2002wh,Holz:2005df} it was shown that for GW sources at redshifts higher than $z\sim 0.5$, WL is the dominant LSS correction and can quickly reach $\sim 5-10 \%$ of the GW strain amplitude.
Given its substantial magnitude, WL can also be a valuable signal to be exploited, rather than only a source of error. Indeed, in this work we use it to probe the growth history and pattern of the LSS in the Universe which depends on the gravitational model taken into consideration. 
Since WL is an integrated effect it builds up along the propagation. For this reason, we focus on GW sources at high redshift, reaching up to $z\simeq 2.5$, for which we can approximate
 \begin{equation}
    \frac{\Delta d_L^{\rm \,gw}}{\bar{d}_L^{\rm \,gw}} \simeq - \kappa_{\rm gw}\,,
    \label{eq:dl_gw_correction}
\end{equation}
where $\kappa_{\rm gw}$ is the lensing convergence field of GWs, that for a populations of sources reads
\begin{equation}
\kappa_{\rm gw} (\hat{n}) = -\frac{1}{2}\nabla^2_{\theta} \phi_L(\hat{n}) = - \int_{0}^{\chi} \frac{d\chi'}{\chi'} \int_{\chi '}^{\infty} d\chi_* \Bigg(\frac{\chi_* - \chi '}{\chi_*}\Bigg) \frac{dn_{\rm gw}}{d\chi_*} \nabla^2_{\theta} \,\Psi_W (\chi '\hat{n}, z')
\label{eq:lensing_convergence}
\end{equation}

where $dn_{\rm gw}/d\chi_*$ is the distribution of the GW sources, $\Psi_W $ is the Weyl potential and $\nabla^2_{\theta}$ indicates the 2D Laplacian with respect to the angle between the image and optical axis.  
Note that we are neglecting the shear deformations of the signal since these are subdominant in linear perturbation theory where WL mainly affects the magnification of the GWs. 
The convergence $\kappa_{\rm gw}$ can probe the growth of matter perturbations in the Universe because it is given in terms of the Weyl potential, which in addition is sensitive to modifications of gravity affecting the Poisson equation.

\subsection{Lensing Convergence estimators for Gravitational Waves}
\label{subsec:theory_estimators}

In the case of bright events, building on  Eq.~\eqref{eq:dl_gw_correction} we can construct an estimator to extract the WL convergence from GW data as follows
  \begin{equation}
    \hat{\kappa}_{\rm gw} (z, \hat n) \equiv 1 - \frac{d_L^{\rm \,gw}(z, \hat n)}{{\bar  d}_L^{\rm \,em}(z)}\,,
    \label{eq:convergence_estimator_bs_def}
  \end{equation}
namely as the fractional difference between the GW luminosity distance $d_L^{\rm \,gw} (z, \hat  n)$, inferred from the measurement of the GW strain, and the electromagnetic background luminosity distance  ${\bar  d}_L^{\rm \,em}(z)$, obtained from the known redshift via Eq.~\eqref{eq:distance-to-z} by choosing a fiducial cosmological model.
This estimator can be biased by three factors: the experimental error on the $d_L^{\rm \,gw} (z, \hat  n)$ measurement, the error on the source redshift, or the choice of the wrong fiducial model. Indeed, the calculation of $\bar{d}_L^{\rm \, em}$ depends not only on the measured $z$, but on a set of fiducial cosmological parameters that enter Eq.~\eqref{eq:distance-to-z} and whose chosen values affect the final result. In other words, the uncertainties on the parameters that characterize the fiducial model propagate, through Eq.~\eqref{eq:distance-to-z}, into an uncertainty on $\bar{d}_L^{\rm \, em}$.
To account for the sources of bias mentioned above we introduce three parameters $\epsilon_{\rm gw}$, $\epsilon_z$ and $\epsilon_c$, respectively representing the uncertainties due to the $d_L^{\rm \,gw}$ measurement error, the redshift error and the fiducial model, and modify the convergence estimator as

\begin{equation}
    \hat{\kappa}_{\rm gw} = 1 - \frac{\sqrt{1+\Omega (z)} \, (1 - \kappa_{\rm gw} + \epsilon_{\rm gw})\, \bar{d}_L^{\rm \,em}}{(1+\epsilon_z + \epsilon_c)\,\bar{d}_L^{\rm \,em}} \sim 1 - \sqrt{1+\Omega(z)} \, (1 - \kappa_{\rm gw} + \epsilon_{\rm gw} - \epsilon_z - \epsilon_c)\,,
  \label{eq:convergence_estimator_bs}
 \end{equation}

where we linearized to first order in $\kappa_{\rm gw}$, $\epsilon_{\rm gw}$, $\epsilon_z$ and $\epsilon_c$, assuming $\kappa_{\rm gw}$ to be small and of the same order as the $\epsilon_i$.\\
The estimator in Eq.~(\ref{eq:convergence_estimator_bs_def}) crucially relies on the availability of the information on the redshift of the source, to disentangle the lensing convergence from the measured luminosity distance. This is one of the reasons that prompted us to restrict the analysis to bright events. Not only, accurate redshift measurements also allow for a very fine binning of sources in redshift space. Exploiting the weak lensing measurement from dark events - which are expected to be significantly higher in number than bright sources - is not impossible, and has been tackled in the context of 3x2 point statstics of GW sources e.g. in~\cite{Namikawa:2015prh}, but it requires to work in luminosity distance space. This carries limitations due to the experimental uncertainty on $d_L$, which imposes a much larger bin size and consequent loss of information, as the lensing correlation gets partially smoothed out in wider tomographic bins (see Appendix~\ref{app:binning}). Arguably, however, the greatest advantage of employing bright events for our probes lays in the accessibility of smaller angular scales, where the lensing correlation is stronger, as we discuss in Sec.~\ref{sec:analysis}.
We show the effectiveness of $\hat \kappa_{\rm gw}$ for cross-correlation purposes in the next section.


\section{Tomographic Observables}
\label{sec:observables}

We consider cross-correlations between the density field of galaxies,  $\delta_{\rm g}$, the weak lensing convergence fields as measured by galaxies, $\kappa_{\rm g}$, and the weak lensing convergence field as measured by GW, $\kappa_{\rm gw}$.  The angular power spectrum for the cross-correlations  is~\cite{Euclid:2019clj}
\begin{equation}\label{eq:generalCl}
      C_{ij}^{XY}(\ell) = \int_0^{z_{\rm max}}{\frac{{\rm d}z}{\chi^2(z)H(z)}W_{X}^i(k(\ell,z),z)W_{Y}^j(k(\ell,z),z)P_{\rm P}\left(k(\ell,z), z \right)}\, ,
  \end{equation}
where $X,Y=[\delta_{\rm g},\kappa_{\rm g},\kappa_{\rm gw}]$ at the $i$th and $j$-th tomographic bin, $P_{\rm p}(k)\propto A_{\rm s}(k/k_*)^{n_{\rm s}-1}$ is the primordial power spectrum, with $A_{\rm s}$ and $n_{\rm s}$ its amplitude and spectral index and $k_*$ a pivot scale, and we applied the Limber and flat-sky 
approximations~\cite{Kaiser:1991qi,LoVerde:2008re,Kitching:2016zkn,Kilbinger:2017lvu,Lemos:2017arq,Matthewson:2020rdt}, which sets $k(\ell,z)=(\ell+1)/\chi(z)$. For brevity, we indicate $k(\ell,z)$ as $k$ throughout the rest of the paper\footnote{Note that, with respect to the approach of~\cite{Euclid:2019clj}, in Eq.~(\ref{eq:generalCl}) we absorb the transfer functions in the window functions $W_{X}^i$, which makes them depend explicitly on the scale $k$ and makes the integral depend on the primordial power spectrum $P_{\rm P}$, rather than on the matter power spectrum $P(k,z)$.}.
$W_{X_i}(k,z)$ is the window function for the observable $X$ in the $i$-th tomographic bin. For galaxies, the window function can be written as
\begin{equation}\label{eq:galaxywin}
    W^i_{\delta_{\rm g}}(k,z) = T_\delta(k,z) b^i_{\rm g}(z)n^i_{\rm g}(z) H(z)\,,
\end{equation}
where $T_\delta(k,z)$ is the matter transfer function, evolving the primordial power spectrum such that $P_{\delta}(k,z)=T^2_\delta(k,z)P_{\rm p}(k)$.  
In Eq.~(\ref{eq:galaxywin}), $n_{\rm g}^i(z)$ and $b_{\rm g}^i(z)$ are, respectively, the galaxy redshift distribution and the linear galaxy bias in the $i$-th redshift bin which we model following~\cite{Euclid:2019clj}, $i.e.$ assuming the bias is constant in each redshift bin, with the values used as a free nuisance parameter in our analysis.

The window function of the GW convergence we adopt is
\begin{equation}\label{eq:kappaGWwin}
    W^i_{\kappa_{\rm gw}}(k,z) = T_{\Psi_W}(k,z)\int_{z}^{\infty} {\rm d}z'\, \frac{\chi(z')-\chi(z)}{\chi(z')} n^i_{\rm gw}(z')\,,
\end{equation}
where $n_{\rm gw}^i(z)$ is the redshift distribution of GW sources in the $i$-th bin, and $T_{\Psi_W}(k,z)$ is the Weyl potential transfer function, which allows to get its power spectrum $P_{\Psi_W}(k,z)=T^2_{\Psi_W}(k,z)P_{\rm p}(k)$. The galaxy WL convergence window function is
\begin{equation}\label{eq:kappaGALwin}
    W^i_{\kappa_{\rm g}}(k,z) = T_{\Psi_W}(k,z) \int_{z}^{\infty} {\rm d}z'\, \frac{\chi(z')-\chi(z)}{\chi(z')} n^i_{\rm g}(z') +W^i_{\rm IA}(k,z) \,,
\end{equation}
where, with respect to Eq.~(\ref{eq:kappaGWwin}) we introduce an extra term $W_{\rm IA}^i(k,z)$ to include the Intrinsic Alignment (IA) systematic effect that we model following~\cite{Euclid:2019clj}, $i.e.$
\begin{equation}\label{eq:IAwin}
    W_{\rm IA}^i(k,z) = -T_\delta(k,z)\frac{A_{\rm IA}C_{\rm IA}\Omega_{\rm m,0}F_{\rm IA}(z)}{D(z)}n_{\rm g}^i(z) H(z)\,,
\end{equation}

where $D(z)$ is the growth factor, $\Omega_{\rm m,0}$ the current matter density, and the subscript IA highlights the terms including the nuisance parameters $A_{\rm IA}$, $\beta_{\rm IA}$ and $\eta_{\rm IA}$, the last two being contained in the $F_{\rm IA}$ function, while $C_{\rm IA}=0.0134$ is a fixed constant. 

Scalar-tensor theories modify the growth pattern of perturbations, affecting differently the different observables. The window functions just described will be correspondingly modified, mainly through the transfer functions $T_\delta(k,z)$ and $T_{\Psi_W}(k,z)$, which encode the evolution of density perturbation and lensing potential, respectively. Modifications of the growth also affect the growth factor $D(z)$ entering the IA contribution to the power spectra. All these effects are included in our angular power spectra, and the cosmological functions needed for the calculations, i.e. the functions describing the background expansion ($H(z)$ and $\chi(z)$) and those related to the perturbation evolution (the growth factor $D(z)$ and the transfer functions $T_\delta(k,z)$ and $T_{\Psi_W}(k,z)$) are obtained numerically via \texttt{EFTCAMB}~\cite{Hu:2013twa, Hu:2014oga}, which modifies the public code \texttt{CAMB}~\cite{Lewis:1999bs,Howlett:2012mh} to include modified gravity models.

Finally, we model the distribution of sources (${\rm src}$) in each redshift bin as
\begin{equation}\label{eq:convsourcedist}
    n_{\rm src}^i(z) = \frac{dn_{\rm src}}{dz} \Bigg[\text{Erf}\Bigg(\frac{z-z_-^i}{\sqrt{2}\sigma^{\rm src}_z(z)}\Bigg) - \text{Erf}\Bigg(\frac{z-z_+^i}{\sqrt{2}\sigma^{\rm src}_z(z)}\Bigg)\Bigg]\,,
\end{equation}
with ${\rm src}=\left[{\rm g},{\rm gw}\right]$, $z_-^i$ and $z_+^i$ the lower and upper limits of the $i$-th bin, $\sigma^{\rm src}_z(z)$ the error on the redshift measurement for the considered source, and $dn_{\rm src}/dz$ is parameterised as
  \begin{equation}
    \frac{dn_{\rm src}}{dz} \propto \left(\frac{z}{z_0}\right)^2 \exp\Bigg[-\Bigg(\frac{z}{z_0}\Bigg)^{3/2}\Bigg]\,.
    \label{eq:sources_distribution}
  \end{equation}
The parameter $z_0$ entering Eq.~\eqref{eq:sources_distribution}, as well as the redshift error $\sigma_z^{\rm src}$ of Eq.~\eqref{eq:convsourcedist}, are survey dependent and we will specify them in the following Section (Sec.~\ref{sec:analysis}) where we introduce the surveys considered in this study.\\

We now focus on the GW lensing auto-correlation, $C^{\kappa_{\rm gw} \kappa_{\rm gw}}({\ell})$,  and its cross-correlations with galaxies, $C^{\kappa_{\rm gw}\delta}({\ell})$ and $C^{\kappa_{\rm gw} \kappa_{g}}({\ell})$.
In the case of GW events with electromagnetic counterpart, to build these correlations we  can use the weak lensing convergence estimator of Eq.~\eqref{eq:convergence_estimator_bs}. 
In this case, since the three sources of error $\{\epsilon_{\rm gw}$,$\epsilon_z$,$\epsilon_c\}$ are not correlated between themselves or with either the convergence or the contrast density field, we have

\begin{subequations}
\label{eq:convergence_brackets}
\begin{align}
\label{eq:convergence_brackets_ini}
\langle \hat{\kappa}_{\rm gw}\, \delta_{\rm g} \rangle  &\simeq \sqrt{1+\Omega} \langle \kappa_{\rm gw} \delta_{\rm g} \rangle\,,\\
\langle \hat{\kappa}_{\rm gw} \, \kappa_{\rm g} \rangle &\simeq \sqrt{1+\Omega} \langle \kappa_{\rm gw} \kappa_{\rm g} \rangle\,,\\
\langle \hat{\kappa}_{\rm gw} \, \hat{\kappa}_{\rm gw} \rangle &\simeq \begin{aligned}[t]
 &(1+\Omega) \left[\langle \kappa_{\rm gw} \kappa_{\rm gw} \rangle + \langle\epsilon_{\rm gw} \epsilon_{\rm gw}\rangle + \langle\epsilon_z \epsilon_z\rangle\right] = \\
  & = (1+\Omega)\langle \kappa_{\rm gw} \kappa_{\rm gw}\rangle\, + N_{\rm gw} + N_z,
\end{aligned}
\label{eq:convergence_brackets_end}
\end{align}
\end{subequations}

where we have kept only terms up to second order in perturbations and used that $\langle \epsilon_{\rm gw}\rangle = \langle \epsilon_z \rangle = 0$ when the average is performed over large volumes. On the contrary, in general $\langle \epsilon_c \rangle \neq 0$, hence in a complete analysis the term proportional to $\langle \epsilon_c \rangle$ should be accounted for by introducing a proper additional term to the correlations noise. However, this correction is typically small and for the purposes of this work we will assume the noise to be dominated by other terms, that we model in Sec.~\ref{sec:analysis}. Hence, we neglected the contributions $\langle \epsilon_c \rangle$, together with those $\propto \langle \epsilon_c \, \epsilon_c \rangle$ which are second order.

Exploiting the results of Eq.~\eqref{eq:convergence_brackets}, the angular power spectrum in Eq.~(\ref{eq:generalCl}) can be re-written as
    \begin{equation}\label{eq:generalCl_modified}
      \hat C^{XY}_{ij} (\ell) = \int_0^{z_{\rm max}}{\frac{{\rm d}z}{\chi^2(z)H(z)}\, \left[\Theta_X(z) \Theta_Y(z)\right] \, W_{X}^i(k,z)W_{Y}^j(k,z)P_{\rm p}\left(k\right)}\, ,
  \end{equation}
where again $X,Y=[\delta_{\rm g},\kappa_{\rm g},\hat{\kappa}_{\rm gw}]$, $\Theta_{\hat{\kappa}_{\rm gw}}(z) = \sqrt{1+\Omega(z)}$ and $\Theta_{[\delta,{\kappa}_{\rm g}]}(z) = 1$.
We point out that now the angular correlations involving GW include factors $\sqrt{1+\Omega(z)}$, which can make these probes promising for capturing deviations from \lcdm{}.
This multiplicative factor is even squared in the case of the auto-correlation of the GW weak lensing field.


\section{Models, analysis method and surveys specifications} \label{sec:analysis}

This section provides details of the gravitational models on which we forecast constraints, as well as the details on the analysis we performed to obtain them. Furthermore, we also provide the galaxy and GW survey specifications that are considered.

\subsection{Models}\label{subsec:analysis_models}

The models we investigate in this work belong to the Horndeski class of theories introduced in Sec.~\ref{sec:theory} and described by the action of Eq.~\eqref{eq:EFTaction}. We have already discussed at length the role of $\Omega(a)$ in the previous Sections. As for $\gamma_1(a)$, past works have shown that it affects only negligibly linear perturbations in the range of our observables~\cite{Pogosian:2016pwr,Frusciante:2018jzw,Bellini:2015xja} and does not enter into the modified equations for the propagation of tensors~\cite{Frusciante:2019xia}. The function $\gamma_2(a)$ is non-zero in models that have a kinetic mixing between the scalar field and the metric~\cite{Frusciante:2019xia}; similarly to $\gamma_1$, it  does not enter explicitly in the linear equation for tensors, but contrary to $\gamma_1$ it has a non-negligible effect on the LSS~\cite{Pogosian:2016pwr,Frusciante:2018jzw,Bellini:2015xja} hence we expect our observables to be sensitive to it.

We  work in the \emph{designer} approach ~\cite{Hu:2014oga} and fix the background expansion history to a choice for the equation of state $w_{\rm DE}$. In this framework, $\Lambda$ and $c$ become superfluous, as they can be determined in terms of the Hubble parameter $\mathcal{H}(a)$ and the non-minimal coupling $\Omega(a)$, and the model is then fully specified by the EFT functions $\left\{w_{\rm DE}, \Omega,\gamma_1,\gamma_2\right\}$.
We model the dark energy equation of state using the CPL parametrization~\cite{Chevallier:2000qy,Linder:2002et}
\begin{equation}
\label{eq:omega_rde_parametrization}
 w_{\rm DE}(a) = w_{0} + w_{a}(1-a)
\end{equation}
With this prescription, the continuity equation dictates that the background DE density evolves as
\begin{equation}
    \rho_{\rm DE}(a)=\rho^0_{\rm DE} a^{-3 (1+w_0+w_a)}e^{-3w_a(1-a)} \, ,
\end{equation}
with $\rho^0_{\rm DE}=\rho_{\rm DE}(a=1)$ its value today.
We choose to parameterize the time dependencies of $\Omega(a)$, $\gamma_1(a)$ and $\gamma_2(a)$ as
\begin{equation}\label{eq:parEFfunction}
 \Omega(a) = \Omega_0 \, \frac{\rho_{\rm DE} }{\rho^0_{\rm DE}}\,, \qquad  \gamma_1(a) = \gamma^0_1 \, \frac{ \rho_{\rm DE}}{\rho^0_{\rm DE}}\,, \qquad  \gamma_2(a) = \gamma^0_2 \, \frac{ \rho_{\rm DE}}{\rho^0_{\rm DE}}\,
\end{equation}
considering that any modification to \lcdm{} is expected to be small when $\rho_{\rm DE}$ is not the dominant contribution to the energy density budget of the Universe.
We consider three scenarios:

\begin{itemize}
\item \textbf{\lcdm{}:} in this case, the cosmological model is determined by the standard set of cosmological parameters. This means that the set of free parameters in the analysis will consist of the matter and baryonic relative energy densities at present time, $\Omega_{\rm m,0}$ and $\Omega_{\rm b,0}$, the reduced Hubble constant $h$, the spectral index of the primordial power spectrum $n_{\rm s}$ and the amplitude  $\sigma_8$ of the linear matter power spectrum measured at $8/h \, {\rm Mpc}$.
We consider all MG parameters fixed to their GR values, i.e. $w_0=-1$ and $w_a = \Omega_0= \gamma_1^0=\gamma_2^0=0 $;

\item \textbf{Model I (M1):} this model allows us to study theories falling within the Generalized Brans-Dicke class. In this class, the background expansion history deviates from that of $\Lambda$CDM and depends on the specific parameters of the selected model. As we aim here for a generic approach we leave the background parameters $w_0$ and $w_a$ free to vary, we allow for a non-minimal coupling ($\Omega_0\neq0$ and free to vary), and keep the standard $\Lambda$CDM free to vary. In these kind of models there is no derivative interaction, thus we set $\gamma_1^0 = \gamma_2^0=0$;

\item \textbf{Model II (M2):} where the set of available parameters includes $\{w_0, w_a, \Omega_0, \gamma_2^0\}$ and those of \lcdm{}. The background expansion history follows the same parametrization of M1, but with a different fiducial for the DE equations of state, and $w_0$ and $w_a$ are free to vary. M2 represents the class of kinetic gravity braiding~\cite{Deffayet:2010qz} models, where we allow for both a conformal coupling and derivative coupling ($i.e.$ $\Omega\neq 0$, $\gamma_2\neq 0$). For this model, we also take a non-zero $\gamma_1(a)$, which is necessary to ensure the mathematical stability of this theory around the assumed fiducial. As past works have shown that LSS are not sensitive to $\gamma_1$, the corresponding parameter is treated as a constant in the analysis.

\end{itemize}

\begin{figure}
    \centering
    \includegraphics[width=0.495\textwidth]{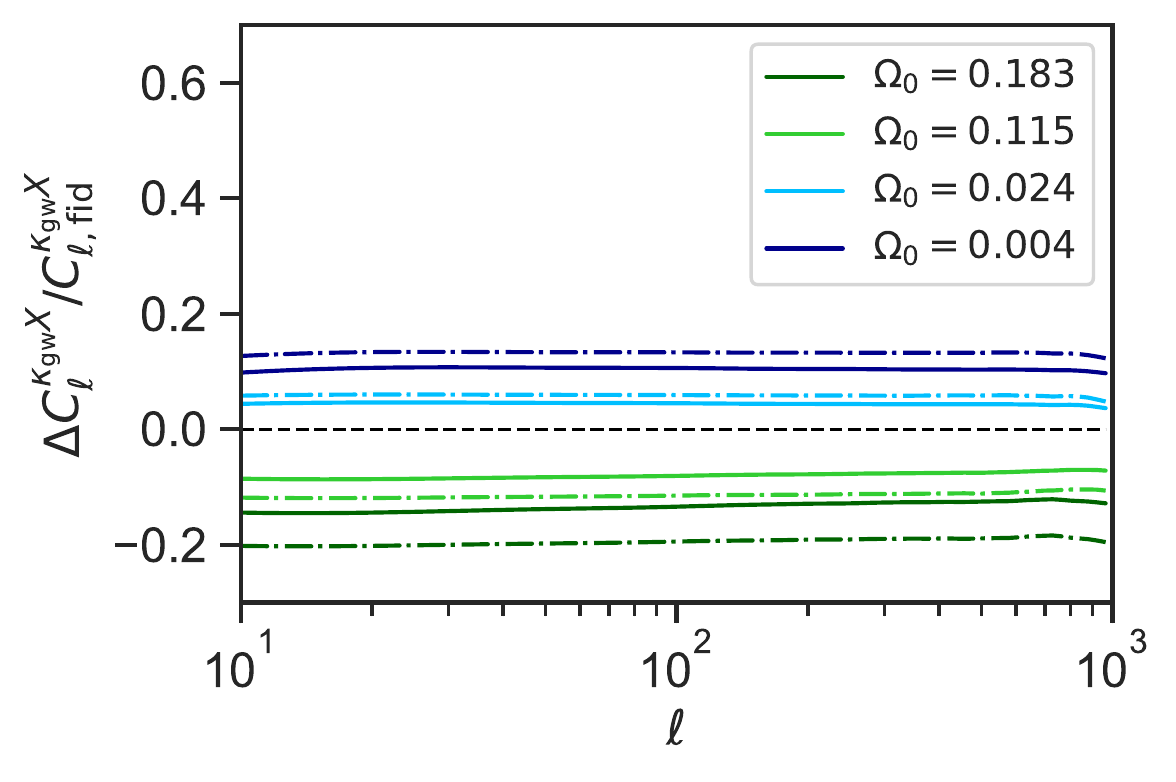}
    \includegraphics[width=0.485\textwidth]{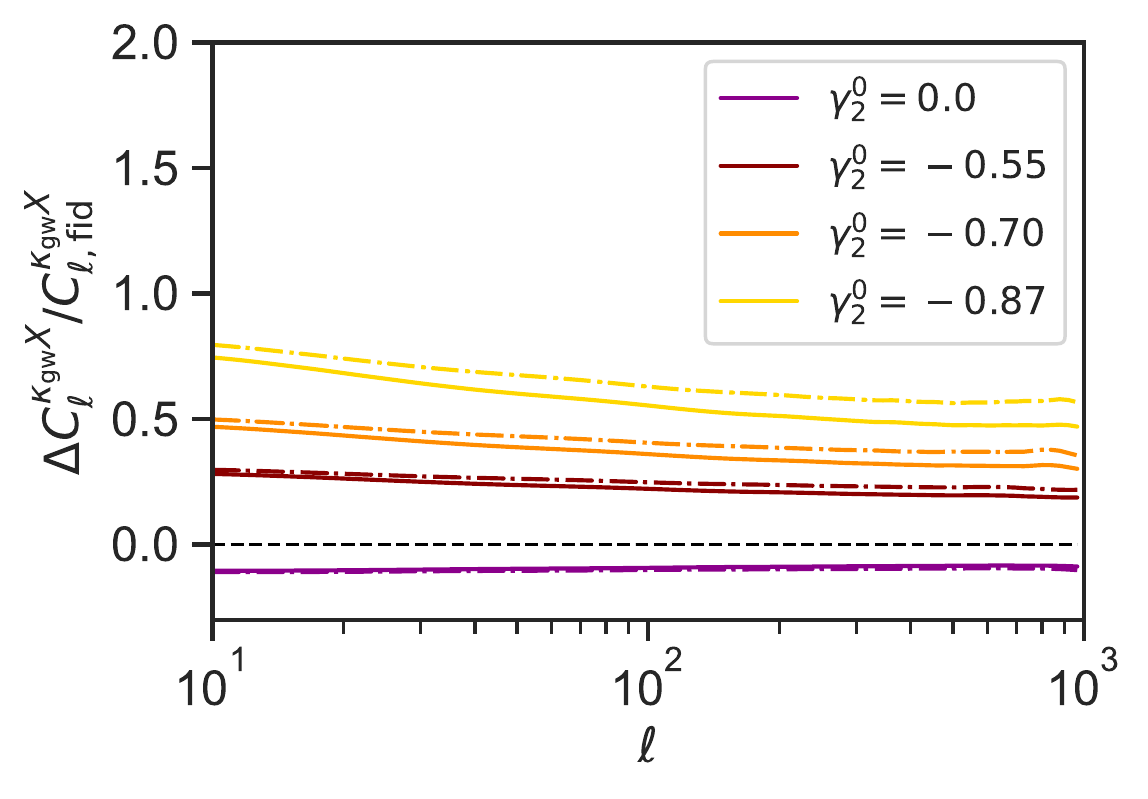}
    \caption{Relative deviations $(C^{\kappa_{\rm gw}X}_{\ell}-C^{\kappa_{\rm gw}X}_{\ell, \rm fid})/C^{\kappa_{\rm gw}X}_{\ell, \rm fid}$ as a function of multipole $\ell$ for model M2, where $C^{\kappa_{\rm gw}X}_{\ell, \rm fid}$ are the correlations computed at the fiducial cosmology. We plot in the left (right) panel how the $C_{\ell}$ vary for different values of $\Omega_0$ ($\gamma_2^0$), while we keep all other parameters fixed at the fiducial value. In each figure, the solid lines correspond to auto-correlations of GW-WL ($X=\kappa_{\rm gw}$), while the dashed lines mark cross-correlations of GW-WL with galaxy WL ($X=\kappa_g$).}
    \label{fig:theoretical_cls}
\end{figure}
\begin{table}
    \renewcommand{\arraystretch}{1.5}
    \centering
    \begin{tabular}{|l|c|c|c|c|c|c|c|c|c|c|}
        \hline
        Model & $h$ & $\Omega_{m,0}$ & $\Omega_{b,0}$ & $n_s$ & $\sigma_8$ & $w_0$ & $w_a$ & $\Omega_0$ & $\gamma_1^0$ & $\gamma_2^0$ \\
        \hline
        \hline
        \lcdm{} & 0.6774 & 0.31 & 0.05 & 0.9667 & 0.8159 & (-1.0) & (0.0) & (0.0) & (0.0) & (0.0) \\
        Model I & 0.6774 & 0.31 & 0.05 & 0.9667 & 0.8159 & -0.946 & -0.098 & 0.018 & (0.0) & (0.0) \\
        Model II & 0.6774 & 0.31 & 0.05 & 0.9667 & 0.8159 & (-0.94) & (-0.31) & 0.047 & (4.4) & -0.23 \\
        \hline
    \end{tabular}
    \caption{Fiducial parameter values for the cosmological models considered in this work. The numbers in parenthesis correspond to parameters that were kept fixed in that specific model, while all other parameters were constrained simultaneously. Along with these parameters, we also let free to vary the galaxy bias in each bin and the IA parameters, for which we use the same fiducial values as ~\cite{Euclid:2019clj}}
    \label{tab:models_fiducial}
\end{table}

We refer the reader to  Tab.~\ref{tab:models_fiducial} for a complete list of the fiducial values chosen for the different parameters. Parameters that are kept fixed in the analysis are indicated in parentheses. Our set of fiducial \lcdm{} parameters are chosen to be compatible with current constraints from LSS and CMB~\cite{Planck:2018vyg,Frusciante:2018jzw}. Existing bounds on the DE equation of state and the MG parameters come from a number of observations, including GW~\cite{Leyde:2022orh, Mancarella:2021ecn, Ezquiaga:2021ayr}, Big Bang Nucleosythesis and CMB measurements~\cite{Uzan:2010pm}, local tests of gravity~\cite{Brax:2012gr,Joyce:2014kja,Perenon:2015sla} and theoretical constraints obtained imposing on MG theories a set of stability requirements~\cite{Perenon:2015sla}.
However, constraints on the non-\lcdm{} parameters, and especially on the MG parameters, are significantly model dependent, as the values of the EFT functions today are partially degenerate with their assumed evolution in time and with the assumed model for the DE equation of state. As an example, in~\cite{Frusciante:2018jzw} $\Omega_0$ is constrained to be positive if its redshift evolution is the same as Eq.~\eqref{eq:parEFfunction}, while negative values for $\Omega_0$ are found to be favoured when the non-minimal coupling is assumed to evolve as a power law of the scale factor. Oftentimes, bounds are also placed on a re-parameterization of the EFT action in Eq.~\eqref{eq:EFTaction} in terms of a different set of EFT functions, commonly dubbed $\{\alpha_i\}$~\cite{Bellini:2014fua, Huang:2015srv, SpurioMancini:2019rxy, Noller:2018wyv, Bellini:2015xja, Traykova:2019oyx, Planck:2018vyg} or on a set of phenomenological functions introduced specifically to capture deviations from \lcdm{} in observations~\cite{Amendola:2007rr, Planck:2018vyg, Simpson:2012ra, Ferte:2017bpf}. These functions - and the resulting bounds - can be mapped into our set $\{\Omega, \gamma_i\}$ (see e.g.~\cite{Hu:2014oga}), though the mapping is not trivial. For a complete overview of the existing cosmological bounds on EFT parameters, encompassing different parametrizations of the EFT action and several models, we refer the interested reader to the extensive review work of~\cite{Frusciante:2019xia}. Our chosen fiducial values correspond to the best fit values obtained in~\cite{Frusciante:2018jzw}, where the non-\lcdm{} parameters were constrained against current cosmological data from CMB, BAO, SNIa and LSS. This work assumes the parametrization~\eqref{eq:parEFfunction} for the EFT functions, and finds (at $2\sigma$ significance) $w_0 = -0.946^{+0.090}_{-0.060}$, $w_a = -0.098^{+0.25}_{-0.28}$ and $\Omega_0 = 0.018^{+0.032}_{-0.019}$ for our Model I, and $w_0 = -0.94^{+0.15}_{-0.13}$, $w_a = -0.31^{+0.48}_{-0.63}$, $\Omega_0 = 0.047^{+0.068}_{-0.051}$, $\gamma_1^0>0.295$ and $\gamma_2^0 = -0.23^{+0.26}_{-0.32}$ for our Model II. These are, to our knowledge, the most stringent constraints to date corresponding to these models.

We sketch the impact of the EFT parameters on the $C^{XY}(\ell)$ in Fig.~\ref{fig:theoretical_cls}, where we plot the relative difference $(C^{\kappa_{\rm gw}X}_{\ell} - C^{\kappa_{\rm gw}X}_{\ell, \rm fid})/C^{\kappa_{\rm gw}X}_{\ell, \rm fid}$ adopting model M2 and choosing as baseline the cross-correlation computed in the fiducial cosmology. We plot curves for $X=\kappa_{\rm gw}$ (solid lines) and $X=\kappa_{\rm g}$ (dashed lines) choosing different values of $\Omega_0$ in the left panel and of $\gamma_2^0$ in the right panel, while all other parameters are kept fixed at the fiducial value. We observe that taking $\Omega_0$ lower than the fiducial value (and closer to $0$) acts to strengthen the signal, while values higher than the fiducial diminish the signal. On the contrary, higher values of $\gamma_2^0$ and closer to $0$ tend to dampen the signal, while lower values enhance it. The values chosen for the left panel of Fig.~\ref{fig:theoretical_cls} span the full range available to $\Omega_0$ in this model, up to twice the $2\sigma$ constraints of~\cite{Frusciante:2018jzw}, while causing variations in the cross-correlations of, at most, $20\%$. Similarly, values for $\gamma_2^0$ reaching down to twice the $2\sigma$ bounds of~\cite{Frusciante:2018jzw} can cause variations of up to $70\%$ in the correlations. In Fig.~\ref{fig:theoretical_cls}, we excluded positive values of $\gamma_2^0$, as we found that, if all other parameters remain set at the fiducial values, a positive $\gamma_2^0$ is theoretically forbidden for this model, giving rise to theories that are mathematically unstable.

\subsection{Analysis method}\label{subsec:analysis_method}

To compute forecasts, we adopt the  Fisher matrix formalism\citep{Tegmark:1996bz, Euclid:2019clj}, which allows us to obtain bounds on the free parameters of the analysis from the information matrix $\mathcal{F}_{\alpha\beta}$. Following~\cite{Euclid:2019clj}, we define the Fisher matrix as
\begin{equation}
    \mathcal{F}_{\alpha\beta} = \sum_{\ell=\ell_{\rm min}}^{\ell_{\rm max}} \frac{2\ell+1}{2} \sum_{i,j,m,n}\sum_{A,B,C,D} \frac{\partial C^{AB}_{ij}(\ell)}{\partial \theta_\alpha} \left[K^{-1}(\ell)\right]^{BC}_{jm} \frac{\partial C^{CD}_{mn}}{\partial \theta_\beta} \left[K^{-1}(\ell)\right]^{DA}_{ni}\,,
    \label{eq:fisher_matrix_lensing}
  \end{equation}
  where $\alpha$ and $\beta$ run over the set of free cosmological parameters $\theta$ (standard and MG), while $A$, $B$, $C$ and $D$ run over the density and convergence fields $[\delta_{\rm g}, \kappa_{\rm g}, \hat{\kappa}_{\rm gw}]$ and finally $i$, $j$, $m$ and $n$ run over all unique pairs of tomographic bins. By the Cramer-Rao inequality, the lower bound on the standard deviation for the parameter $\theta_\alpha$ is
  \begin{equation}
    \sigma_\alpha = \sqrt{\Sigma_{\alpha\alpha}} \,  \quad \text{with} \quad \Sigma_{\alpha \alpha} \equiv [\mathcal{F}^{-1}]_{\alpha\alpha}\,.
  \end{equation}
  To compute our matrices we use {\tt CosmicFish} ~\cite{Raveri:2016xof,Raveri:2016leq}\footnote{\texttt{CosmicFish} is publicly available at \url{https://cosmicfish.github.io/}. However, the version we are using is an update of the public one, which will be released in the near future.}, that we extended to include GW weak lensing as an additional cosmological probe, incorporating also the MG contributions.
The covariance matrix $K$ is defined as
  \begin{equation}\label{eq:fishcovariance}
    K^{AB}_{ij}(\ell) = \frac{C^{AB}_{ij}(\ell) + N^{AB}_{ij}(\ell)}{\sqrt[4]{f^{A}_{\rm sky} f^{B}_{\rm sky}}}\,,
  \end{equation}
where $f^{A}_{sky}$ is the sky fraction covered by the detector measuring the observable $A$. Note that all currently planned GW interferometers present a full-sky area coverage ($f^{\kappa_{\rm gw}}_{sky} = 1$), so there will always be total overlap with the region covered by EM surveys. $N^{AB}_{ij}$ is the noise of the correlation considered, which we model as
\begin{align}
    N^{\delta_g\delta_g}_{ij}(\ell) &= \frac{1}{\bar{n}^i_{\rm g}}\delta_{ij}\,, \label{eq:gc_noise_corr} \\
    N^{\kappa_{\rm g}\kappa_{\rm g}}_{ij}(\ell) &= \frac{\sigma_\epsilon^2}{\bar{n}^i_{\kappa_{\rm g}}}\delta_{ij}\,,\label{eq:wl_noise_corr} \\
    N^{\kappa_{\rm gw}\kappa_{\rm gw}}_{ij}(\ell) &= \frac{1}{\bar{n}^i_{\rm gw}} \Bigg(\frac{\sigma^2_{d_{L}}}{d^2_{L}} + \frac{\sigma^2_s}{d^2_{L}}\Bigg)e^{\frac{\ell^2 \theta^2_{\rm min}}{8 \ln 2}}\delta_{ij}\,,\label{eq:gw_noise_corr}
\end{align}
where $\delta_{ij}$ is the Kronecker delta and $\bar{n}^i_A$ is the number of sources in the $i$-th redshift bin for the probe $A$. Here we assume that  the instrument noise terms of the different probes are not correlated, $i.e.$ $N^{AB}_{ij}(\ell)=0$ for $A\neq B$, which is a reasonable assumption as GW detectors operate in a totally different manner than galaxy surveys. However, the terms in Eq.~\eqref{eq:fishcovariance} related to cosmic variance are still correlated between different probes, $i.e.$ $C^{AB}_{ij}(\ell)\neq0$ even when $A\neq B$. In these noise terms, $\sigma_{\epsilon}$ represents the intrinsic ellipticity affecting shear measurements, $\sigma_{d_{L}}$ represents the average experimental error on the luminosity distance of the GW sources, while $\sigma_s = (\partial d_{L}/\partial z) \, \sigma^{\rm gw}_z$ is the contribution to the luminosity distance error brought by the uncertainty on the merger redshift $\sigma^{\rm gw}_z$, where the propagation is obtained assuming a fiducial cosmology. Lastly, $\theta_{\rm min}$ is the sky-localization area of the GW event, which also dictates the maximum available multipole for the analysis.

Finally, the combination of GW and galaxy tracers can in general be exploited in cosmic variance cancellation techniques for partially canceling the cosmic variance noise in the common Fourier modes~\cite{Schmittfull:2017ffw}. This could further improve the predicted constraining power, though it is currently not included in our Fisher analysis and we leave it for future investigations. 

\subsection{Galaxy and GW surveys}\label{subsec:analysis_surveys}

The above discussion of the observables and methodology is quite generic and can be applied to nearly all of the upcoming galaxy surveys and GW observatories. We describe below the surveys specifications adopted to perform our forecasts. 

For what concerns the galaxy distribution (Eq.~\eqref{eq:sources_distribution}) and the fraction of sky observed, in this work we use the specifications of ~\cite{Euclid:2019clj}, as representative of a Stage IV galaxy survey. We instead take a simpler approach for the photometric redshift error, using Eq.~\eqref{eq:convsourcedist} with $\sigma^{\rm g}_z=0.03(1+z)$, from which we obtain the distribution of sources in ten equipopulated redshift bins with $z\in[0,2.5]$. For the limiting multipoles, we follow the "pessimistic" settings of ~\cite{Euclid:2019clj}, and we use a minimum multipole $\ell_{\rm min}=10$ for both galaxy clustering and weak lensing, while we fix the maximum multipole to $\ell_{\rm max}^{\rm GC}=750$ for the former and $\ell_{\rm max}^{\rm WL}=1000$ for the latter. The $\ell_{\rm min}$ truncation eliminates the large scales, at which the Limber approximation breaks down and further relativistic effects become relevant alongside lensing and should be considered (see e.g.~\cite{Martinelli:2021ahc}). The $\ell_{\rm max}$ limit allows to account for both the uncertainties brought by the modelling of nonlinear scales, and the effects of non-Gaussian terms in the data covariance, which have a different impact on the two observables. However, with respect to the settings in~\cite{Euclid:2019clj}, we further limit the weak lensing multipoles because of the lack, in the MG models considered, of a consolidated recipe for treating small scales, where the perturbations evolution enters the non-linear regime. 

\begin{figure}
\includegraphics[width=0.49\textwidth]{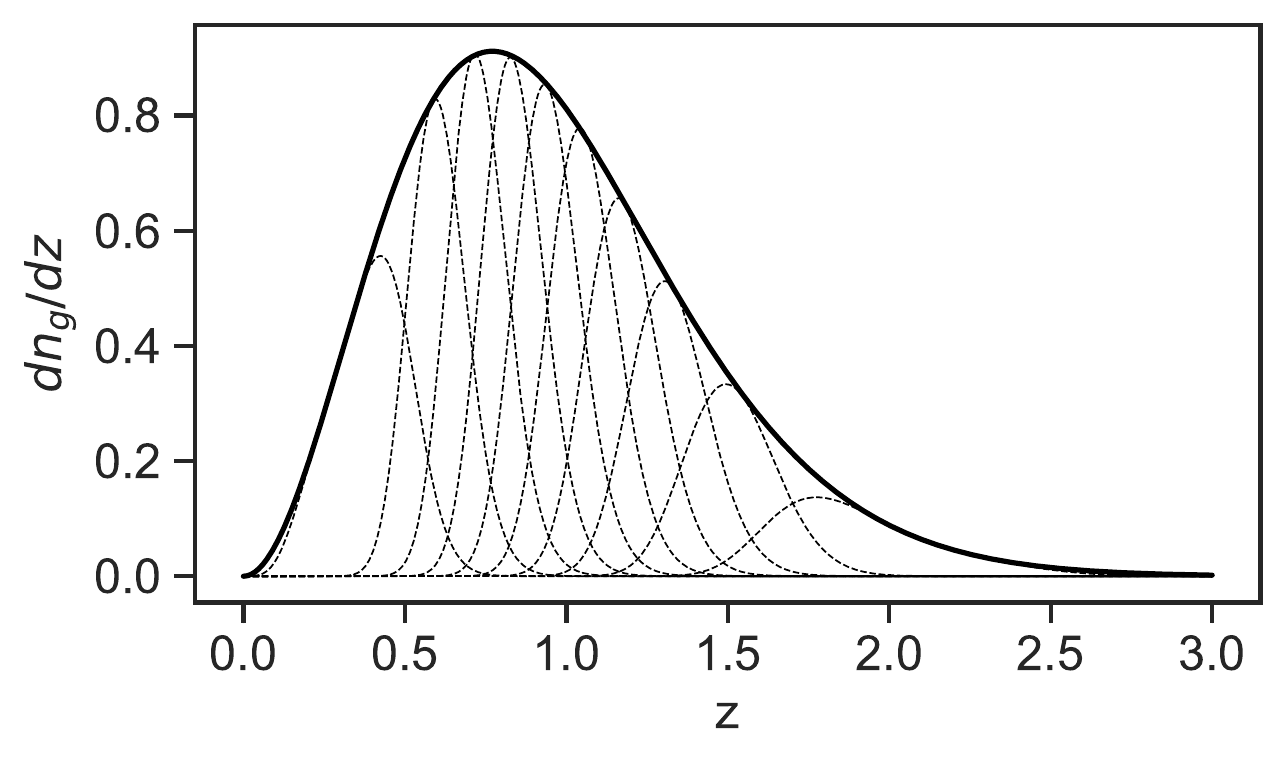}
\includegraphics[width=0.49\textwidth]{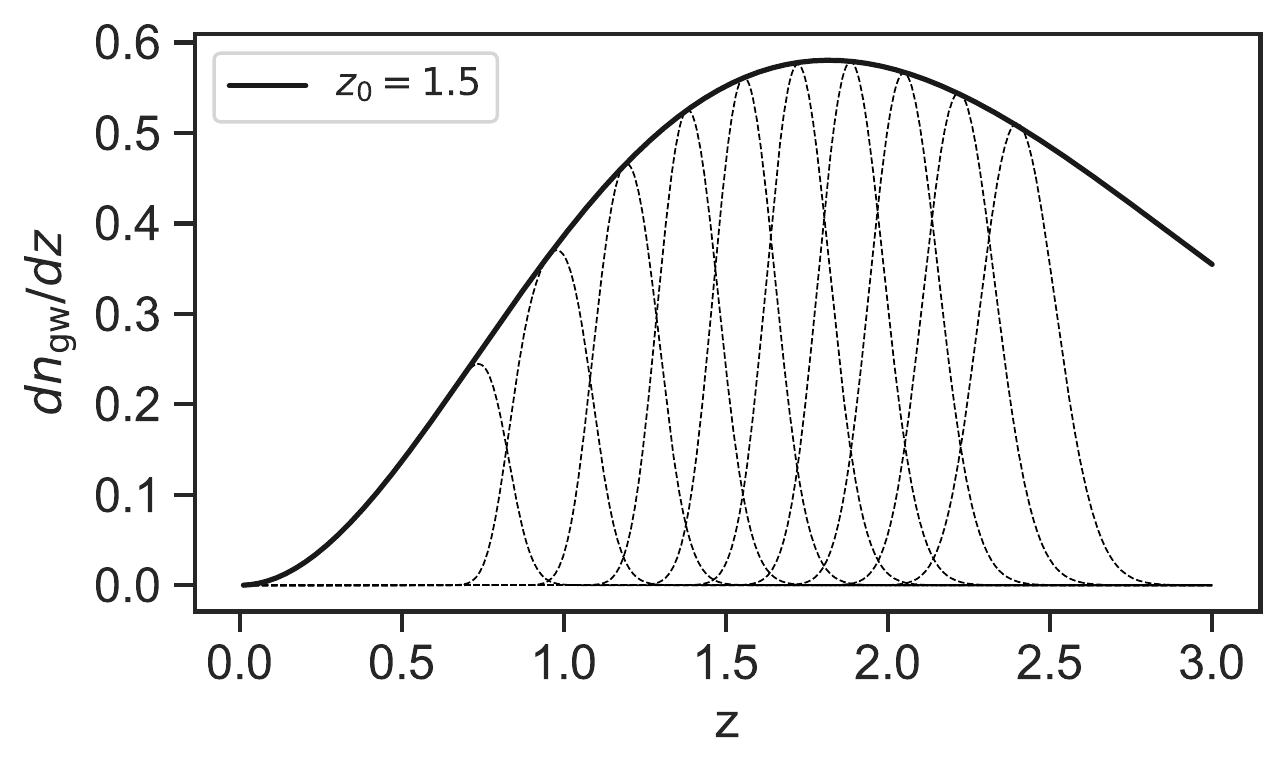}
\caption{Normalized redshift distribution of galaxies and bright GW sources (left and right panel respectively). The dashed lines illustrate the redshift binning applied in our analysis.}\label{fig:distributions}
\end{figure}

Concerning the GW survey, we take the parameter entering the source distribution in Eq.~\eqref{eq:sources_distribution} as $z_0=1.5$. This corresponds to a phenomenological parametrization for the normalized distribution of observed sources that we make compatible with the forecasts of~\cite{Ronchini:2022gwk} for the BNS observed by a network of ET in combination with two CE detectors. This parametrized redshift distribution is also similar to the one obtained in~\cite{Tamanini:2016uin} for LISA luminous MBHB observations (i.e. MBHB with a detectable counterpart), though in that case the events number count peaks at higher redshifts, $z_0\sim 2.5$. Even if our parametrization fails to accurately describe LISA sources, this is not extremely concerning for the purposes of the present study, as we expect - and will demonstrate below - that GW-WL will be effective in constraining cosmologies only if a high statistics of events is available. As current LISA forecasts at best predict only a few hundreds of total MBHB merger events over the span of 10 years~\cite{LISA:2022yao}, stellar mass binaries are likely to remain the primary target for GW-WL studies for the foreseeable future. Nonetheless, the class of experiments sensitive to these sources still encompasses most of the currently planned GW detectors, thus we feel safe in choosing $z_0=1.5$ without signifcant loss in generality. We characterize the impact on our analysis of choosing different values for $z_0$ in App.~\ref{app:redshifts}.

As in the case of galaxies, we consider $z \in [0,2.5]$ and bin the GW sources in ten equipopulated redshift bins. We also explore different binning choices: we consider the cases with only eight and six equipopulated bins. We observe a very small deterioration in the cosmological constraints when decreasing the number of tomographic bins, as larger bins tend to smooth out the lensing signal, as we show in Appendix~\ref{sec:binning}. However, our conclusions on the impact of GW-WL on the constraining power remains the same regardless of binning, thus in what follows we will report results only for the most favourable choice of 10 bins.

The main quantities that can impact the weak lensing estimation from bright GW sources are the uncertainty on the luminosity distance, the error on the source redshift and the total number of bright GW events. The peculiar velocity correction for sources at high redshift are going to be a less significant contamination~\cite{Mukherjee:2019qmm}. Without specializing on the specifics of any particular detector, we consider several possibilities for the total number of detected GW events, $N_{\rm gw}$, ranging from $10^3$ to $10^6$, while we vary the average precision on the luminosity distance, $\sigma_{d_L}/d_L$, from $10 \%$ down to $0.5 \%$. For what concerns the redshift error, we make the requirement that our sources will either have a detectable electromagnetic counterpart or an identifiable host galaxy, from which we can measure the redshift photometrically, with error $\sigma^{\rm gw}_z = 0.03 (1+z)$, or spectroscopically, with error $\sigma^{\rm gw}_z = 0.0005 (1+z)$~\cite{DESI:2016fyo, Laureijs2011, LSSTDarkEnergyScience:2018jkl, Salvato:2019}.
In Fig.~\ref{fig:distributions} we show the normalized source distribution of galaxies (left panel) and bright GW events (right panel) adopted in this analysis, together with a representation of the redshift binning that is applied, and we summarise the specifications considered for galaxy and GW surveys in Table~\ref{tab:specifications}.

As $N_{\rm gw}$ is the total number of events, $\bar n^i_{\rm gw}$ in the noise Eq.~\eqref{eq:gw_noise_corr} can be written as the product between $N_{\rm gw}/4\pi$ and the integral of the normalized source distribution over each bin.
Since we are focusing on bright sources, we set $\theta_{\rm min}=0$ in Eq.~\eqref{eq:gw_noise_corr}, thus implying that the event is perfectly localized by identifying the EM counterpart. This implies, in principle, that the GW analysis could include very small scales (very high values of $\ell$), since the exponential term in Eq.~\eqref{eq:gw_noise_corr} which dampens the signal-to-noise ratio at high multipoles has no effect.
Nevertheless, in our analysis we truncate the summation in Eq.~\eqref{eq:fisher_matrix_lensing} at $\ell_{\rm max}= 1000$ to match the limiting multipoles of the galaxy survey, for analogous reasons. This choice is equivalent to limit angular scales to $\theta_{\rm min}\sim 11 \, {\rm arcmin}$.
Going  up to $\ell_{\rm max} > 1000$ would result in a further enhancement of the signal-to-noise ratio. This on one hand would imply tighter constraints on cosmological parameters when all sources are combined. At the same time though, the GC and WL signal themselves would benefit from the accessibility of higher multipoles, resulting in tighter galaxy-only constraints to begin with. Thus, we can expect that, qualitatively, the impact of GW on the galaxy bounds remains similar even if higher multipoles are accessible.\\

Pushing $\ell$ to such high values is a significant advantage of considering bright sirens. Including dark events limits significantly the highest reachable multipole, depending on the sky-localization $\Omega$ of the GW events as $\ell_{\rm max} \sim 180/\sqrt{\pi \Omega}$. For example, for a 3G network composed by one ET and one or two CE detectors the localization of most sources is forecasted to be within $1 \, {\rm deg}^2  < \Omega < 100 \, {\rm deg}^2$~\cite{Maggiore:2019uih, Ronchini:2022gwk, Pieroni:2022bbh}, which implies $10 \leq \ell_{\rm max} \leq 100$. These scales are barely below the threshold at which the lensing correlations signal becomes relevant, thus reducing hugely the signal-to-noise. On the other hand, dark detections are expected to be much more numerous and to reach higher redshift depth, two factors which in contrast increase the SNR. It is not trivial to predict to which extent these advantages can counterbalance the lost multipoles. Preliminary to this work, we have conducted approximate explorations suggesting that the gain in constraining power coming from the higher statistics and redshifts might not be enough to compensate for the loss due to the inaccessibility of the smaller scales, at least until 4th generation detectors with very small localization volumes are considered. However, more work needs to be dedicated to properly characterize the potentialities of dark sirens, and we leave the extension of our analysis to include dark events for future investigation.

\begin{table}
    \centering
    \renewcommand{\arraystretch}{1.5}
    \begin{tabular}{|c|c|c|c|c|c|c|}
    \hline
    \multicolumn{7}{|c|}{Galaxy Clustering}\\
    \hline
     $f_{\rm sky}^{_{\rm g}}$ & $\sigma^{\rm g}_z$ & \multirow{2}{*}{-} & $z_0$ &  $\bar n_{g}$ [arcmin$^{-2}$] & $\ell_{\rm min}$ & $\ell_{\rm max}$\\
     $0.35$ & $0.03(1+z)$ &  & $0.9/\sqrt{2}$& 30 & 10 & 750 \\
    \hline
    \hline
    \multicolumn{7}{|c|}{Galaxy Weak Lensing}\\
    \hline
     $f_{\rm sky}^{\kappa_{\rm g}}$ & $\sigma^{\rm g}_z$ &  $\sigma_\epsilon$ & $z_0$ & $\bar n_{\kappa_{\rm g}}$ [arcmin$^{-2}$] & $\ell_{\rm min}$ & $\ell_{\rm max}$\\
     $0.35$ & $0.03(1+z)$ & $0.3$ & $0.9/\sqrt{2}$ & 30 & 10 & 1000 \\
    \hline
    \hline
    \multicolumn{7}{|c|}{Bright GW Weak Lensing}\\
    \hline
     $f_{\rm sky}^{\kappa_{\rm gw}}$ & $\sigma^{\rm gw}_z$& $\sigma_{d_L}/d_L (\%)$ & $z_0$ & $N_{\rm gw}$ & $\ell_{\rm min}$ & $\ell_{\rm max}$\\
     \multirow{2}{*}{$1$} & $0.03(1+z)$  & \multirow{2}{*}{$\in[0.5,10]$} & \multirow{2}{*}{1.5} & \multirow{2}{*}{$\in[10^2,10^6]$} & \multirow{2}{*}{10} & \multirow{2}{*}{1000} \\
     & $0.0005(1+z)$ & & & & & \\
    \hline

    \end{tabular}

    \caption{Summary of the parameters entering the noise modeling for the probes considered and the binned source distributions of Eq.~(\ref{eq:convsourcedist}). Parameters whose values change through the different analyses are reported with their range of variation.}
    \label{tab:specifications}
\end{table}

\section{Results}\label{sec:results}

Using the specifications, models and methodology described in Sec.~\ref{sec:analysis}, we include in the Fisher matrix the contributions from all correlators $C_{\ell}^{XY}$ with $X$ and $Y$ in $[\delta_{\rm g}, \kappa_{g}, \hat{\kappa}_{\rm gw}]$, to explore  the joint constraining power of GW-WL and galaxy surveys over cosmological parameters. We collect the fiducial values chosen for the parameters of the different models in Tab.~\ref{tab:models_fiducial}. We choose the fiducial cosmology in such a way that the parameters' values fall within the stable region of the models considered (identified via the stability sampler of \texttt{EFTCAMB}) and that the numerical derivatives, needed to obtain the Fisher matrices using Eq.~\eqref{eq:fisher_matrix_lensing}, can be performed without exiting this region.

In our analysis, we account for degeneracies between parameters by always attempting to constrain simultaneously the highest number of parameters possible for each model. For example, we always keep all the standard cosmological parameters free even when the GW-WL is used as a single probe, even though, since WL is mostly sensitive to the combination of $\Omega_{\rm m}$ and $\sigma_8$, we know a priori that it will not be much constraining of other parameters such as, e.g., $\Omega_{\rm b}$. We choose to marginalize over these parameters rather than fixing them, not to neglect any possible degeneracy, as we will show and discuss below. However, in a few cases we instead decide to fix a parameter to its fiducial value, and put bounds only on the remaining set. Additionally, we always let free to vary the galaxy bias in each bin $b^i_{\rm g}$ and the IA parameters $A_{\rm IA}$, $\beta_{\rm IA}$ and $\eta_{\rm IA}$, for which we use the same fiducial values as ~\cite{Euclid:2019clj}.

\subsection{\lcdm{}}
\label{subsec:results_lcdm}

\begin{figure}
    \includegraphics[width=0.5\textwidth]{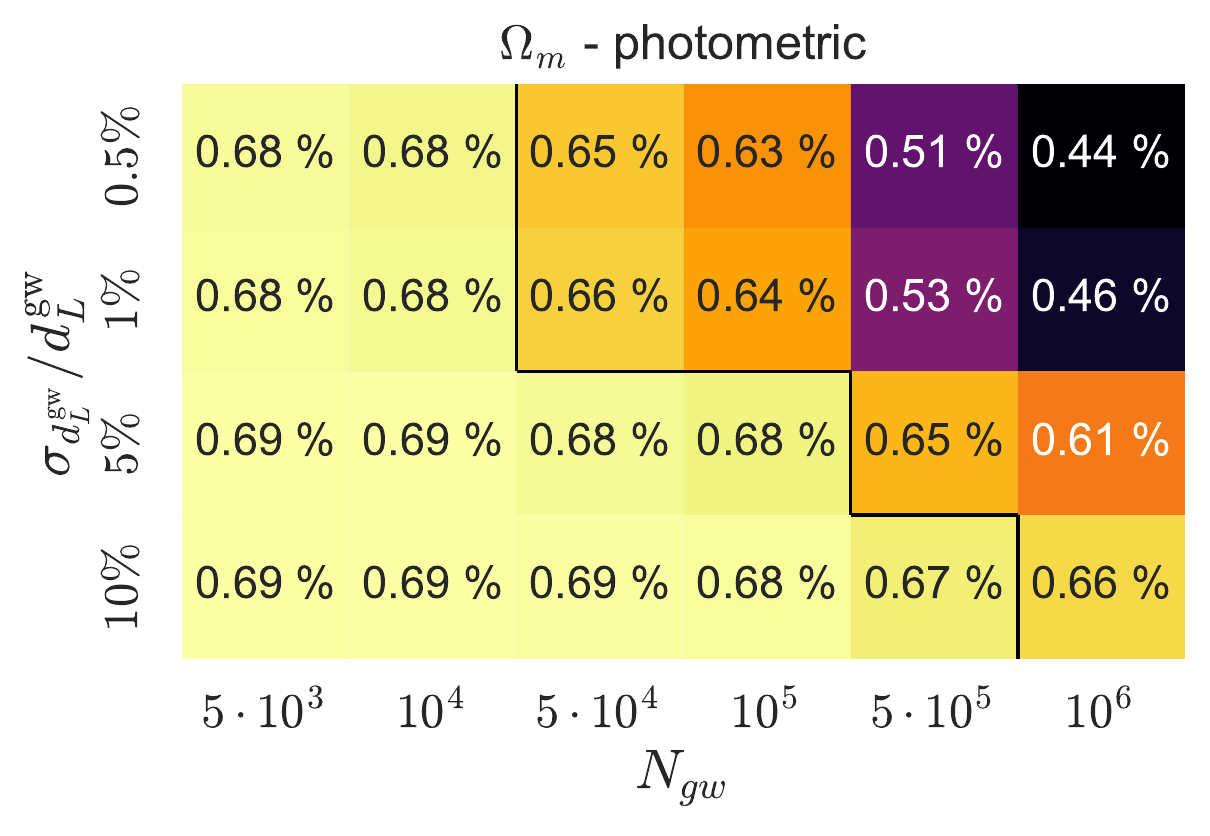}
    \includegraphics[width=0.5\textwidth]{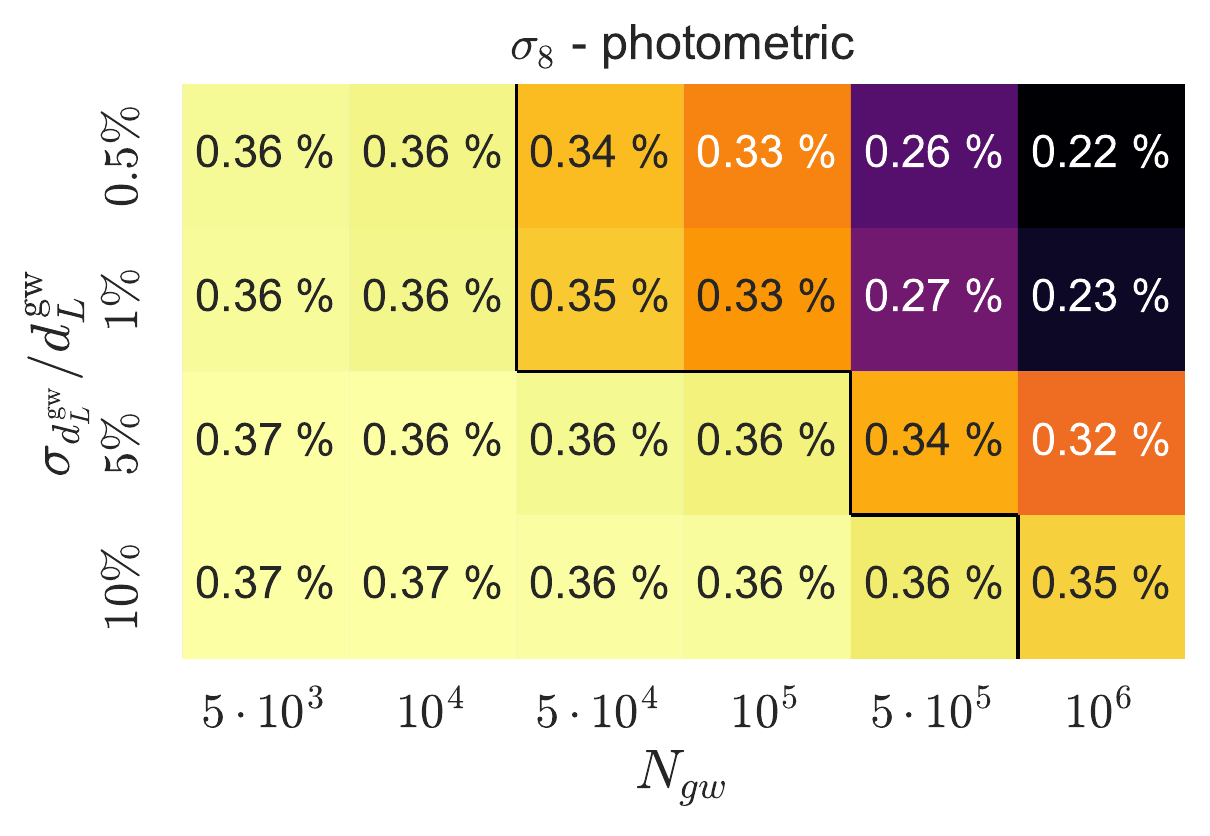}    
    \includegraphics[width=0.5\textwidth]{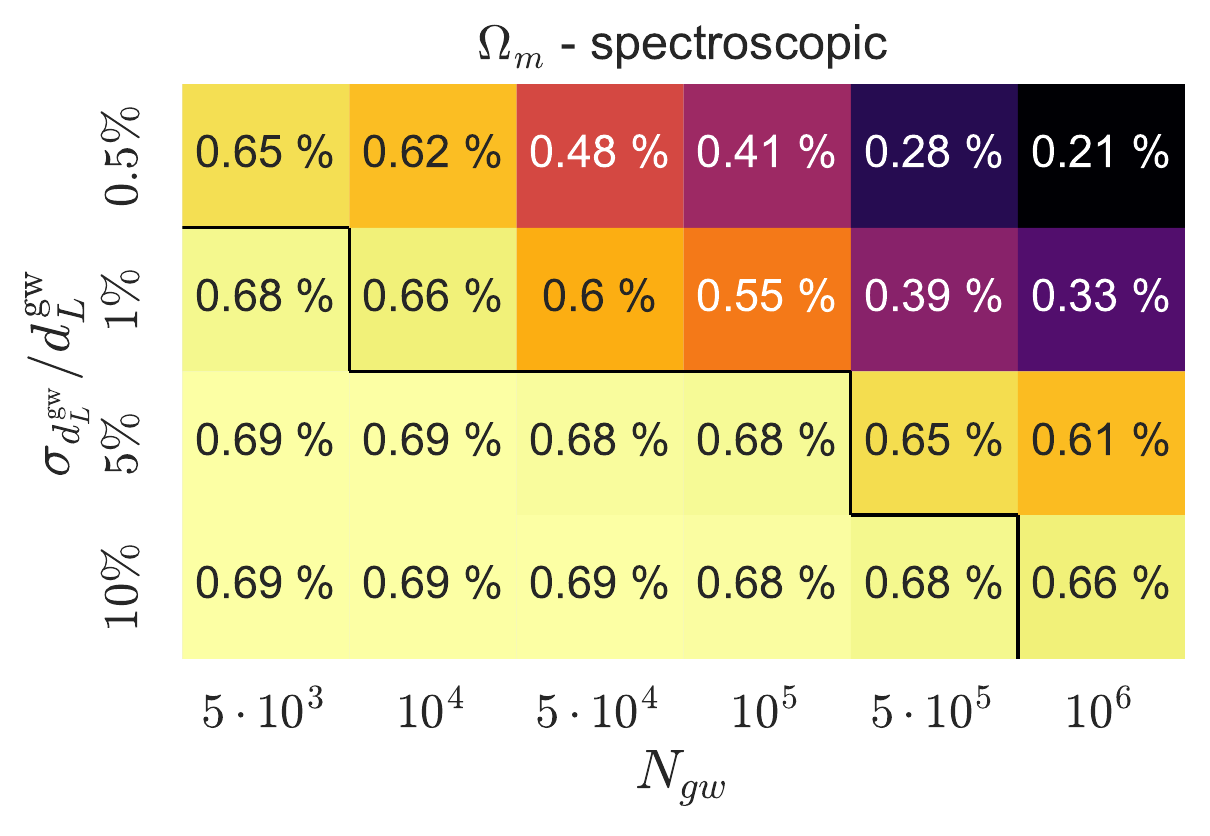}
    \includegraphics[width=0.5\textwidth]{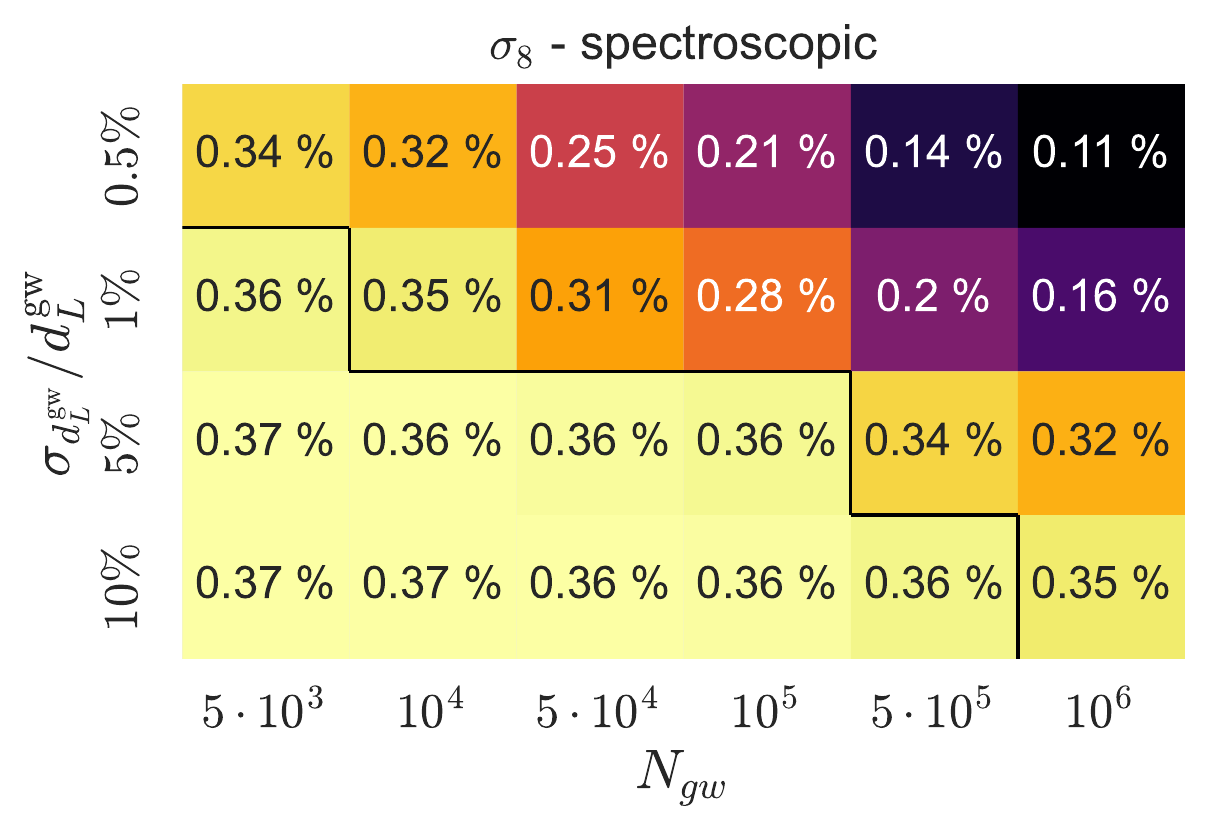}
     \caption{Marginalized $1\sigma$ relative confidence bounds for $\Omega_m$ (left panel) and $\sigma_8$ (right panel) in our \lcdm{} model. In the top row, we consider bright sirens with EM counterparts observed photometrically, while in the bottom row we assume spectroscopic observations of the counterparts. In all tables are presented bounds from the combination of GW weak lensing, galaxies weak lensing and galaxy clustering. We vary the number of GW detections $N_{\rm gw}$ and the precision on the luminosity distance determination $\sigma_{d_L}$.}
    \label{fig:lcdm_varying_specs_bright}
 \end{figure}

Let us start with the \lcdm{} scenario. 
In Fig.~\ref{fig:lcdm_varying_specs_bright} we show the marginalized $1\sigma$ relative bounds obtained on $\Omega_m$ and $\sigma_8$ varying the total number of observations $N_{\rm gw}$ and the luminosity distance precision $\sigma_{d_L}/d_L$. The first row shows the results for photometric observations of the electromagnetic counterparts or of the GW host galaxy, while the results with spectroscopic observations 
are shown in the second row. 

For photometric observations we notice that, when GWs and galaxy surveys are combined, the bounds on both $\Omega_m$ and $\sigma_8$ do not change for about half of the configurations considered (yellow regions in the plots), regardless of the choices made for the GW sector. In these regions, the bounds on cosmological parameters are strongly dominated by galaxies, which contribute almost entirely to the constraining power. The black line marks the point at which GW events start to weigh in significantly. A high number of events is required for that: about $5\times10^5$ bright sources with luminosity distance determined at least at $1\%$ precision, or $10^6$ sources with $d_L$ determined at $5\%$ precision or better are necessary to be seeing a significant impact of GW-WL on the galaxy constraints.

The spectroscopic case is analogous, though there is one case where $5\times10^4$ GW events could already be sufficient to tighten the galaxy-only bounds on both $\Omega_m$ and $\sigma_8$ of $\sim 0.5\%$ and $0.25\%$ respectively. For this to happen, however, the average $\sigma_{\rm d_L}$ is required to be no higher than $0.5\%$. Other than that, even in the spectroscopic scenario GW becomes complementary to galaxies only for $N_{\rm gw} \geq 10^5$ for $\sigma_{\rm d_L} \leq 1\%$ (or $N_{\rm gw} \geq 10^6$ for $\sigma_{\rm d_L} \leq 5\%$). We can though recognize the reduced noise affecting GW lensing in the overall better performance of the spectroscopic sirens: for all configurations in which GW contributes significantly to the constraining power, spectroscopic sources always provide bounds up to $0.1\%$ tighter than the photometric ones. For example, the $0.7\%$ ($\sim 0.4\%$) bound placed on $\Omega_m$ ($\sigma_8$) by galaxies can be reduced to $0.45\%$ ($0.2\%$) by factoring in the contribution of $10^6$ photometric sirens measured with $\sigma_{\rm d_L} = 1\%$, and further shrinked to $\sim 0.3\%$ ($0.15\%$) considering instead the same number of spectroscopic events.

\begin{figure}
  \centering
    \includegraphics[width=0.49\textwidth]{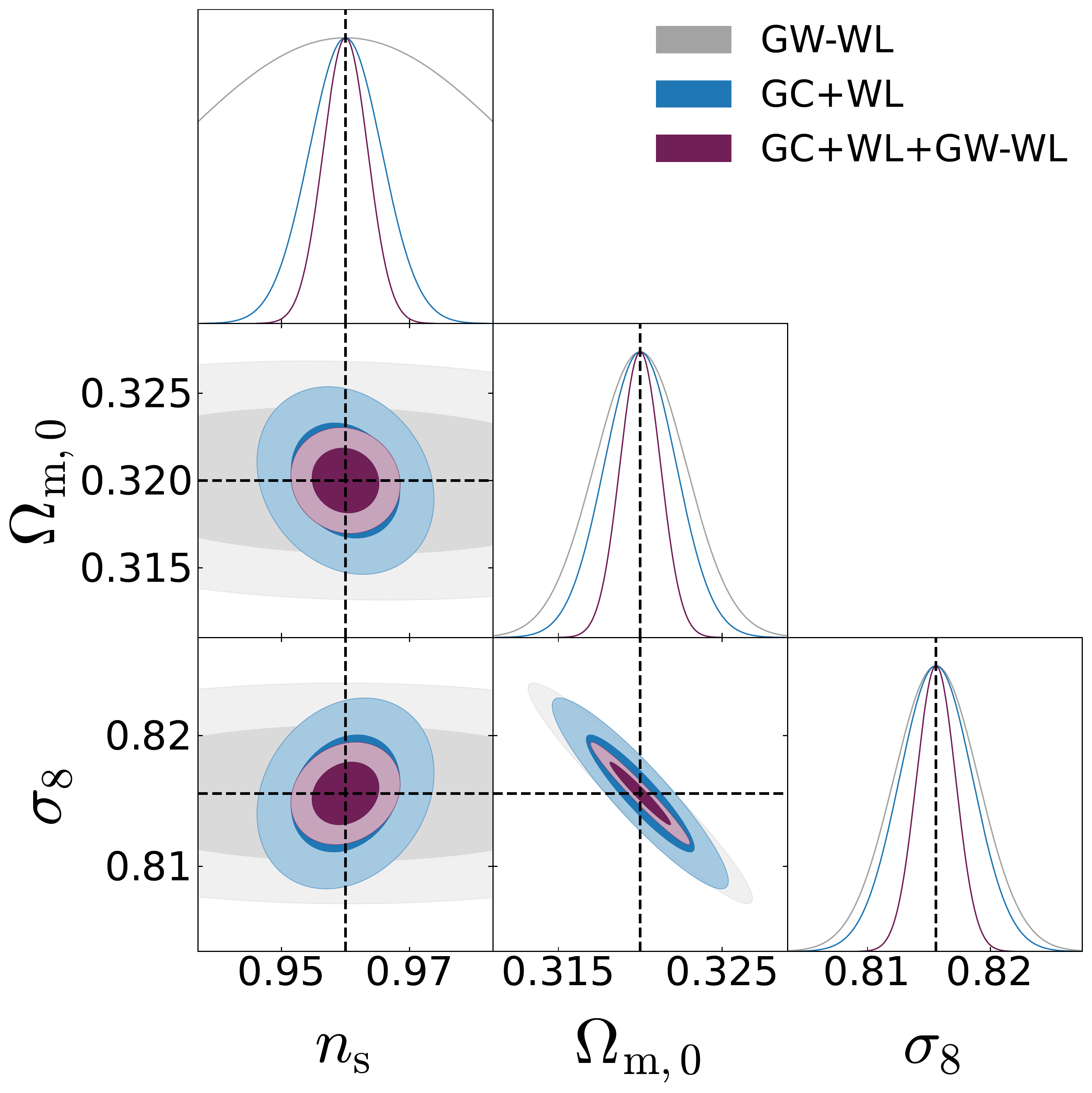}
\caption{Marginalized forecasts for \lcdm{} parameters in an idealistic scenario where $5\cdot10^5$ GW events are detected, with luminosity distance measured with $1\%$ precision. All counterparts are assumed to be observed spectroscopically. Constraints come from GW weak lensing only (grey), galaxies WL and clustering (blue), and GW and galaxy probes combined, including contributions from the cross correlations (dark red).}
    \label{fig:lcdm_triangle_plot}
  \end{figure}

While for brevity we do not reproduce them here, we have performed the same analysis for the other free \lcdm{} parameters. However, for the cases of $h$ and $\Omega_b$ the impact of GW-WL remains mild for all configurations explored in terms of $N_{\rm gw}$ and $\sigma_{d_L}/d_L$. This is direct consequence, as anticipated, of GW-WL not being particularly sensitive to those parameters, which can instead be constrained much more efficiently through other GW probes, such as standard sirens. A different story holds for $n_s$, for which we find that GW-WL impacts the constraints in a way that is qualitatively absolutely analogous to $\Omega_m$ and $\sigma_8$. To better understand this behaviour, we point to Fig.~\ref{fig:lcdm_triangle_plot}

In Fig.~\ref{fig:lcdm_triangle_plot} we show the triangular plots for three free parameters of \lcdm{} marginalising over the remaining two, in an idealistic scenario for which we choose $N_{\rm gw} = 5\cdot10^5$ and $\sigma_{d_L}/d_L=1\%$, showing the results from spectroscopic observations of the counterparts.
We plot bounds obtained considering GW-WL only, GC and galaxies WL, and the joint contribution of galaxy surveys and GW-WL.
In the spectroscopic case the GW-WL constraints on $\Omega_{m,0}$ and $\sigma_8$ are comparable with those coming from galaxies, and they half when all probes are combined, also thanks to the reduced impact of nuisance parameters. Interestingly though, the GW-WL only constraints on $n_s$ are quite large, showing that this probe is not particularly sensitive to variations of the spectral index. However, when all probes are combined, the impact of GW-WL cross-correlations with galaxies manifest by shrinking the bounds on $\sigma_8$ and $\Omega_m$, in turn breaking the mild degeneracy that exists in the galaxy-only constraints between the couple of parameters $(n_s, \Omega_m)$ and $(n_s, \sigma_8)$. The net result is a strong reduction also of the bound on $n_s$, that still feels the impact of the new observables even while not being directly probed by GW-WL.

The analysis performed above shows that GW sources have the advantage of providing a completely independent measurement affected by different systematics and hence can shed light on the standard model of cosmology in a completely independent way. However, the precision of the measurement of the cosmological parameters can effectively improve over the constraints coming from galaxies only when the statistics of measured GW sources are large and the error on luminosity distance is small. 
Indeed, in \lcdm{} where the non-minimal coupling $\Omega(a)=0$, the estimator defined in Eq.~\eqref{eq:convergence_estimator_bs_def} corresponds to the same convergence measured by galaxy surveys. Hence, GW-WL and its cross-correlations with galaxies contributes to the constraining power only by effectively strengthening the WL statistics.
Furthermore, we should emphasize that these constraints are compared with the measurement forecasted for galaxy survey only observations (GC+WL). However, as we show in Fig.~\ref{fig:lcdm_triangle_plot}, GW-WL can provide an independent measurement of the cosmological parameters that is, in some cases, comparable to those of galaxy surveys. Such independent measurement can be done even with fewer GW sources, though the constraining power on cosmological parameters will be weak in comparison to the accuracy reached by next generation galaxy surveys.

\subsection{Scalar-tensor models}
\label{subsec:results_mg}

\begin{figure}
    \includegraphics[width=0.5\textwidth]{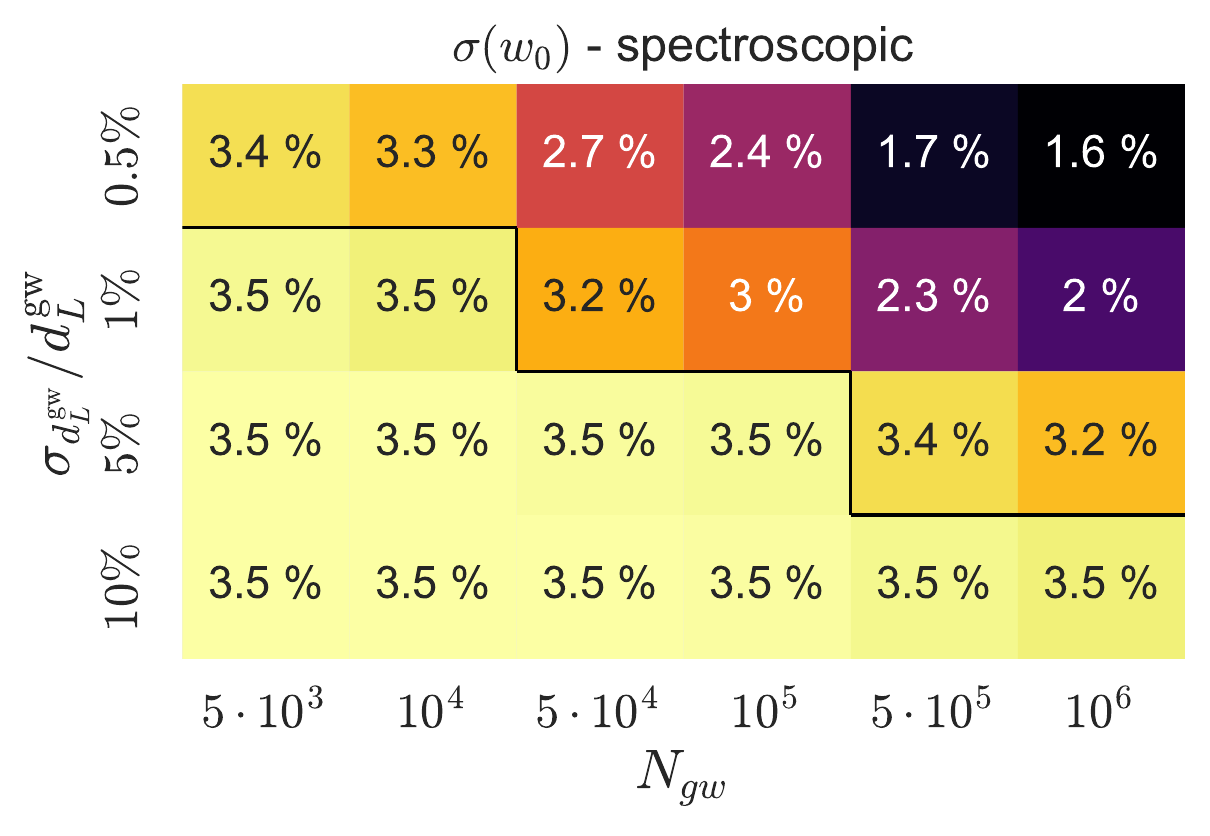}
    \includegraphics[width=0.5\textwidth]{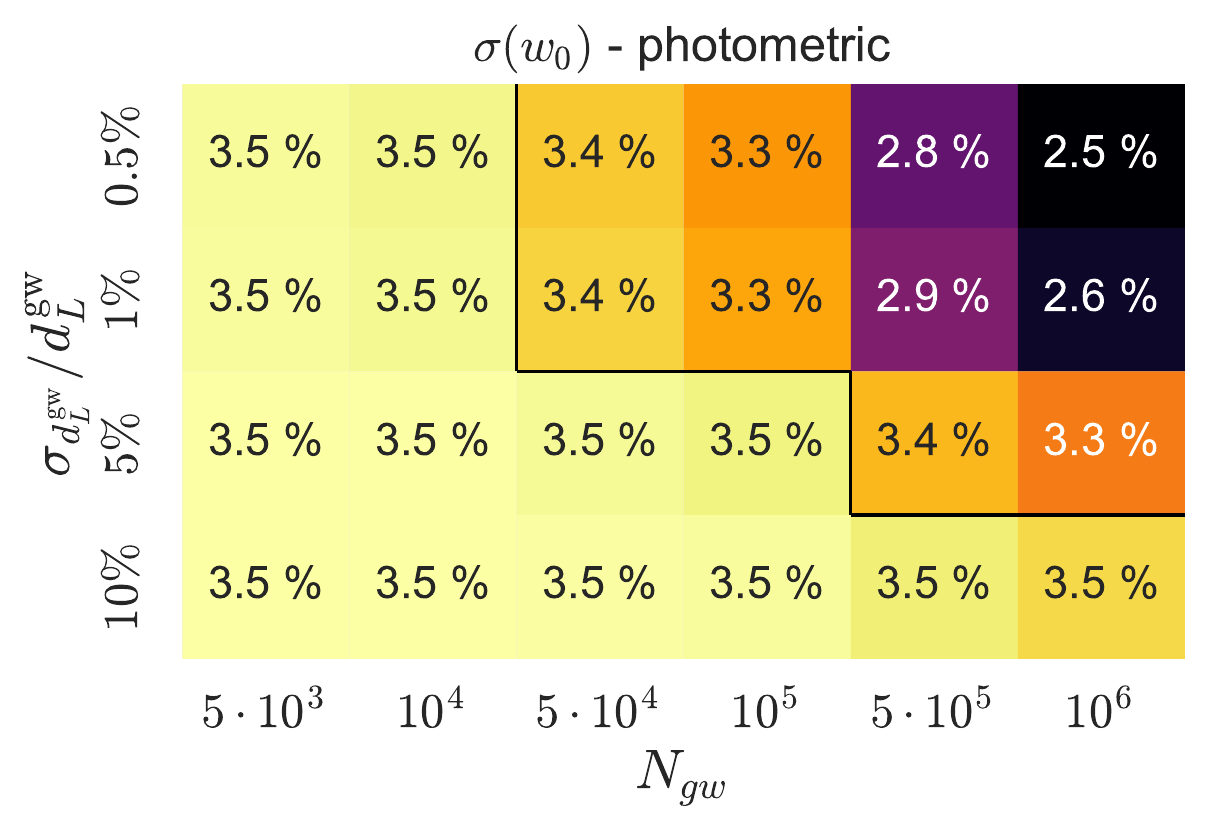}
    \includegraphics[width=0.5\textwidth]{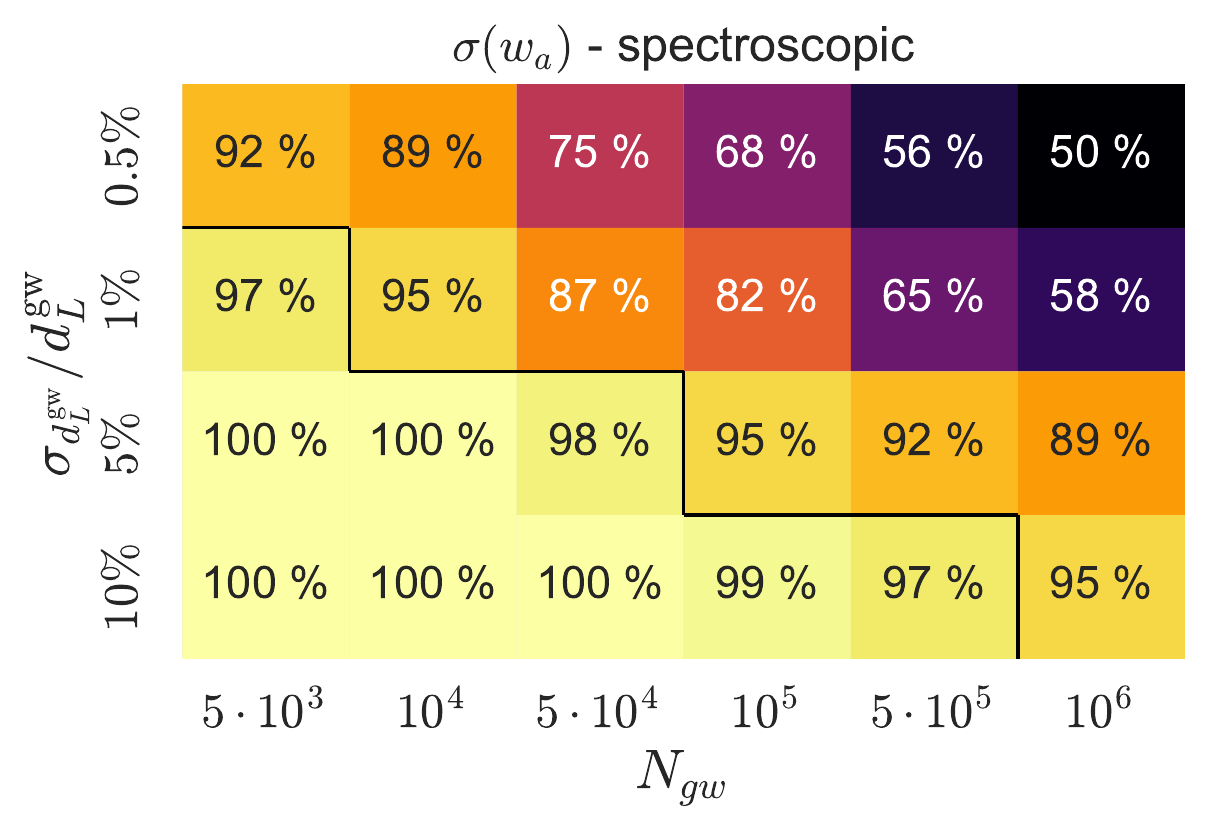}
    \includegraphics[width=0.5\textwidth]{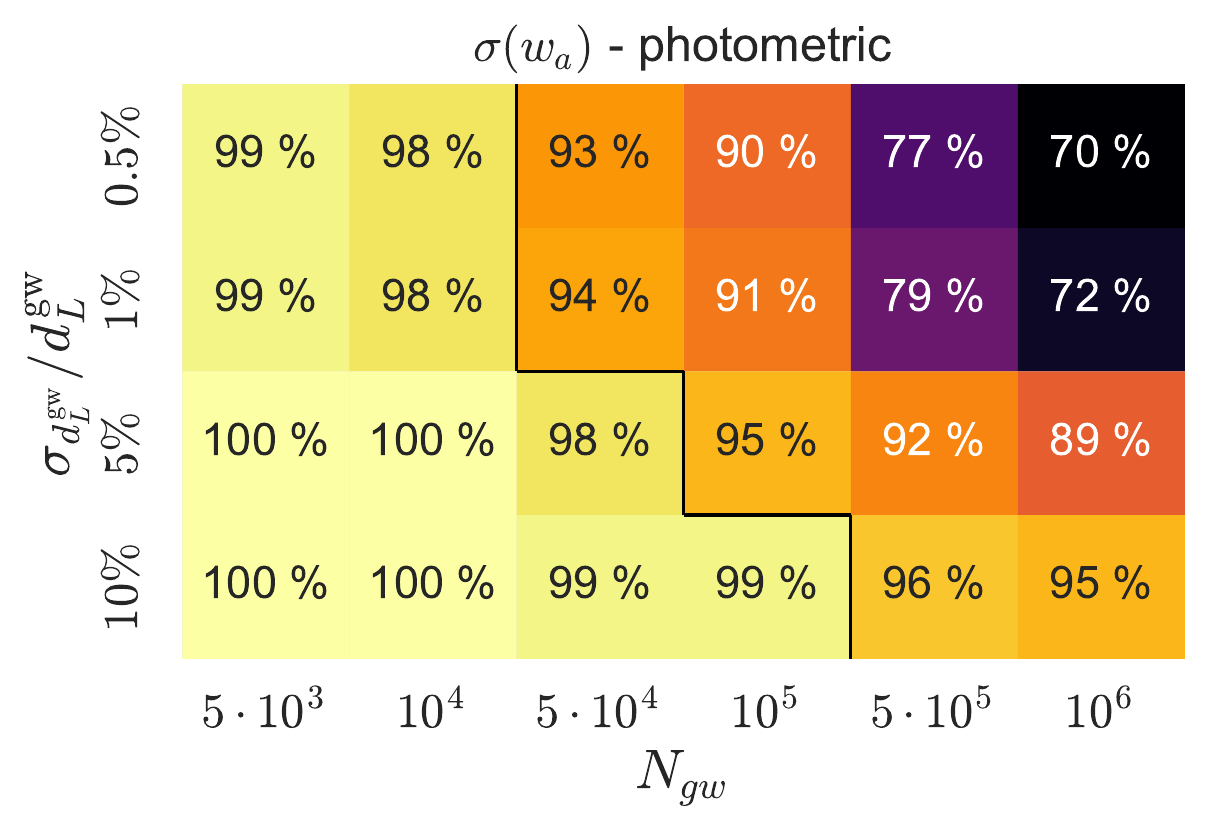}
    \includegraphics[width=0.5\textwidth]{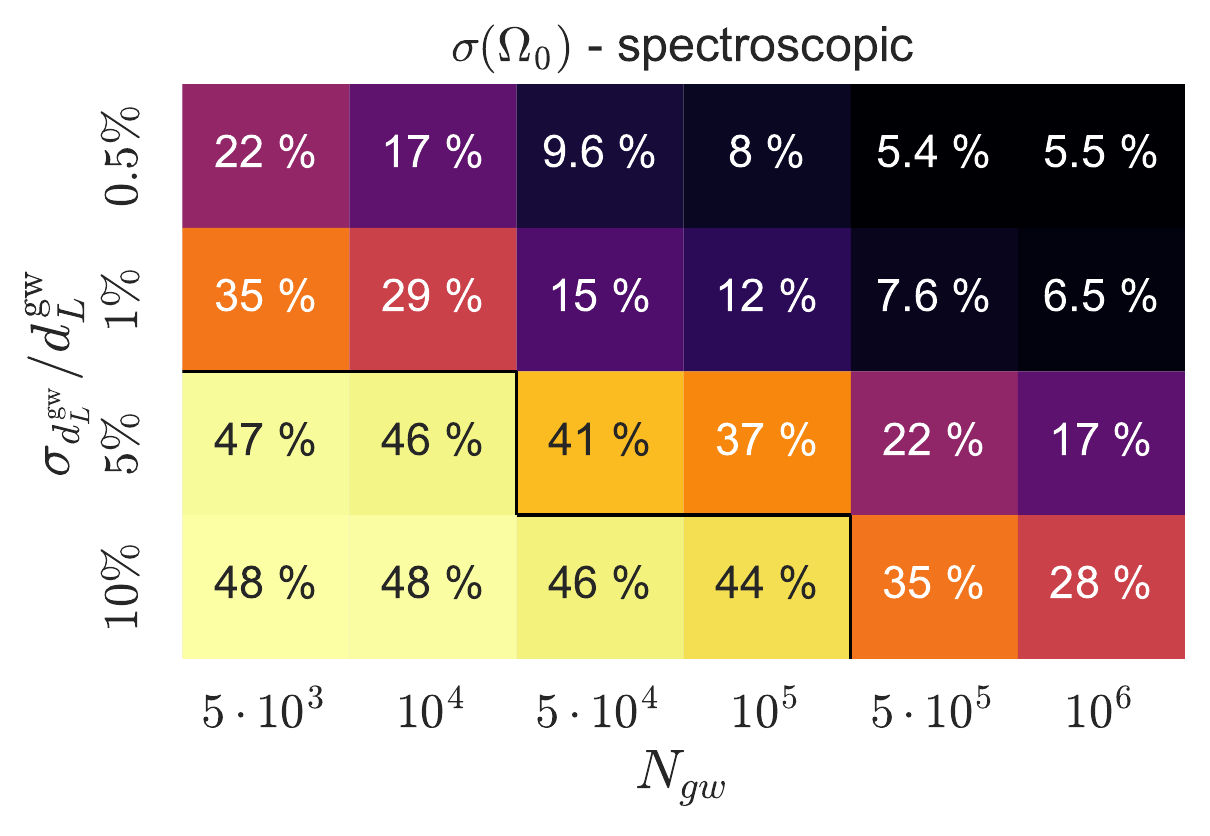}
    \includegraphics[width=0.5\textwidth]{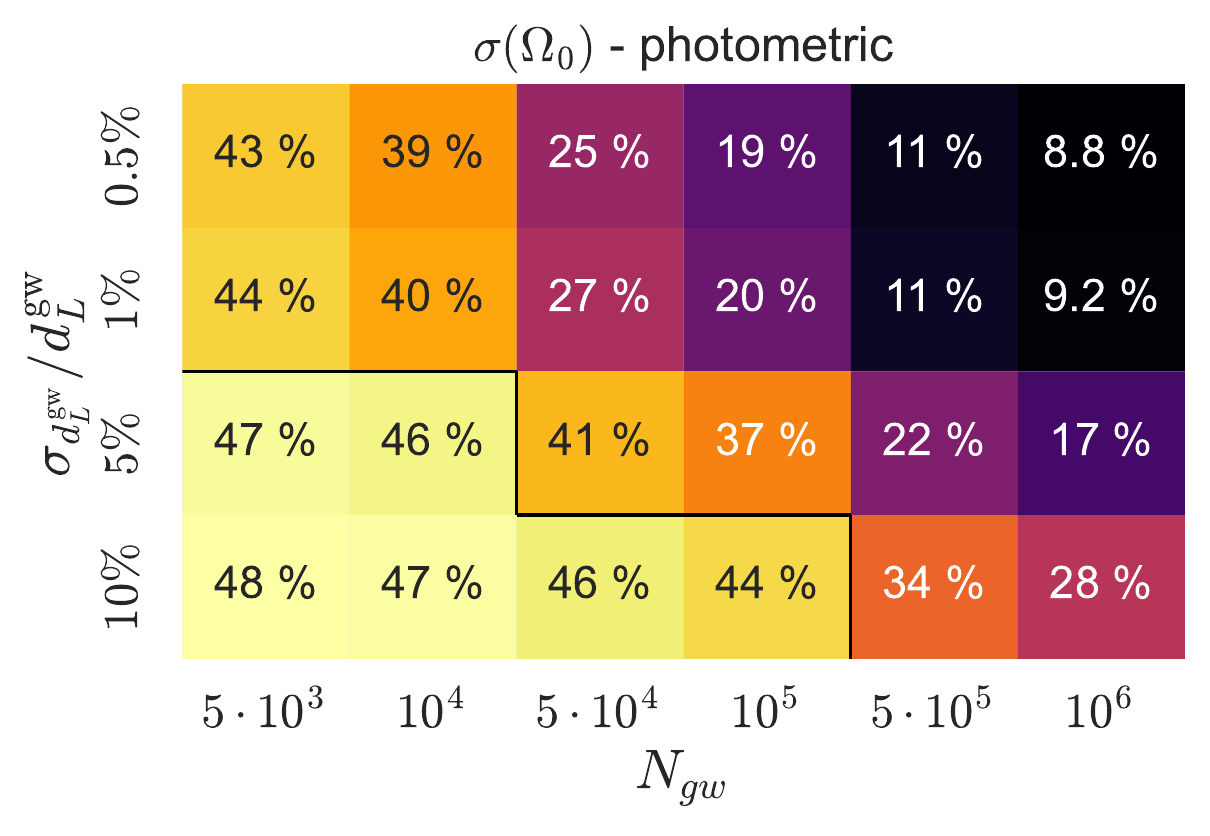}
   \caption{Marginalized $1\sigma$ confidence bounds on the MG parameters of Model I obtained from combining GW weak lensing with galaxy weak lensing and clustering. We consider bright sirens and vary the number of GW detections $N_{\rm gw}$ and the precision on the luminosity distance determination $\sigma_{d_L}$. Bounds are reported with the assumption that all GW sources will have a spectroscopic (left panels) or photometric (right panels) counterpart.}
    \label{fig:m1_varying_specs}
 \end{figure}

 \begin{figure}
  \includegraphics[width=0.5\textwidth]{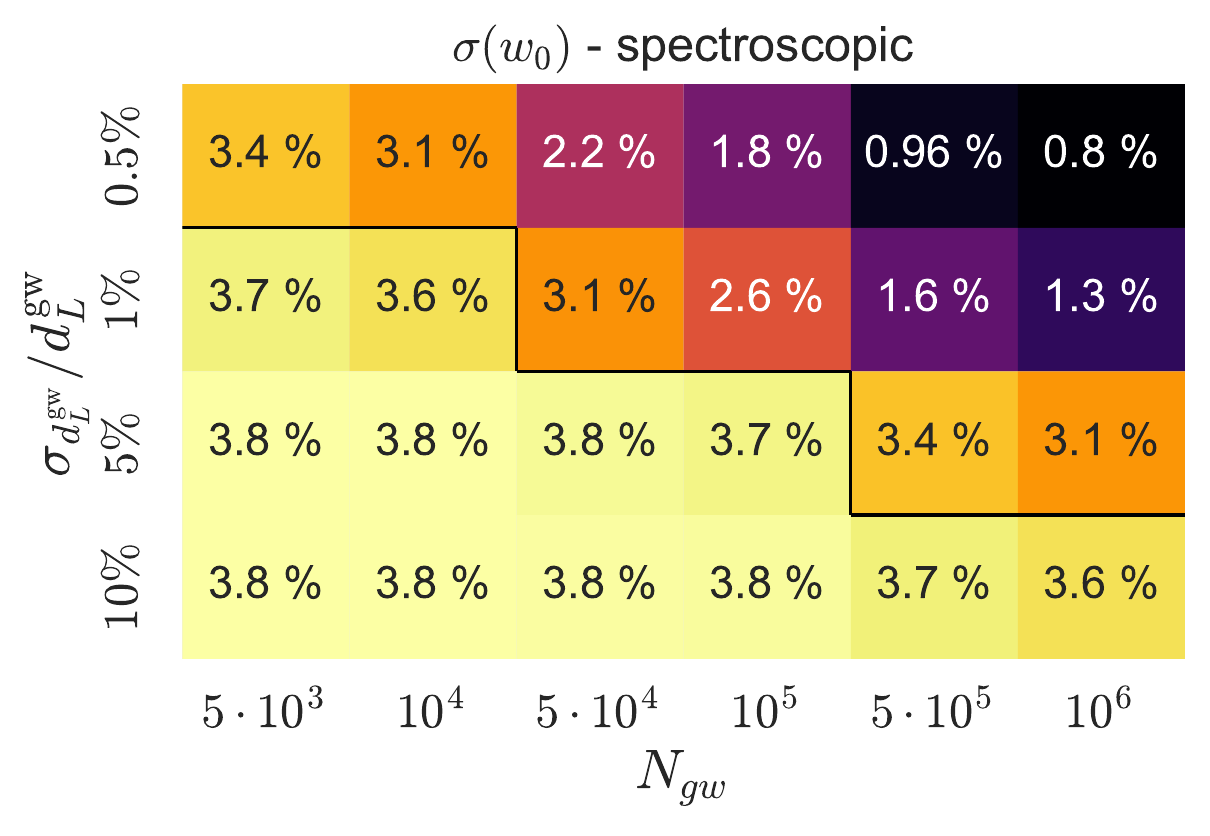}
  \includegraphics[width=0.5\textwidth]{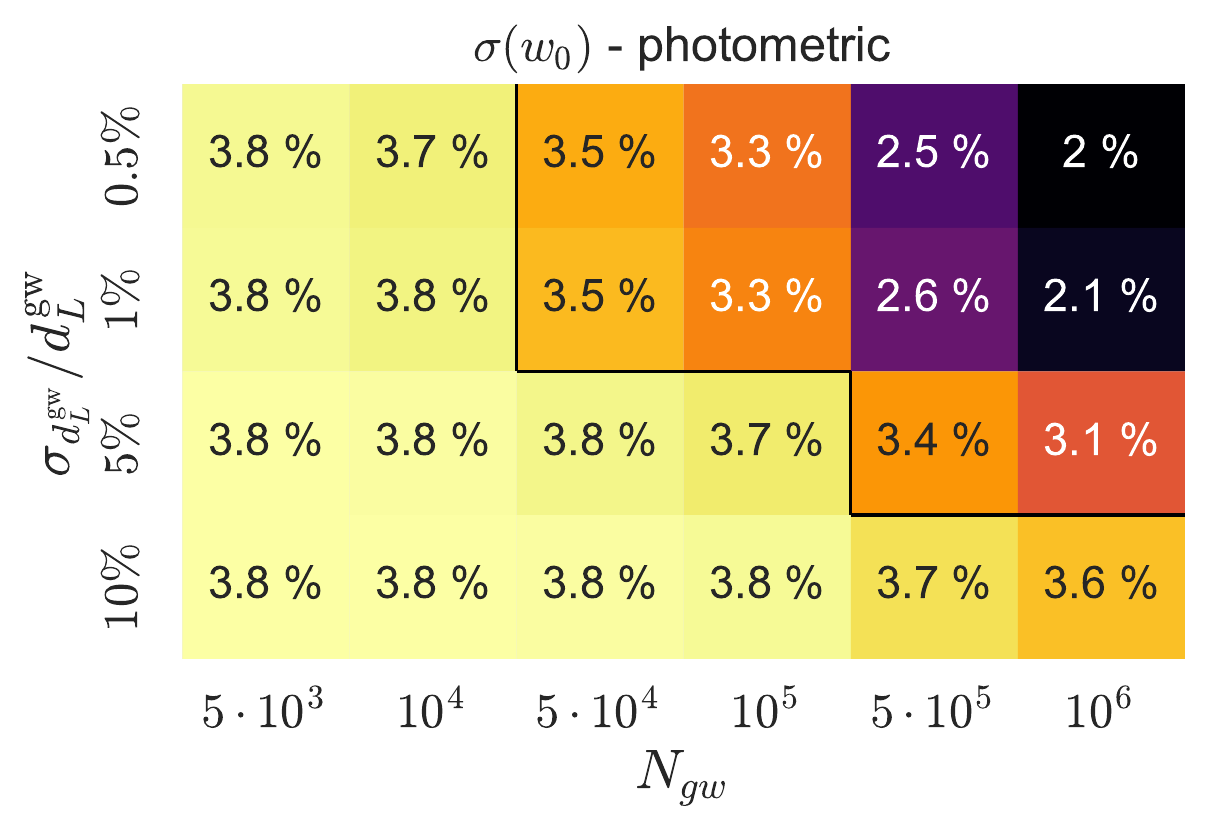}
  \includegraphics[width=0.5\textwidth]{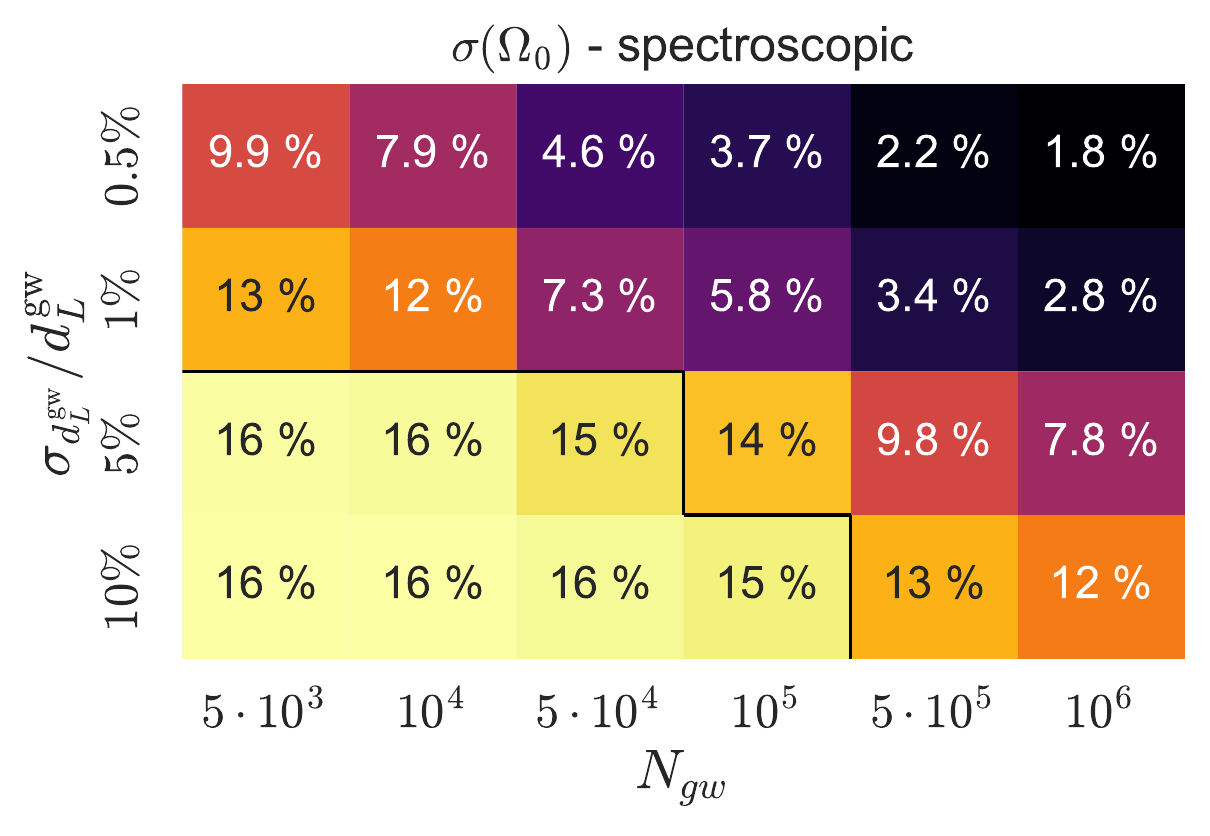}
  \includegraphics[width=0.5\textwidth]{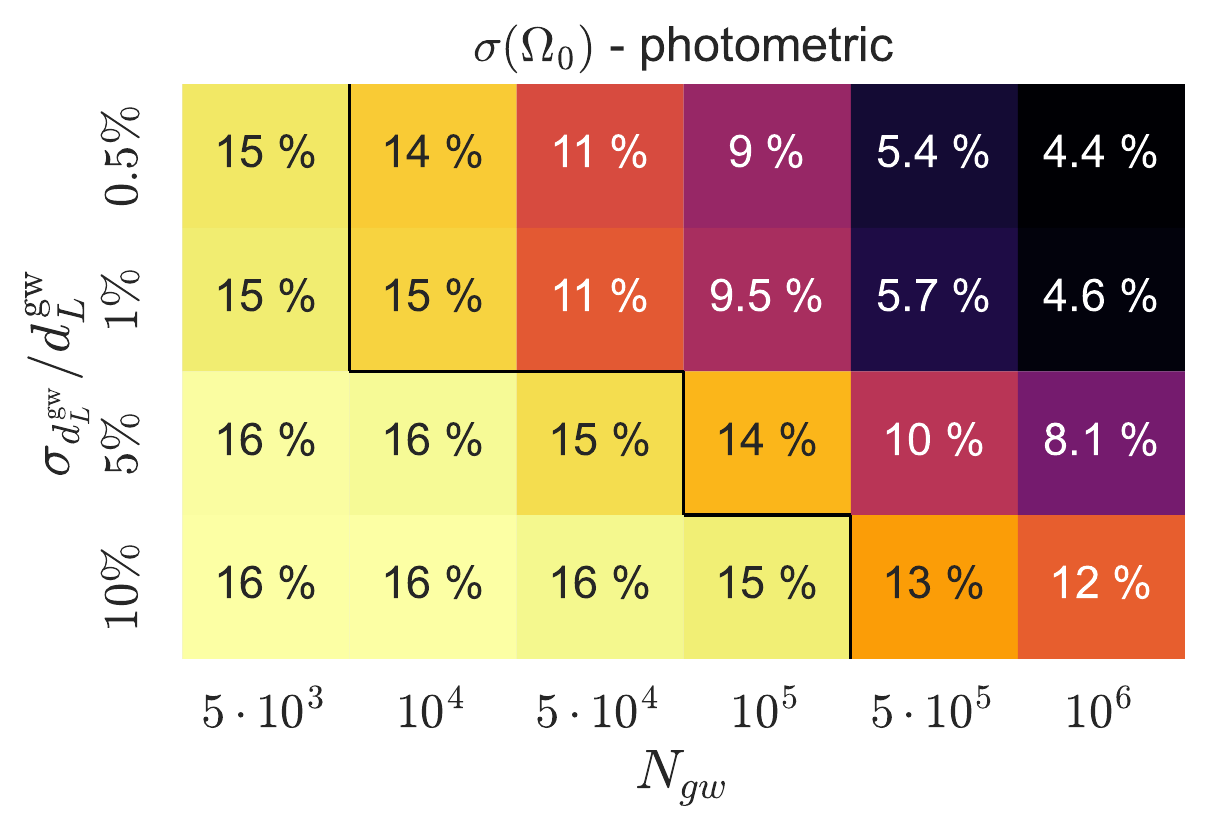}
   \includegraphics[width=0.5\textwidth]{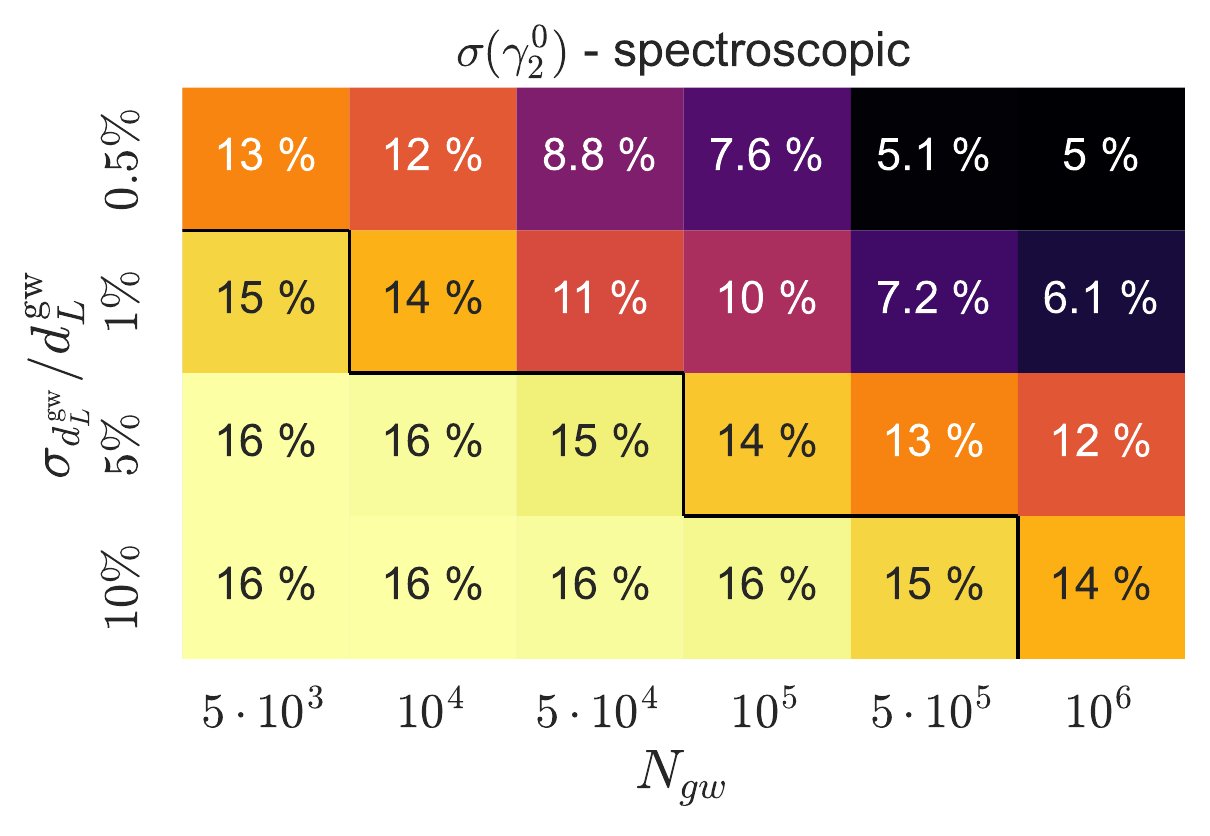}
  \includegraphics[width=0.5\textwidth]{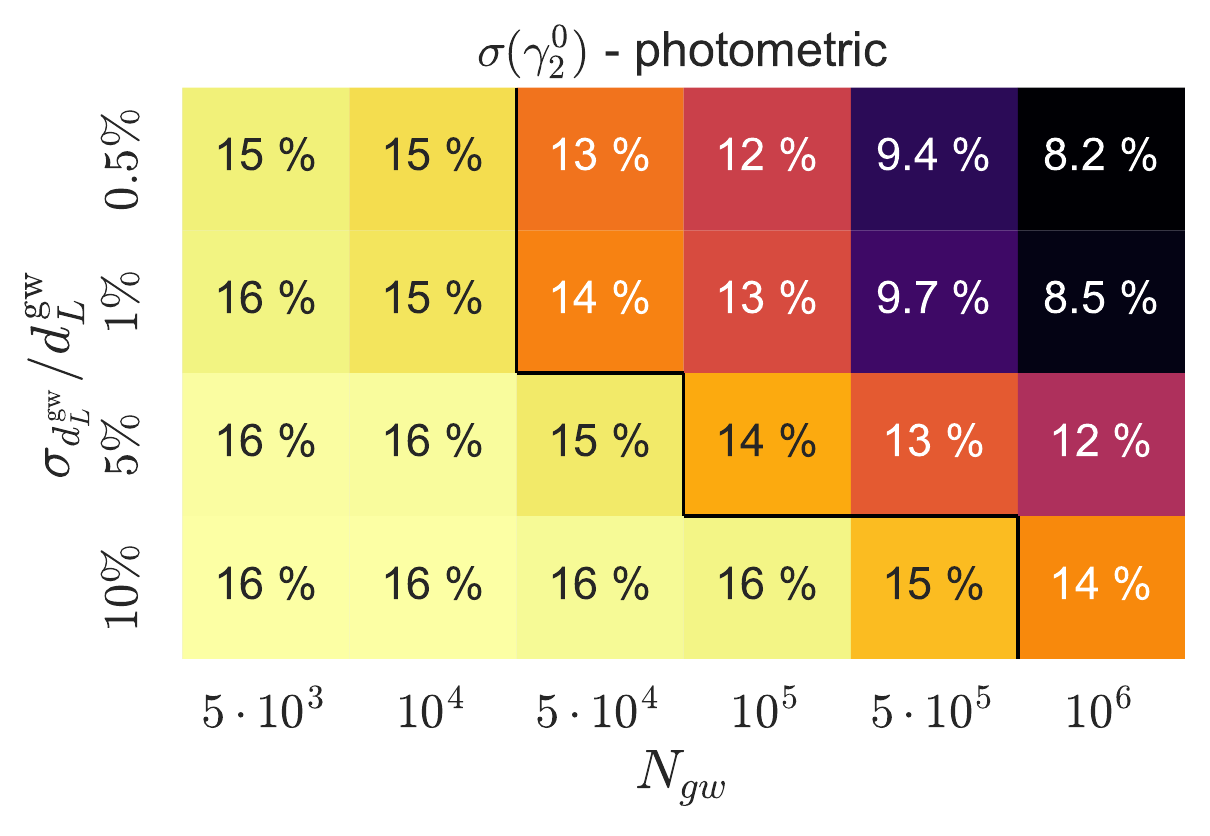}
 \caption{Marginalized $1\sigma$ confidence bounds on the MG parameters of Model II obtained from combining GW weak lensing with galaxy weak lensing and clustering. We consider bright sirens and vary the number of GW detections $N_{\rm gw}$ and the precision on the luminosity distance determination $\sigma_{d_L}$. Bounds are reported with the assumption that all GW sources will have a spectroscopic (left panels) or photometric (right panels) counterpart.}
  \label{fig:m2_varying_specs}
\end{figure}

We not turn our attention to MG models, for which the estimator~\eqref{eq:convergence_estimator_bs_def} receives explicit contributions from the conformal coupling $\Omega(a)$. We wish to explore to which extent this helps in breaking degeneracies between $\Omega_0$ and other cosmological parameters. Again, we vary $N_{\rm gw}$, $\sigma_{d_L}/d_L$ and the error on the sources redshift determination $\sigma_z$. For each combination, we compute the Fisher forecasts on the cosmological parameters, including once again all correlators of the kind $C^{XY}_{\ell}$, $X, Y \in [\delta_g, \kappa_g, \hat \kappa_{\rm gw}]$.

 In Fig.~\ref{fig:m1_varying_specs} we show the marginalized $1 \sigma$  bounds on $w_0$, $w_a$ and $\Omega_0$ in M1, for the case of photometric (top row) and spectroscopic (bottom row) redshifts. Similarly to the \lcdm{} case, there are regions in which the constraining power comes almost exclusively from the galactic probes. In these regions, the number of available GW events is too low and the accuracy of their luminosity distance determination is too poor for GW to improve on galaxies constraints. However, GWs start having an impact sooner this time,  for $N_{\rm gw}$  $\sim1$ order of magnitude lower than what we were observing in the \lcdm{} scenario. In the case of photometric redshift, $5\times10^4$ GW sources determined with $\sigma_{\rm d_L} \leq 1\%$ are already sufficient to detect a tightening of the constraints on all MG parameters, that becomes significant at all $\sigma_{\rm d_L}$ for $N_{\rm gw} = 5\times10^5$ sources or more. The results are similar for $w_0$ and $w_a$, though for equal setups the improvement in constraining power is much lower than in the $\Omega_0$ case.

 Even more impressive are the spectroscopic results, for which at all $N_{\rm gw}$ there is at least one setup in which the GWs contribute to improving importantly the constraining power. In particular, with $10^4$ GW events measured with $\sigma_{\rm d_L}\leq 1\%$ the constraints on $\Omega_0$ are already almost halved. However, $\sim 5\times10^5$ sources are needed to reach $< 10\%$ level precision on $\Omega_0$.

 Fig.~\ref{fig:m2_varying_specs} shows analogous tables for $w_0$, $\Omega_0$ and $\gamma_2^0$ in the M2 case. Here, while in both photometric and spectroscopic scenarios our findings for the DE equation of state parameters are analogous to those of M1, the galaxy-dominated constraints are tighter on the EFT parameters than in the M1 case. This is in line with the findings of~\cite{Frusciante:2018jzw}, which also finds tighter relative constraints on the EFT parameters for models in which $\gamma_1^0$ and $\gamma_2^0$ are not zero. However, the absolute bounds on $\Omega_0$ do not change significantly between the two models: indeed as the percentage error on $\Omega_0$ is smaller for M2 of a factor of $\sim 3$, the fiducial value of $\Omega_0$ in that model increases of approximately the same amount.
 
 In the photometric scenario, about $10^5$ sources or higher are required to reach $<10\%$ level precision on $\Omega_0$, while the number drops to $N_{\rm gw} \sim 10^4$ in the spectroscopic case. Again, spectroscopic sources perform better than photometric ones: with optimal luminosity distance determination ($\sigma_{\rm d_L} \leq 1\%$), $5\times10^4$ sources are enough for GW-WL to meaningfully contribute to the galaxy surveys constraints on all MG parameters. In both the photometric and spectroscopic cases, a large enough statistic of events can allow the determination of both $\Omega_0$ and $\gamma_2^0$ at least at a few $\%$ level.

 \begin{figure}
    \includegraphics[width=0.5\textwidth]{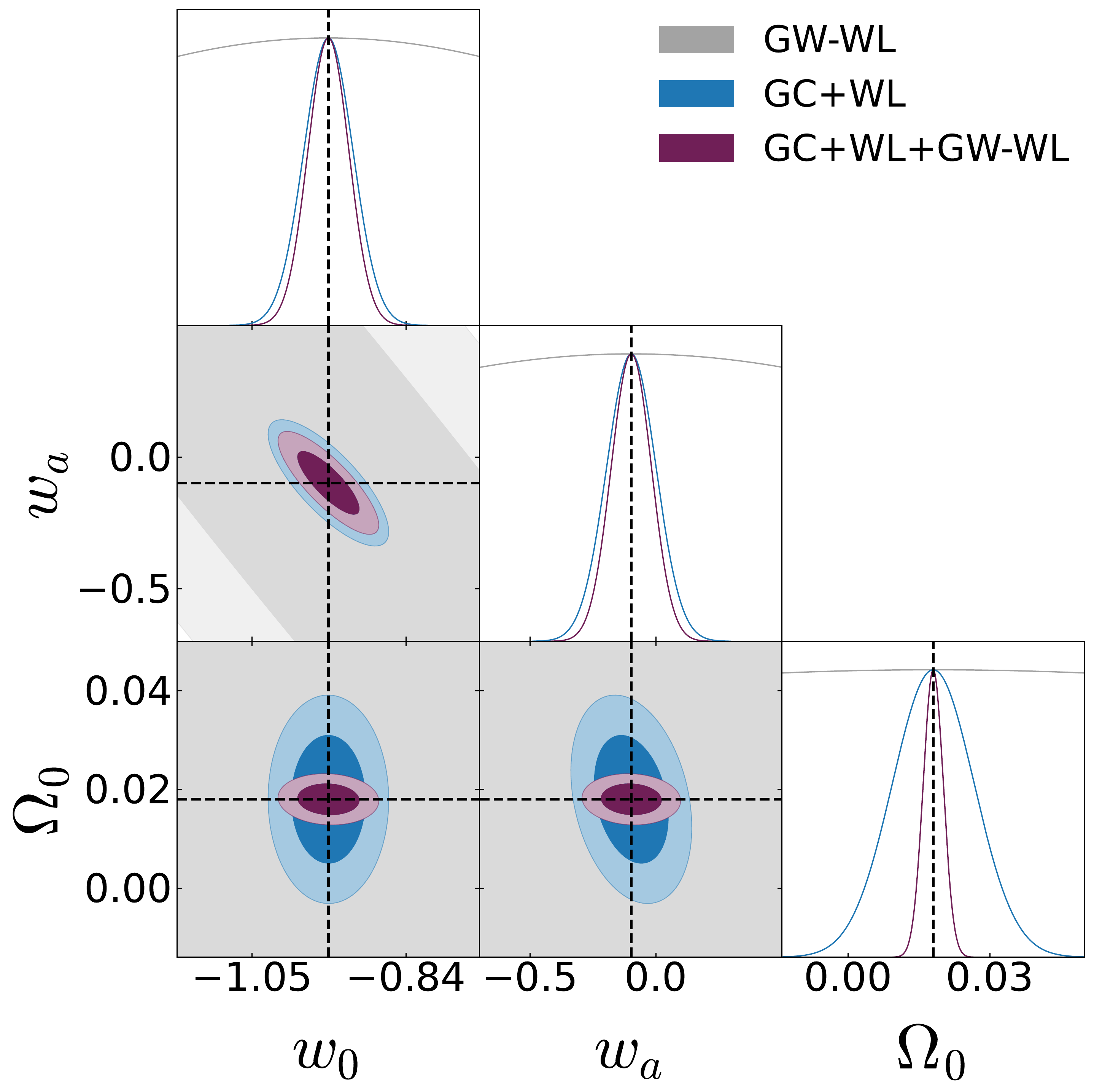}
    \includegraphics[width=0.5\textwidth]{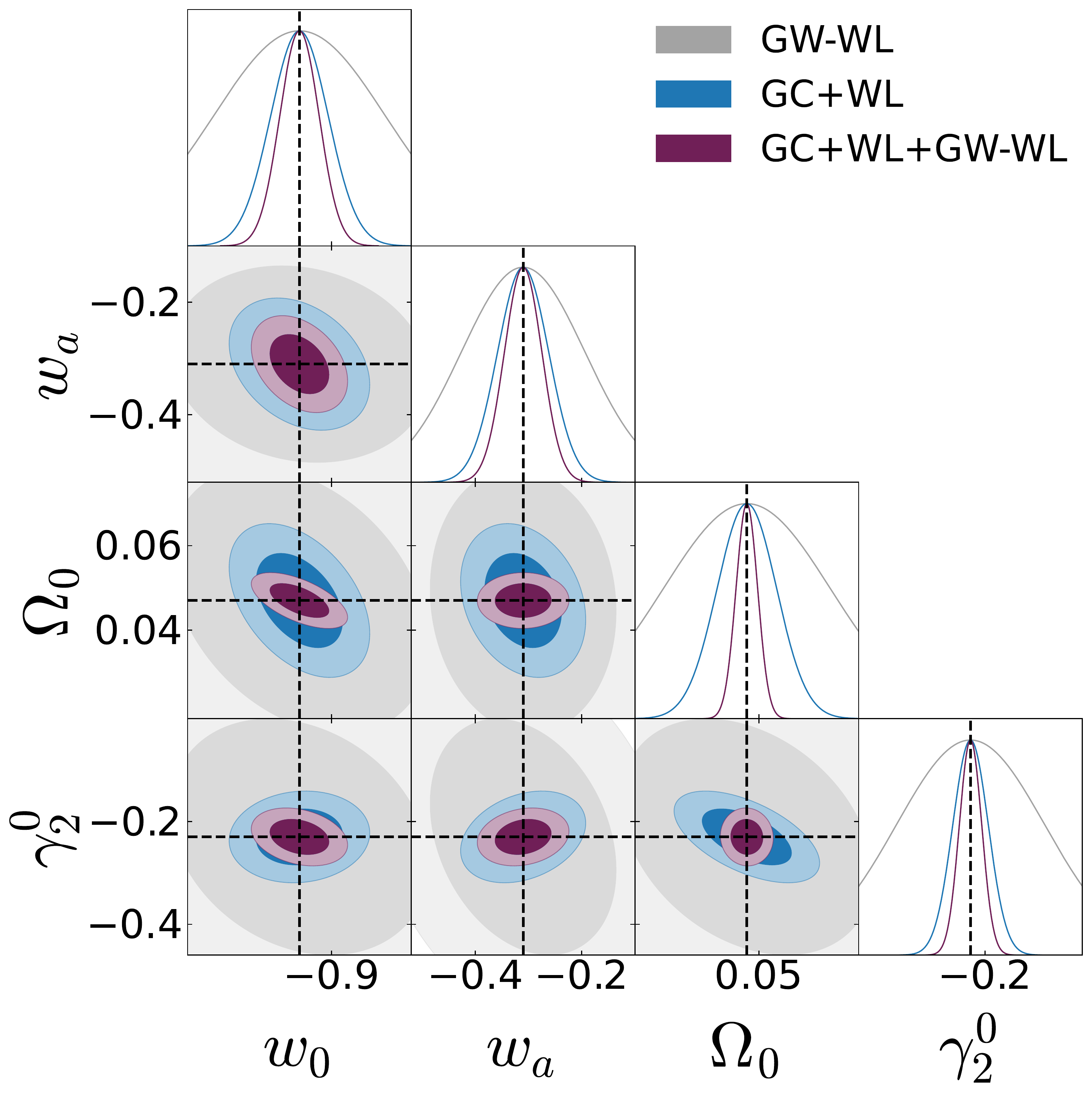}
\caption{Marginalized forecasts for the MG parameters of model M1 (left panel) and M2 (right panel) in an idealistic scenario where $1\cdot10^5$ GWs events are detected with a bright counterpart measured spectroscopically and luminosity distance measured with $1\%$ precision. Constraints come from GW weak lensing only (grey), galaxies WL and clustering (blue), and GW and galaxy probes combined, including cross-correlations between the probes (purple).}
\label{fig:mg_triangle_plots}
\end{figure} 

\begin{figure}
  \centering
    \includegraphics[width=0.49\textwidth]{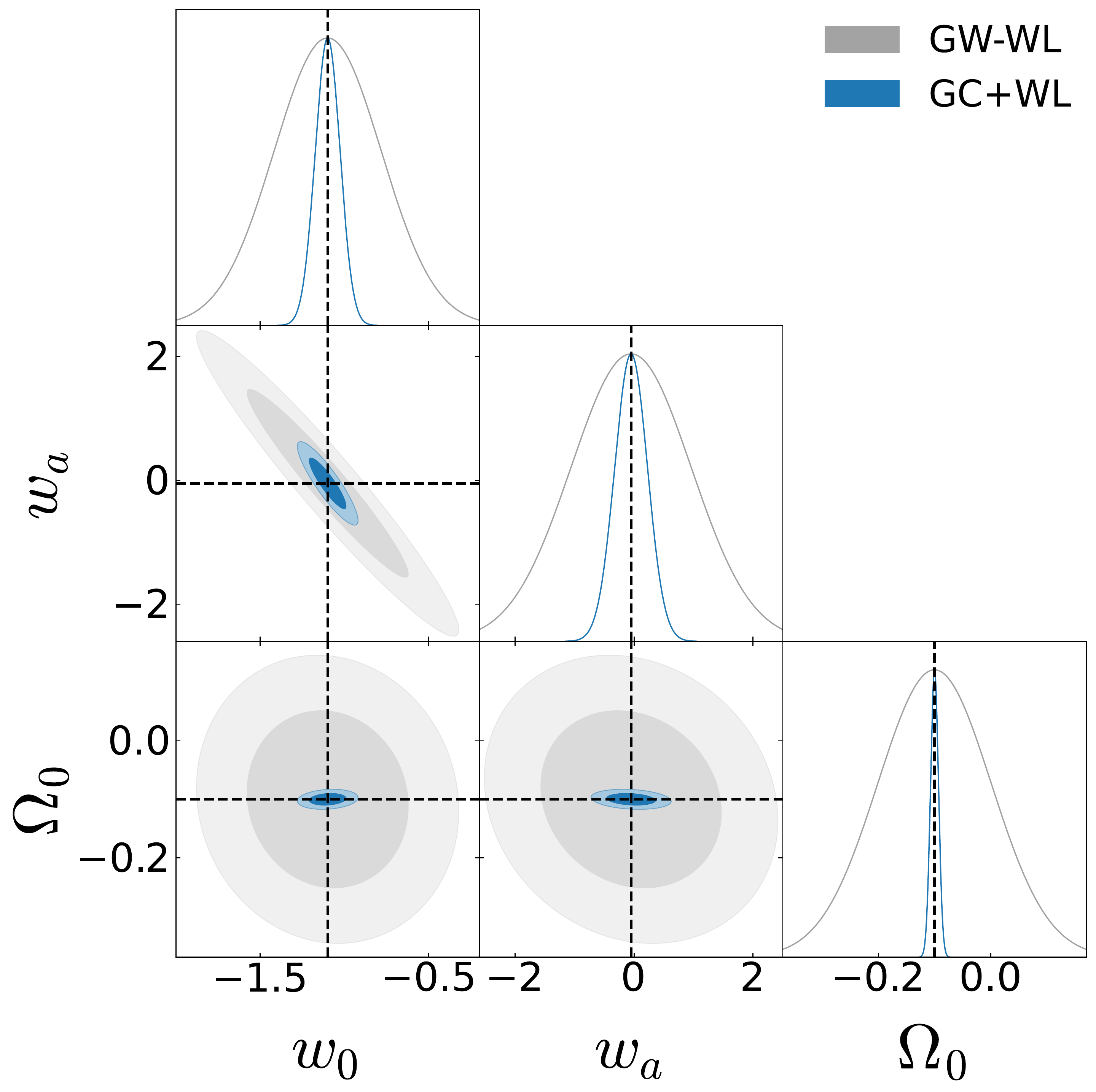}
    \includegraphics[width=0.49\textwidth]{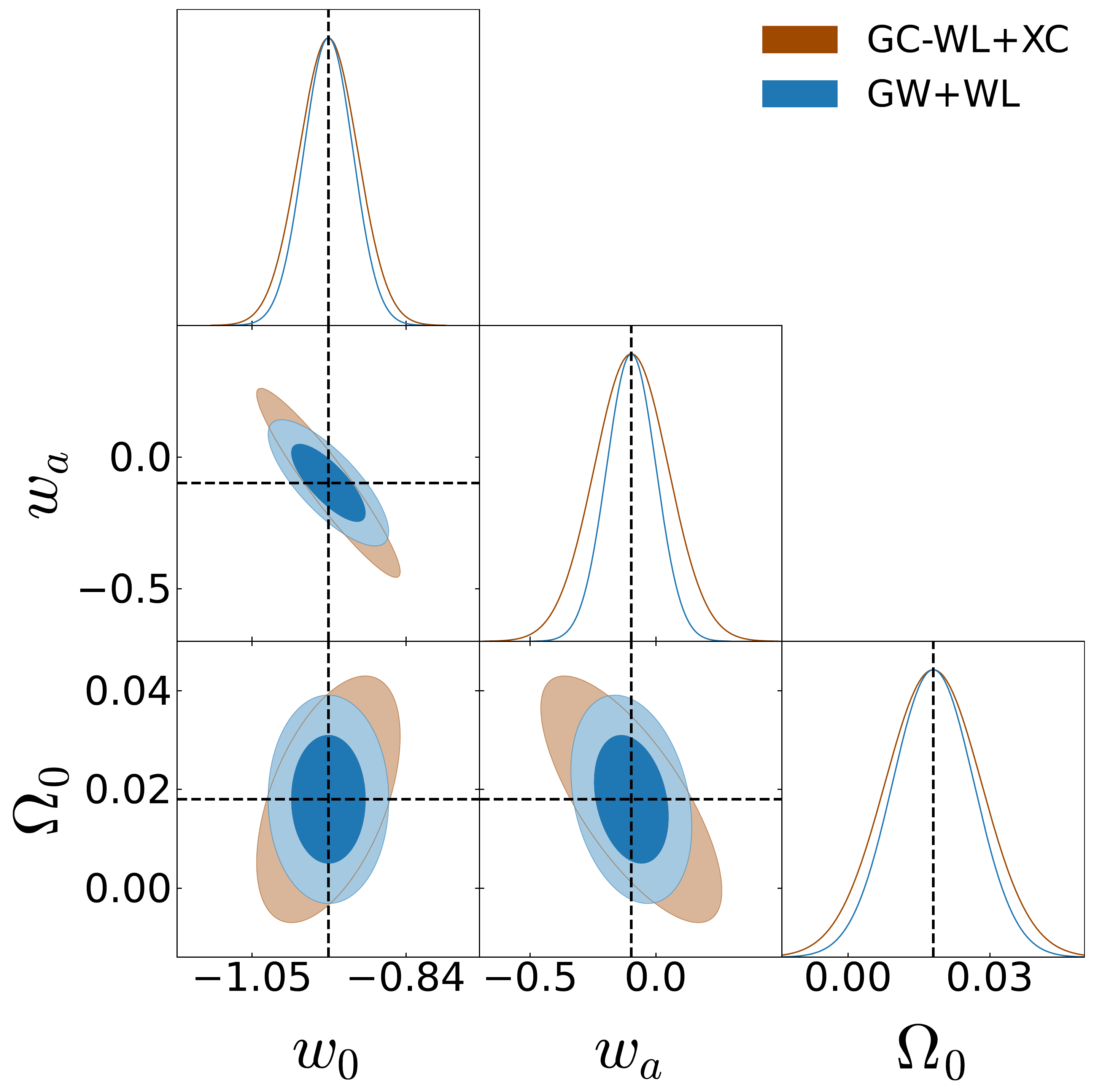}
\caption{Comparison of the marginalized 1$\sigma$ bounds on the MG parameters of M1. We consider the same setup of Fig.~\ref{fig:mg_triangle_plots}, i.e. $N_{\rm gw} = 5\cdot10^5$, $\sigma_{\rm d_L} = 1\%$ and spectroscopic counterparts. In both panels, the blue contours represent bounds placed through GC+WL only, matching the same blue contours of the left panel of Fig.~\ref{fig:mg_triangle_plots}. In the left panel, these are compared with GW-WL-only bounds, zooming out with respect to Fig.~\ref{fig:mg_triangle_plots} to include the full extent of the grey contours. In the right panel instead, the blue contours are confronted with bounds obtained considering only the auto-correlation of GW-WL and its cross-correlations with galaxies (brown regions), while not including GC+WL auto and cross-correlations.}
\label{fig:XC_triangle_plots}
\end{figure}

In Fig.~\ref{fig:mg_triangle_plots} we report the triangular plots with marginalized constraints for the three MG parameters of M1 (left panel) and the four MG parameters of M2 (right panel). These results are obtained for an idealistic scenario in which $10^5$ GW events are detected with a spectroscopic bright counterpart and their luminosity distance is determined at $1\%$ accuracy. Besides confirming the findings reported above, the left panel of Fig.~\ref{fig:mg_triangle_plots} highlights that the GW-only contours are still remarkably wide with respect to the galaxy bound, in particular concerning $\Omega_0$. We show this more evidently in the left panel of Fig.~\ref{fig:XC_triangle_plots}, where we plot the GW-WL-only constraints in gray compared to the GC+WL constraints in blue, zooming out to include the full extent of the GW bounds.

We impute these large constraints primarily to the degeneracy between $w_0$ and $w_a$ that enters weak lensing self-correlations. As a result, GW-WL alone is not a good probe to constrain the DE equation of state, which in this model also determines the time evolution of the conformal coupling. The poor determination of $w_{\rm DE}$ leaves the coupling evolution unconstrained, resulting in very loose bounds on $\Omega_0$.
Quite impressively though, the combined GW+galaxy constraints (dark red contours in Fig.~\ref{fig:mg_triangle_plots}) on the MG parameters of M1 are narrower than the galaxy-only bounds, which must mean that the increased constraining power relies heavily on the cross-correlations $C_{\ell}^{\kappa_{\rm gw}\delta}$ and $C_{\ell}^{\kappa_{\rm gw}\kappa_{\rm g}}$. We verify this hypothesis in the right panel of Fig.~\ref{fig:XC_triangle_plots}, where we plot again in blue the marginalized galaxy bounds, this time compared with brown contours obtained including only GW-WL auto-correlation and its two cross-correlations with the galaxy fields, which we dub 'XC', while leaving out contributions from GC and WL auto and cross-correlations. We see that now the GW constraints become comparable with the galaxy bounds as the cross-correlation of GW-WL with, in particular, GC is breaking (or mitigating) the degeneracies in the MG sector.
This confirms the essential role that GW-WL cross-correlations can have in impacting cosmological bounds.\\

As for model M2 (right panel in Fig.~\ref{fig:mg_triangle_plots}), we see that the GW-WL only contours are still pretty broad, though way tighter than in the M1 case.
This is possibly because non-null $\gamma_1$ and $\gamma_2$ introduce additional EFT functions whose time evolution we also model as dependent on $w_0$ and $w_a$. If lensing is a probe overall efficient in constraining $\gamma_2$, this might break some existing degeneracies between $\Omega(a)$ and the DE equation of state.
Again, the cross-correlations play a striking role in tightening the bounds on the MG parameters, especially on $\Omega_0$, as the combination of GW and galaxies contributes to reducing its degeneracies with $w_0$ and $\gamma_2^0$. Constraints on $w_0$, $w_a$ and $\gamma_2^0$ also improve with respect to the galaxy only case, tough in the case of these parameters the effect is milder. Instead, we observe that GW-WL is not very efficient in breaking the remaining degeneracy in the $(w_0,w_a)$ plane, as the constraints on these two parameters are still largely dominated by the galaxies. We can however deduce from Fig.~\ref{fig:m2_varying_specs} that this will eventually change if a higher number of bright GW events is considered. \\

Lastly, repeating the same analysis of Figs.~\ref{fig:m1_varying_specs} and~\ref{fig:m2_varying_specs} for the standard cosmological parameters $\{h, \Omega_{m,0}, \Omega_{b,0}, n_s, \sigma_8 \}$ in both models, we find very similar results to those of subsection~\ref{subsec:results_lcdm}, $i.e.$ O($10^5$) or more GW sources are needed to even to impact significantly the bounds placed on those by galaxies alone, O($10^6$) in the case of $h$ and $\Omega_b$. This is because these parameters are already determined by the galaxy probes with much more accuracy than the MG sector, and they benefit only very mildly from the increased constraining power introduced by the GWs. \\

\begin{figure}
  \includegraphics[width=0.5\textwidth]{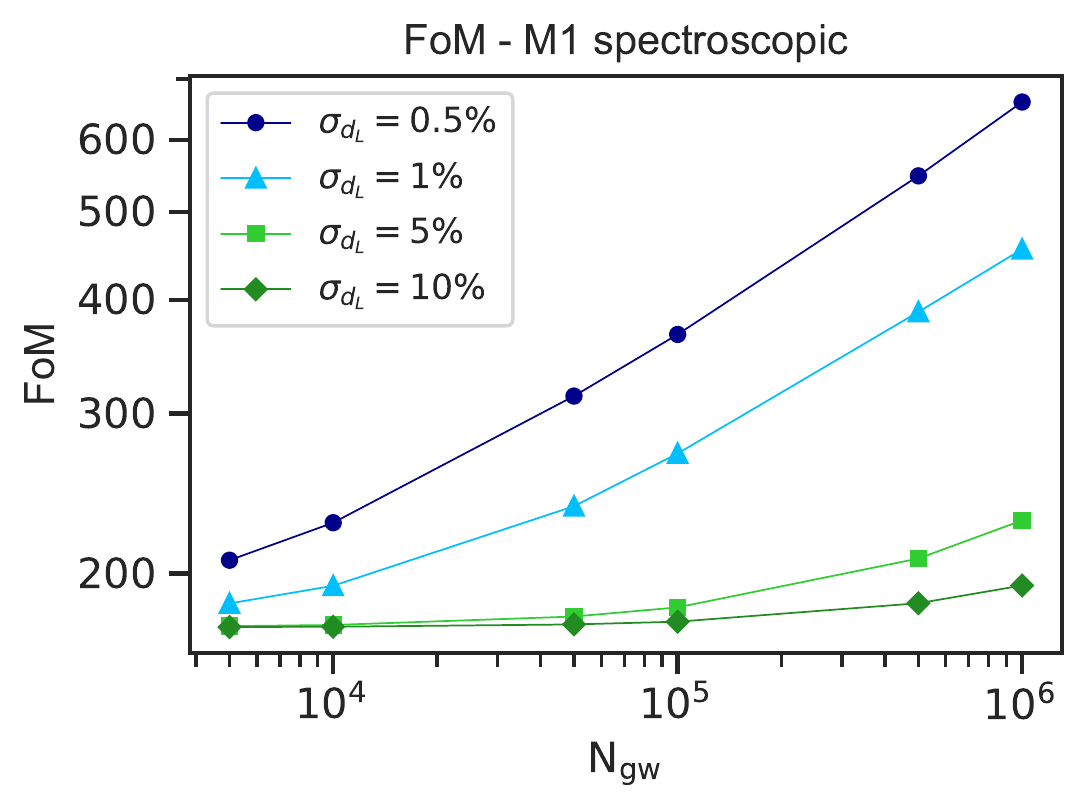}
  \includegraphics[width=0.5\textwidth]{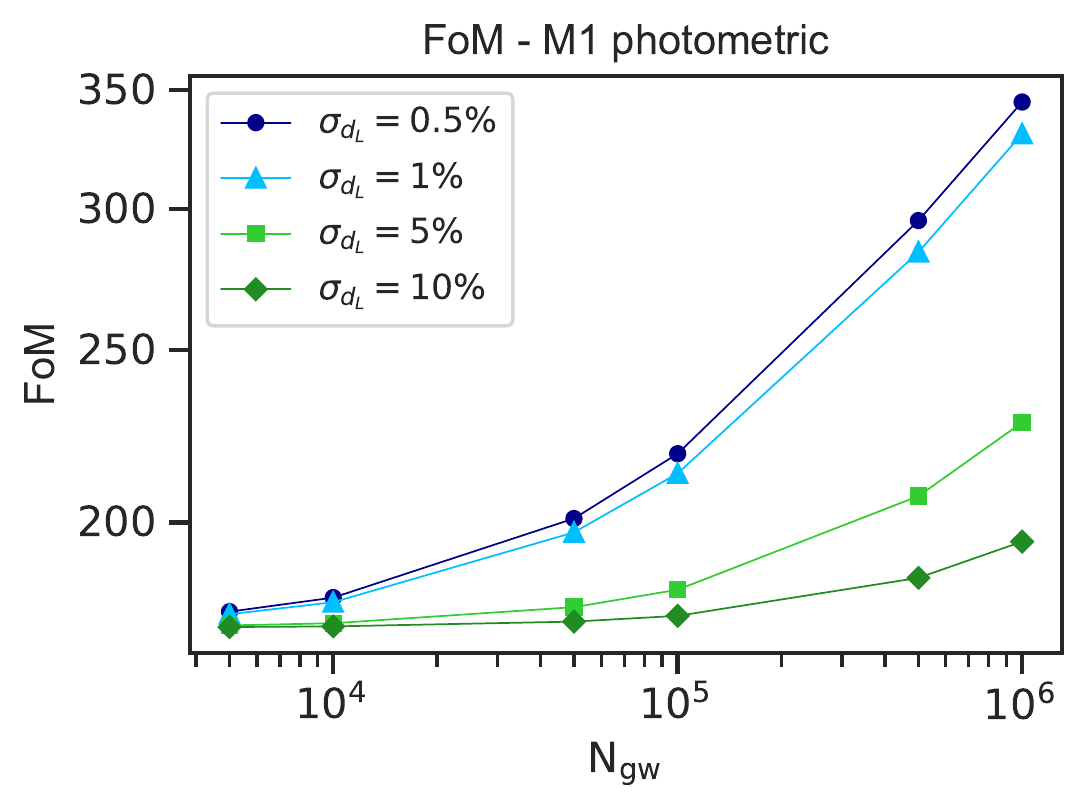}
  \includegraphics[width=0.5\textwidth]{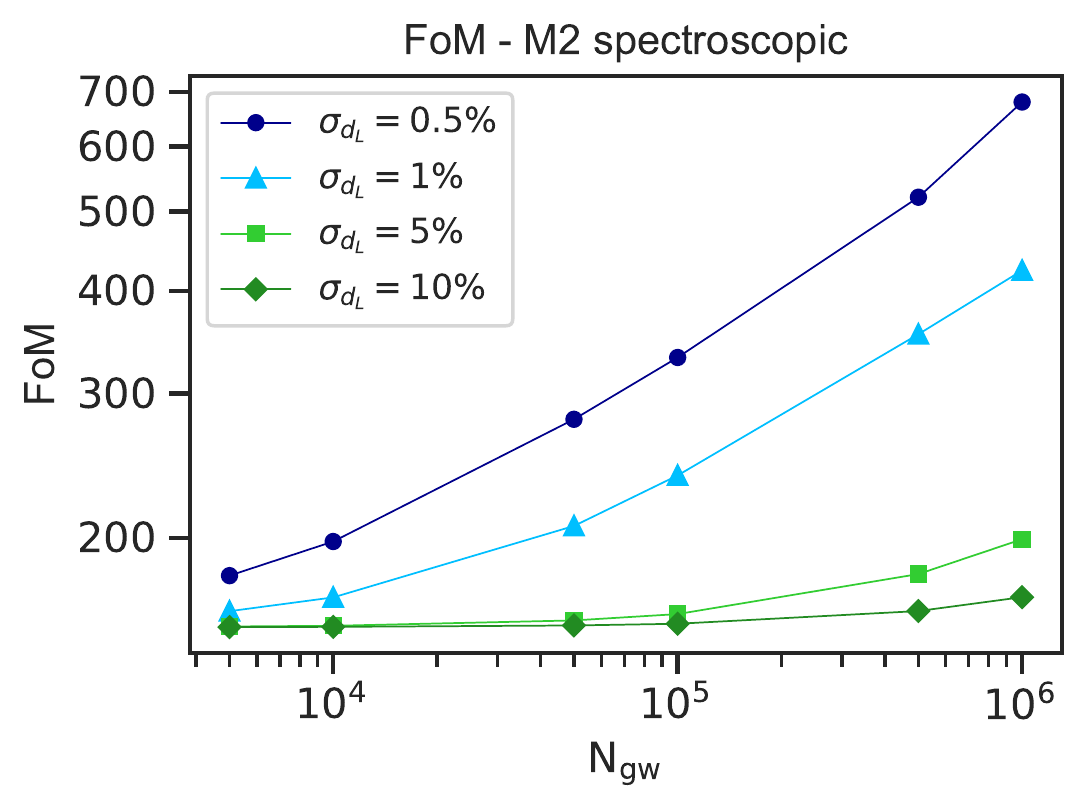}
  \includegraphics[width=0.5\textwidth]{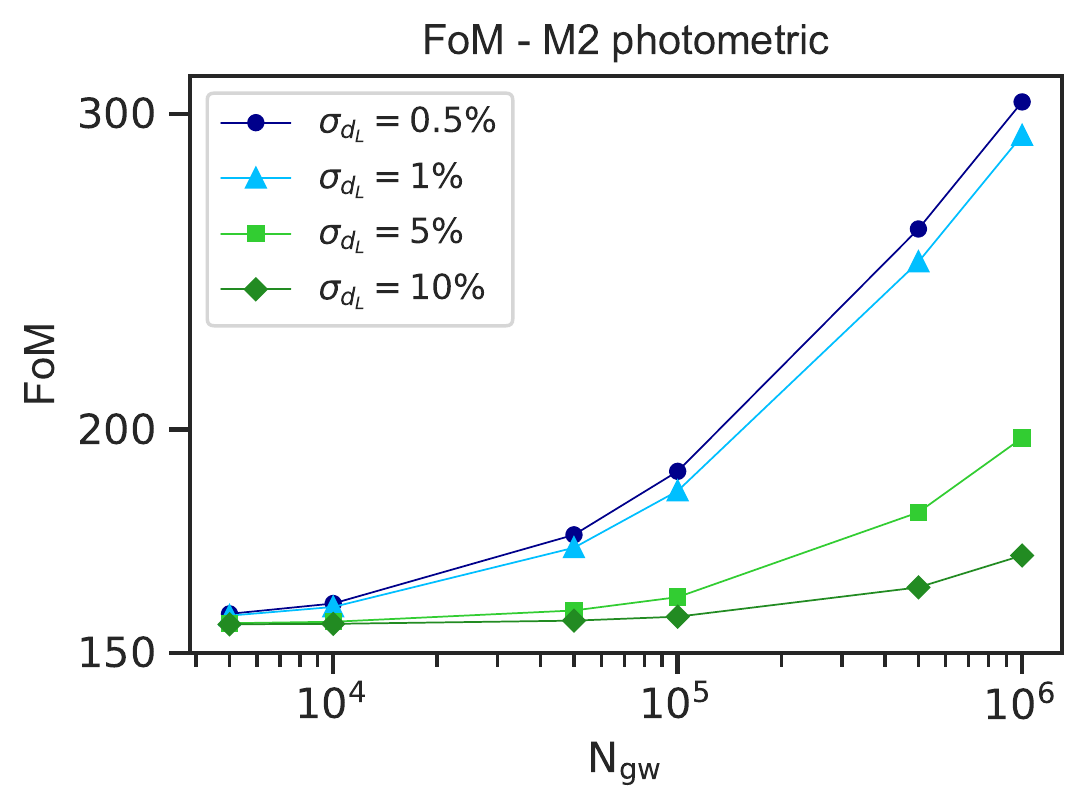}
\caption{Figures of merit for different numbers of total detected events $N_{\rm gw}$ and several choices of the percentage luminosity distance uncertainty $\sigma_{\rm d_L}$. We plot the FoM for model M1 in the top row and M2 in the bottom row, while in the left and right panels are grouped the scenarios with spectroscopic and photometric counterparts respectively.}
\label{fig:fom}
\end{figure}

Lastly, for a better comparison of the results obtained with different setups and on the two MG models considered, we define a Figure of Merit (FoM) on the line of what done in~\cite{Amara:2011}:

\begin{equation}\label{eq:figure_of_merit}
  {\rm FoM} = \rm{det}(\mathcal{F}_{\alpha,\beta})^{\frac{1}{2N}},
\end{equation}
where $\mathcal{F}_{\alpha,\beta}$ is the Fisher matrix of Eq.~\eqref{eq:fisher_matrix_lensing} marginalized over the bias and IA noise parameters, and N is the total number of parameters of the model. The FoM defined in this way is the 2N-th root of the product of the Fisher matrix eigenvalues. The squared root of this product is in turn inversely proportional to the volume of the N-dimensional ellipsoid delimiting the 1$\sigma$ confidence region in the parameter space.
The FoM thus allows us to investigate and quantify not only how the bounds on any individual parameter are affected by GW-WL, but more globally how the volume of the bounds in the parameter space is reduced. We compute the FoM for the Fisher matrices~\eqref{eq:fisher_matrix_lensing} of models M1 and M2, including all auto and cross-correlations of GW-WL, GC and galaxy WL and for all scenarios explored above in terms of $N_{\rm gw}$ and $\sigma_{\rm d_l}$. We also considered separately the cases in which GW events redshifts are determined photometrically or spectroscopically. The results are plotted in Fig.~\ref{fig:fom}.

We notice that the FoM confirms all of our findings above and more. In the photometric cases, for $N_{\rm gw} \leq 10^4$ the overall improvement of bounds in the parameter space is only very mild. However, the FoM start increasing significantly from $N_{\rm gw} = 5\times10^4$ or higher, for all choices of $\sigma_{d_L}$ though significantly better performances occur for $\sigma_{d_L} \leq 1\%$. We also notice that there is only a small difference in the FoM curves for the configurations with $1\%$ and $0.5\%$ average uncertainties on the luminosity distance. This suggests that the photometric redshift error is starting to dominate the correlation noise in Eq.~\eqref{eq:gw_noise_corr}, thus further improving the $d_L$ measurement of phometric events beyond the $1\%$ accuracy level only implies a minor increase in the GW-WL constraining power: the real game-changer becomes in this case the number of available sources.

On the contrary, because the spectroscopic redshift error is much lower, for spectroscopic events the correlation noise remains dominated by $\sigma_{d_L}$, and improvements to the FoM happen for every choice of $\sigma_{d_L}$. We observe, however, that for $\sigma_{d_L} \geq 5\%$ the values of the FoM of both photometric and spectroscopic scenarios are comparable. Only for lower percentage errors on $d_L$ the spectroscopic FoM are higher that the photometric ones, almost doubling their value if the luminosity distance can be measured with sub-percent accuracy.\\

In summary, we conclude that GW-WL is a powerful probe to measure MG parameters and can in some cases outperform the measurements from the upcoming galaxy surveys. Moreover, GWs provide an independent measurement of the MG parameters, affected by completely different systematics and uncertaintes, which can shed new lights on the physics of late-time cosmology.

\section{Discussion and Conclusions}
\label{sec:conclusions}

We have investigated the synergy of the next generation of galaxy surveys and future GW observatories in constraining gravity on cosmological scales, with a focus on the role of weak lensing of gravitational waves. The latter represents a probe complementary to galaxy surveys but characterized by different systematics and thus able to contribute to cosmological constraints in an independent way.

At redshifts higher than $z\sim 0.5$, WL is the dominant contribution to the (linear) corrections to the luminosity distance of GWs induced by large-scale structure; building on this, we have constructed an estimator for the lensing convergence that in the case of bright GW events receives contributions from the conformal coupling of scalar-tensor theories $\Omega(z)$. In the \lcdm{} case, where $\Omega(z)=0$, GW lensing and galaxies eccentricities probe the same quantity and the contribution of GW is primarily that of increasing the WL statistics.

We have explored extensively cross-correlations between $\hat\kappa_{\rm gw}$ and standard galaxy density fields $\delta_g$ and galaxy weak lensing $\kappa_g$, with the aim of determining the observational requirements for future GW detectors to make stringent constraints on late time cosmological models, reaching beyond the limits that will be placed by the next generation of galaxy surveys. To this extent, we did not restrict to any particular GW experiment, rather we considered several scenarios varying the total number of detected GW sources $N_{\rm gw}$, the accuracy of such detection in terms of luminosity distance $\sigma_{d_L}$ and the uncertainty on the estimation of the redshift of the GW event. We have considered separately, and compared, the case in which the GW counterparts (broadly including identified galaxy hosts) can be observed with a photometric survey and an optimistic case in which all redshifts can be determined spectroscopically.

As expected, we find that in the \lcdm{} case GW-WL starts to improve on galaxy constraints only once a very high number of events is observed. We find that $O(10^5)$ detections are needed to start noticing an impact even on parameters such as $\sigma_8$ and $\Omega_m$ to which GW weak lensing is most sensitive, and $O(5\times10^5)$ for this impact to become meaningful. For ST theories with a non-zero conformal coupling, we show that introducing the cross-correlators of $\hat{\kappa}_{\rm gw}$ with other probes can strongly break degeneracies among cosmological parameters, already impacting the constraints on non-\lcdm{} parameters for a relatively small number of detected events. The order of magnitude of the necessary $N_{\rm gw}$ depends again on the accuracy of the luminosity distance measurement, but also non negligibly on the redshift determination and the chosen ST model.

The GW contribution can have a great impact on models in which the time evolution of EFT functions such as the non-minimal coupling is not fixed. These models, such as M1 explored in this work, tend to be more difficult to constrain with galaxies alone, and we find that if $d_L^{\rm \,gw}$ can be measured with an accuracy $\leq 1\%$, then $O(1-5\times10^4)$ GW photometric events are already enough for GW-WL to start improving on galaxy constraints when cross-correlated with the galaxy observables. This contribution becomes significant when a total of $10^5$ sources, or higher, are detected. When spectroscopic counterparts are considered, the required $N_{\rm gw}$ reduces by about one order of magnitude, with $O(10^{4-5})$ sources already doubling the accuracy on the EFT parameters, according to the accuracy on the measurement of $d_L$.

For MG models with all three EFT functions open, such as our M2, we find that the addition of GW-WL to galaxy probes gives, in both spectroscopic and photometric case, results qualitatively similar to M1. We find that in the spectroscopic case GW-WL strongly contributes to reducing degeneracies in the MG sector already when $O(10^{4-5})$ are measured, again depending on the accuracy over the luminosity distance of such sources.

To put the presented numbers in perspective, LISA's most optimistic estimates~\cite{Tamanini:2016zlh} foresee only a few hundreds of GW events over five years of observation, of which only a few tens with a bright counterpart, while recent estimates for a network combining ET and one or two CE detectors place the number of observed BNS at a few tens (or even hundreds) of thousands of events per year~\cite{Pieroni:2022bbh,Ronchini:2022gwk, Maggiore:2019uih}. 
Similarly NS-BH systems are also expected to present electromagnetic counterparts~\cite{Foucart:2012nc, Foucart:2018rjc}. Based on the event rate currently measured by LVK ~\cite{LIGOScientific:2021djp}, it is possible to forecast an NS-BH total rate of a few thousand events per year with the next-generation detectors.
Over the span of 10 years of observations, up to $\sim 10^{5-6}$ GW events could be detected in total.
Therefore, if these events can be associated with the detection of an electromagnetic counterpart, they will potentially be the best candidates to constrain late time cosmology through GW-WL. We must however remark that, according to the cited estimates, third generation ground based detectors are not expected to reach the average $\sim 1\%$ accuracy on the luminosity distance determination required by our more promising setups (see e.g. Figs.~\ref{fig:m1_varying_specs} and~\ref{fig:m2_varying_specs}), likely not even operating as a network. The forecasted error on luminosity distance for a fraction of BNS is expected to be around $10\%$ (slightly lower for nearby sources, but higher for most events)~\cite{Ronchini:2022gwk,Pieroni:2022bbh}. The error bar on the luminosity distance for NS-BHs is estimated to be similar or slightly better (due to higher chirp mass and the relevance of higher-order modes due to higher mass ratios) ~\cite{Vitale:2018wlg, Feeney:2020kxk}.
An average accuracy over $d_L$ of $\sim 10\%$ could still lead to improvements on the galaxy-only cosmological constraints, but in that case we found that $O(10^6)$ GW bright sources are required. This number is likely beyond the capabilities of 3G ground-based detectors, as it is forcasted that in practice it will be possible to associate a counterpart only to a small fraction ($\sim 10^{3-4}$) of all ET and CE events~\cite{Ronchini:2022gwk}, greatly reducing the available statistics. Nonetheless, independent measurements of cosmological parameters through GW-WL can still be carried out with the events from ET and CE, though the reached precision will not be competitive with that of galaxy surveys.

On the bright side, our forecasts could in principle be further improved by considering the cross-correlation signal up to higher $\ell_{\rm max}$, provided that a solid method to treat non-linearities is available, and by exploiting cosmic variance cancellation techniques. Hence, the chance that ET+2CE sources can contribute meaningfully in constraining cosmologies through the cross-correlations of GW-WL with galaxies is not yet completely ruled out, and deserves further investigation.

Moreover, stacking a large statistics of highly accurate GW events will be possible with proposed far future observatories like the space-based Big Bang Observer (BBO), DECi-hertz Interferometer Gravitational-wave Observatory (DECIGO) and Advanced Laser Interferometer Antenna (ALIA). These experiments are expected to reach sub-percent precision in the determination of the luminosity distance and an angular resolution of orders of fractions of degrees for ALIA~\cite{Baker:2019ync} and arcseconds for BBO~\cite{Crowder:2005nr}, with a total number of detected events of several hundreds of thousands. Such a high angular resolution will have the additional advantage of facilitating significantly the task of pinpointing host galaxies, and hence increase the fraction of available bright events\footnote{Note that the number of detected direct EM counterparts is limited, among other things, by the finite observational time that EM surveys can dedicate to the follow-up of GW detections: at present it seems unrealistic that orders of millions of detections per year can be efficiently followed up individually. Hence, host identification will become particularly important in the era of precision cosmology with GW.}.

In conclusion, we find that the cross-correlations of galaxies and GW-WL have the potential to become, with time, crucial probes of cosmology and modified gravity, complementary to galaxy surveys and other cosmological observables. Where sufficient statistics is available, these new probes can both help tighten constraints and make an independent measurement on some \lcdm{} parameters, and strongly reduce existing degeneracies in the MG sector.

While in this paper we have focused on GW weak lensing, GW measurements of the luminosity distance-redshift relation, exploiting both bright (\cite{Tamanini:2016zlh, Caprini:2016qxs, Baker:2020apq,Belgacem:2019tbw,Belgacem:2018lbp, LISACosmologyWorkingGroup:2019mwx}) and dark (\cite{Oguri:2016dgk,Mukherjee:2020hyn,Diaz:2021pem,Mukherjee:2022afz,Laghi:2021pqk}) events, will provide tight bounds on parameters affecting the cosmic expansion history. This potentially helps in breaking some degeneracies between background and growth parameters as measured by GW-WL, as explored in~\cite{Congedo:2018wfn,Mpetha:2022xqo}. These works have shown that the consequence of adding the standard sirens contributions is to effectively tigthen the constraints on some of the parameters to which only GW-WL is sensitive. In this perspective, even with lower statistics GW-WL can play an essential role when one wishes to exploit exclusively GW observations to provide cosmological constraints that are independent of EM probes. We will explore the interplay of standard sirens with galaxy surveys and GW-WL applied to constrain ST theories in a companion work, currently in preparation~\cite{Balaudo:2022}.

\appendix

\section{Impact of GW binning}\label{app:binning}
\label{sec:binning}

\begin{figure}
  \includegraphics[width=0.5\textwidth]{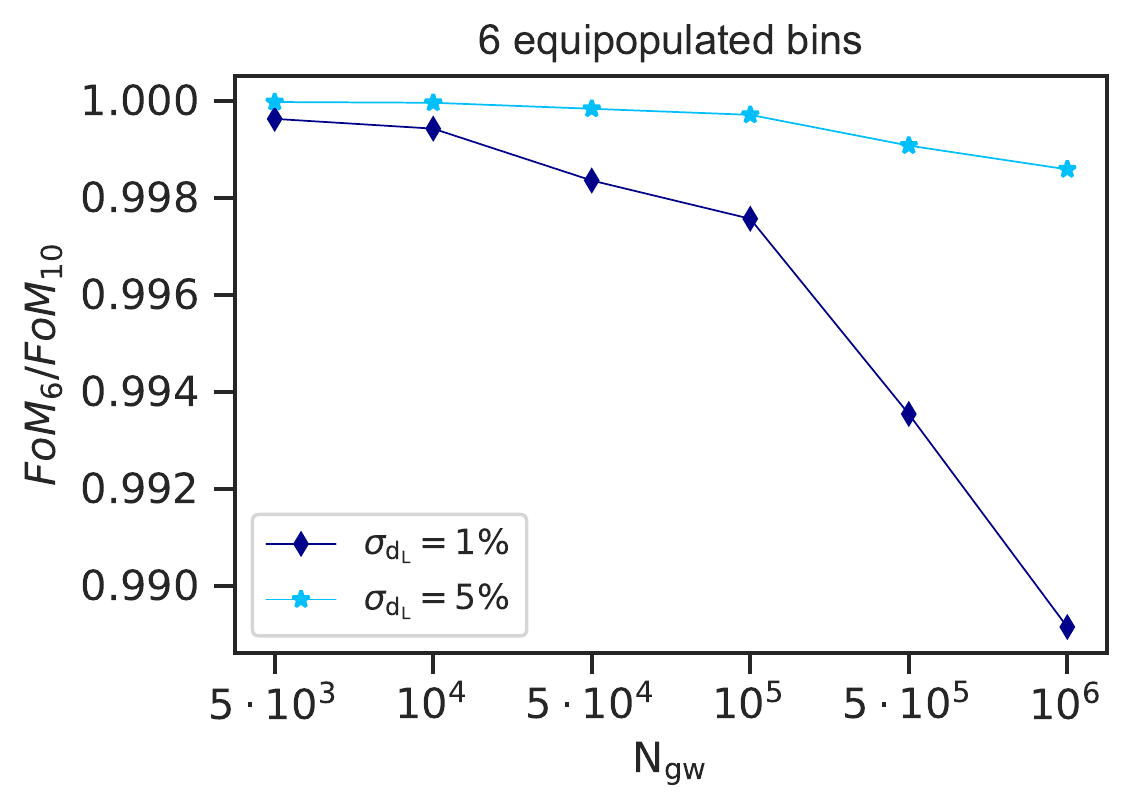}
  \includegraphics[width=0.5\textwidth]{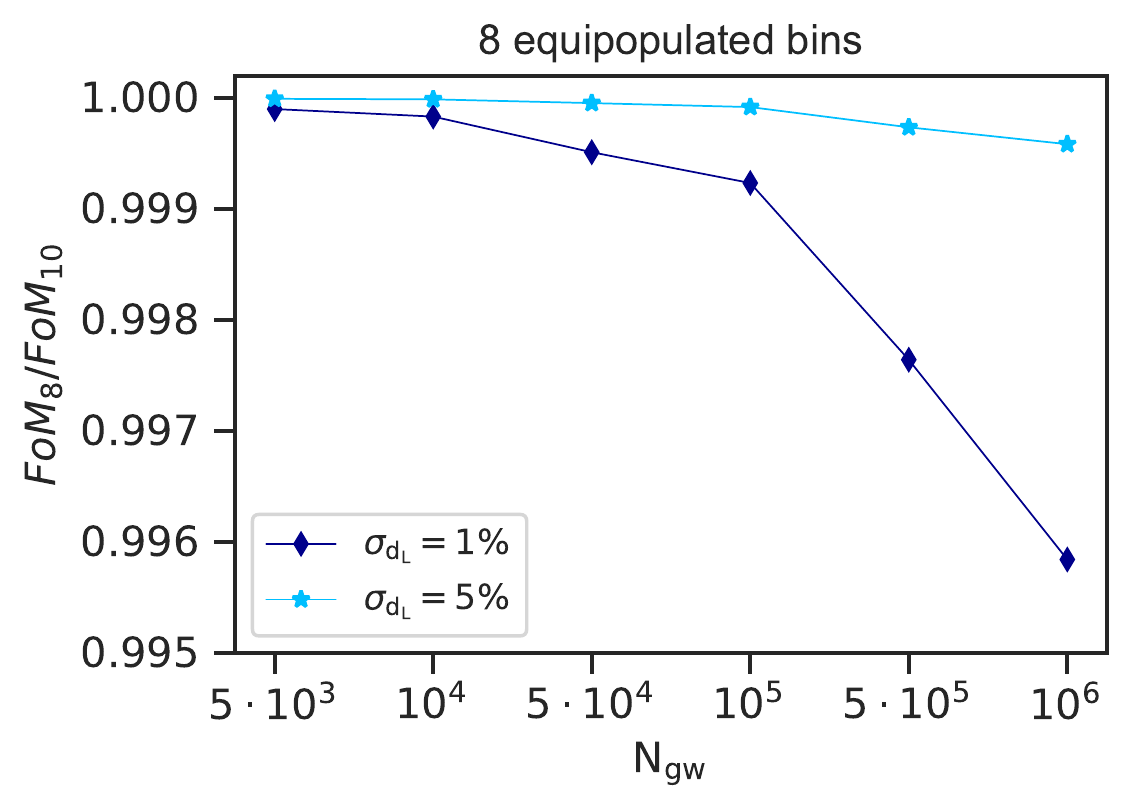}
    \caption{Ratio ${\rm FoM}_{[6,8]}/{\rm FoM}_{10}$ of the FoM computed binning the GW sources in 10 equipopulated tomographic bins and the FoM obtained for the same configurations in terms of number of GW sources $N_{\rm gw}$ and percentage uncertainty on the luminosity distance $\sigma_{d_L}$, but with a different number of bins (6 and 8 in the left and right panel respectively). The results regard Model I.}
    \label{fig:binning}
\end{figure}

We address here the impact that our tomographic binning choice for the GW sources has on the final results, investigating if the additional constraining power brought by GW-WL and its cross correlation with galaxy survey is affected changing the number of bins.

We choose the figures of merit (FoM) defined in Sec.~\ref{sec:results} as indicators of such constraining power, and we compute the FoM for several observational configurations, characterized by the total number of detected GW events with a spectroscopic counterpart ($N_{\rm gw} \in [5\times10^3, 10^6]$) and by the average percentage uncertainty on the GW luminosity distance ($\sigma_{d_L} \in [1\%, 5\%]$). We repeat our computations binning the GW sources respectively in 6, 8 and 10 equipopulated tomographic bins. In Fig.~\ref{fig:binning}, we plot the ratio ${\rm FoM}_n/{\rm FoM}_{10}$ for all observational scenarios, with subscripts indicating the number of bins and $n=(6,8)$ in the left and right panel respectively. We use the 10 bins case as a reference, as it is the most optimistic scenario and also the one we chose to present our results in the main body of this work. In Fig.~\ref{fig:binning} we present the FoM for Model I. We checked that similar results still hold for \lcdm{} and Model II.

We observe that, reducing the numbers of bins in all cases implies lower FoM, which in turns signifies a deterioration of the constraints on cosmological parameters. This happens because decreasing the bins number, and thus increasing the bins size, has the effect of smoothing out the correlation signal in any given bin. However, this effect is mostly compensated by the augmented number of events falling in each bin, that attenuates the correlation noise of Eq.~\eqref{eq:gw_noise_corr}. Indeed, we observe that in all scenarios the FoM depends only very mildly on the number of bins. Higher deviations occur for higher numbers of detected events and better accuracy on the luminosity distance measurement, though we notice that even in the most extreme case of 6 tomographic bins and $10^6$ GW events with $d_L$ detected at $1\%$ accuracy, ${\rm FoM}_6$ is lower than ${\rm FoM}_{10}$ only of $\sim 1\%$.

Consequently, we can conclude that our choice of the number of bins for the GW sources impacts only very mildly the overall constraining power on cosmological parameters gained through GW-WL, and does not affect significantly the results presented here or our conclusions.

\section{Impact of GW sources redshift distribution}\label{app:redshifts}
\label{sec:redshifts}

In Sec.~\ref{sec:analysis}, we made a choice about the assumed redshift distribution of sources of bright gravitational waves events. This redshift distribution results, in general, from a convolution of the underlying distribution of merging binaries in the Universe below $z=2.5$, with the selection function of the GW detector considered for their observation and an additional selection function dependent on the redshift identification method. This last function accounts for the fact that it will in practice be impossible to associate a redshift to all of the observed binaries, and the number of bright sources will differ if the redshift inference is based on a direct counterpart rather than on the identification of a host within a galaxy catalogue. Moreover, in the case of direct counterparts, the final redshift distribution of bright sources also depends on the considered channels for the EM counterpart emission.

Various combinations of all different possibilities generate a mosaic of viable observational scenarios.
To preserve a certain degree of generality, in the main body of this work we decided not to make any direct assumption on the observational setup concerning the redshift. Rather, we took a phenomenological approach, modelling the number count of bright GW sources as a function of redshift as

  \begin{equation}
    \frac{dn_{\rm gw}}{dz} \propto \left(\frac{z}{z_0}\right)^2 \exp\Bigg[-\Bigg(\frac{z}{z_0}\Bigg)^{3/2}\Bigg]\,.
    \label{eq:gw_sources_distribution}
  \end{equation}

with $z_0 = 1.5$. This distribution was chosen to match the population of injected BNS to be observed by a network of ET and two CE detectors in~\cite{Ronchini:2022gwk}. Observationally, this choice is closest to a particularly favorable scenario: it considers BNS - which are expected to be the most numerous binary population observable in future GW detectors - and it assumes that almost all merging binaries can be (i) detected by the GW observatory and (ii) matched with a redshift measurement.
Condition (i) is close to what happens in a 4th generation GW detector with very low signal-to-noise threshold, such as e.g. BBO~\cite{Crowder:2005nr}, while (ii) is likely to be satisfied observationally for high numbers of sources only through host identification, again in principle possible thanks to the high accuracy of 4th generation GW detectors.\\

In the rest of this Appendix, we characterize the impact of these assumptions on the derived cosmological constraints. We do so, by looking at how the constraints change if the underlying GW source distribution is chosen to peak at progressively lower redshifts, i.e. we considered two alternative scenarios with $z_0=1.0$ and $z_0=0.5$. The former is observationally closest to a more realistic 4th-generation detector situation: the distribution again assumes that the low SNR of the GW events in the detector is such that a consistent statistic of sources can be detected up to relatively high redshifts. However, it accounts for the fact that it is more difficult to associate hosts to high redshift events ($z\geq1.5$), both because the sky-localization of the GW event partially deteriorates at high z and because the galaxy catalogues become progressively less complete.
Finally, the case with $z_0=0.5$ is closest to what could realistically be expected for 3rd-generation detectors, such as ET and CE: the SNR threshold, especially if a network of detectors is considered, is low enough that many events can be detected up to $z\sim2.5$. In turn, the sky-location is so poor that individual host identification is impossible for most of the events. Still, BNS are expected to emit direct EM counterparts, and for 3G experiments that is likely to be the primary channel of redshift measurement, with direct counterparts being detectable only at relatively low redshifts. Astrophysical models concerning EM emission and detection vary significantly, though many current forecasts place the peak of measurable bright events around $z\sim0.5-0.7$, with the total number of events dropping quickly for $z>1.0$~\cite{Pieroni:2022bbh,Ronchini:2022gwk, Maggiore:2019uih}.

\begin{figure}
    \centering
    \includegraphics[width=0.49\textwidth]{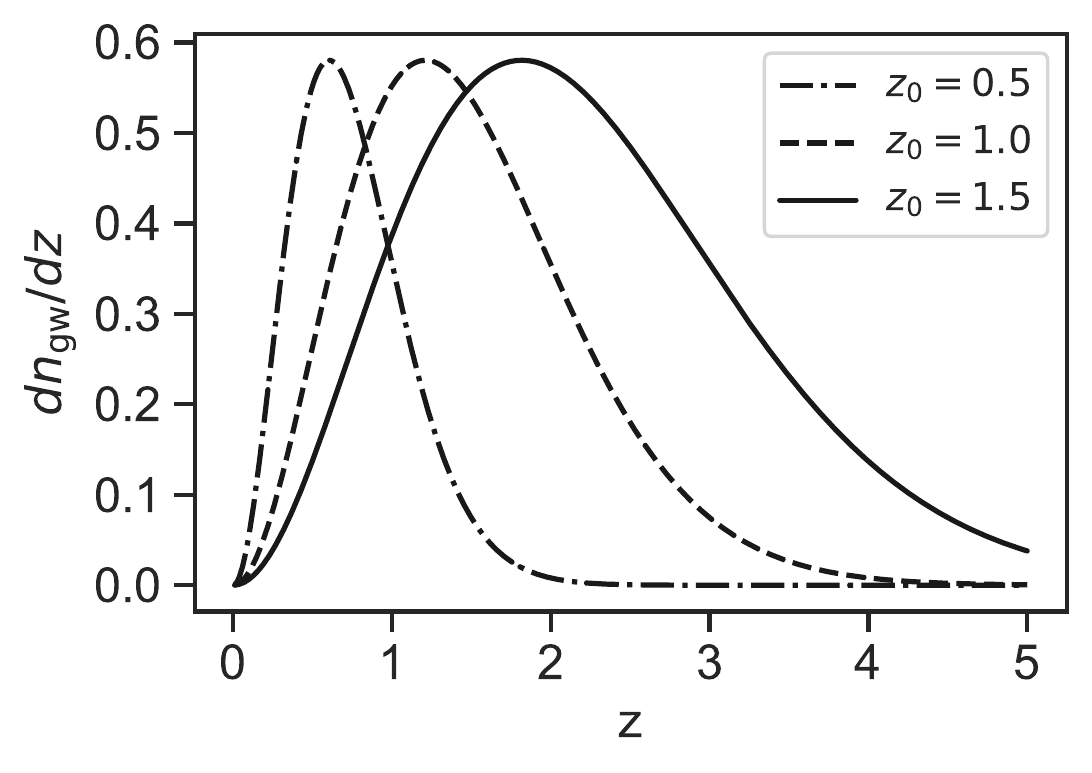}
    \includegraphics[width=0.49\textwidth]{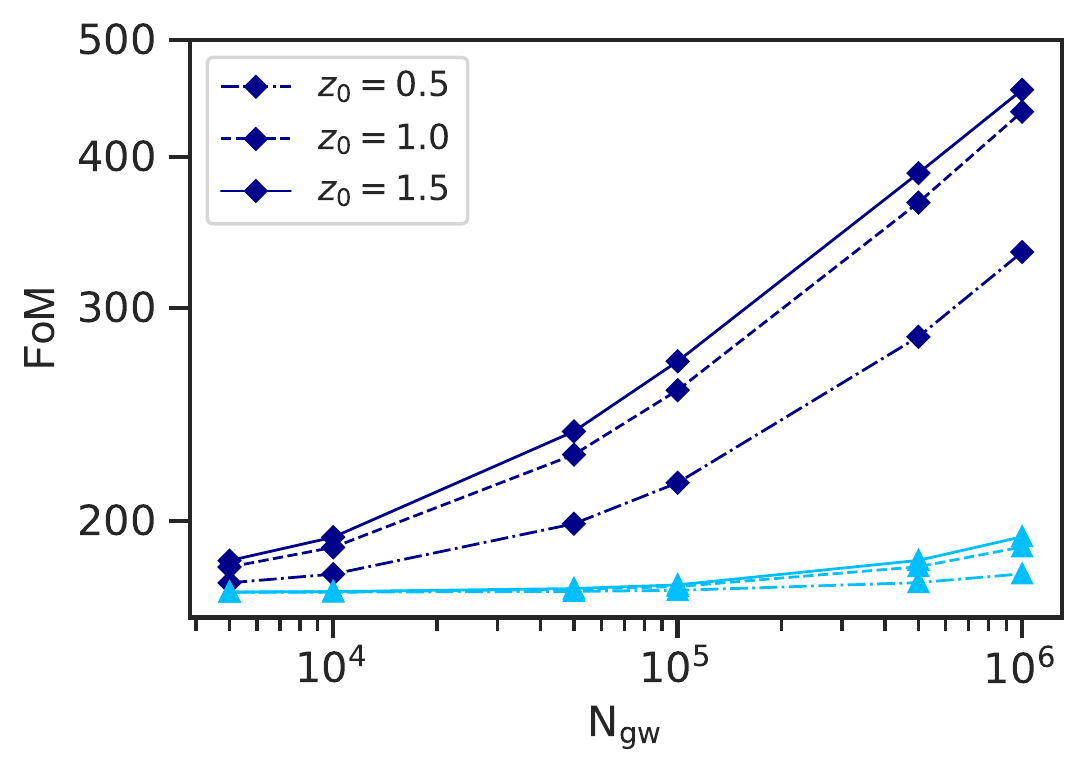}
    \caption{Left panel: comparison between three different choices for the normalized redshift distribution of gravitational wave sources. All distributions assume Eq.~\eqref{eq:gw_sources_distribution} with $z_0=1.5$ (solid line), $z_0=1.0$ (dashed line) and $z_0=0.5$ (dash-dotted line). Right panel: FoM as a function of the total number of bright GW events, plotted for different choices of $z_0$ and assuming $\sigma_{d_L} = 1\%$ (dark blue diamonds) or $\sigma_{d_L} = 10\%$ (light blue triangles).}
    \label{fig:gw_distribution_comparison}
\end{figure}

Our three different choices for the source distributions are compared in the left panel of Fig.~\ref{fig:gw_distribution_comparison}.
The right panel shows instead the evolution of the figures of merit (FoM) of our Model I defined in Sec.~\ref{subsec:results_mg}, as a function of the total number of observed events and for different choices of $z_0$ (solid, dashed and dotted lines) and different average uncertainty on the luminosity distance ($\sigma_{d_L} = 1\%$ for dark blue lines and $\sigma_{d_L} = 10\%$ for light blue lines). The FoM for Model II are not shown, as we find no evident qualitative difference with respect to Model I.

We notice that the FoMs for the cases $z_0=1.5$ and $z_0=1.0$ are still very similar, and there is only a mild deterioration of the overall bounds in the latter case. This can be interpreted in the light of our findings of Sec.~\ref{subsec:results_mg}, i.e. that most of the increase in constraining power obtained when adding GW to the LSS does not come from the self-correlation of GW-WL, but from the cross-correlation of GWs with the galaxies. These cross-correlations are the highest when there is a significant overlap between the redshift distributions of GW binaries and galaxies, which is the case for both $z_0=1.5$ and $z_0=1.0$. The case $z_0=1.5$ remains however slightly more performing, as it presents higher statistics of GW sources at high z, which enhances the constraining power of the GW-WL self-correlations terms. On the contrary, the distribution with $z_0=0.5$ has very few sources at $z\geq1.0$, thus reducing the overlap with the galaxy distribution and affecting the constraining power of the cross-correlations terms. The overall cosmological constraints are looser as a result.

 \begin{figure}
    \centering
    \includegraphics[width=1.\textwidth]{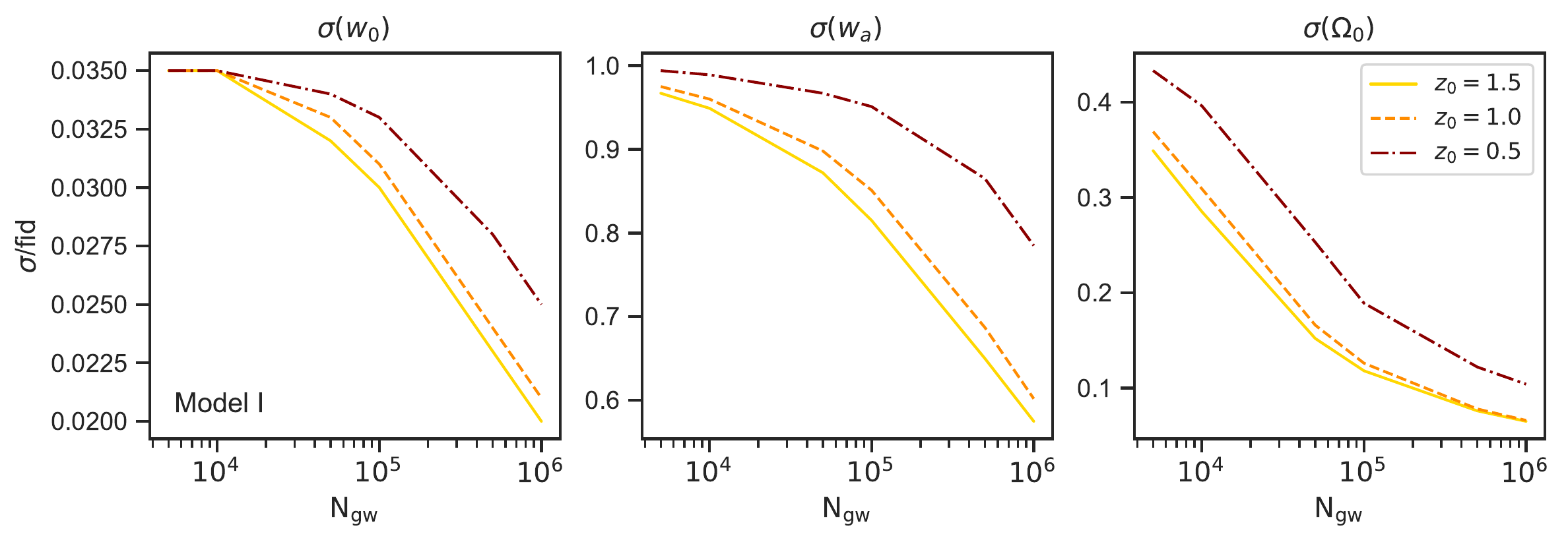}
    \includegraphics[width=1.\textwidth]{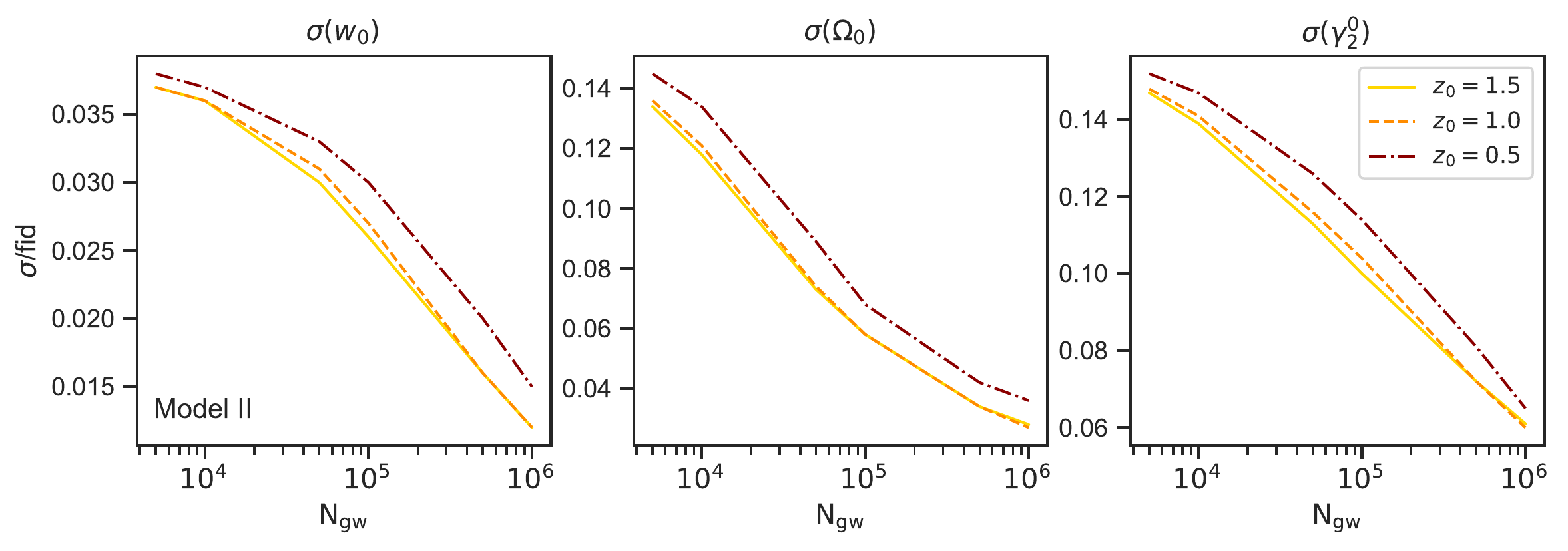}
    \caption{Relative $1\sigma$ bounds on different MG parameters $\theta_i$ computed as $\sigma(\theta_i)/\theta_{i, \rm fid}$ as a function of the total number of bright GW sources, for Model I (top row) and Model II (bottom row). Solid, dashed and dash-dotted lines correspond to $z_0 = 1.5$, $1.0$ and $0.5$ respectively.}
    \label{fig:sigmas_sd}
\end{figure}    

 In Fig.~\ref{fig:sigmas_sd}, we show the evolution of the relative 1$\sigma$ bounds ($\sigma(\theta_\alpha)/\theta_{\alpha,{\rm fid}}$) on the individual MG parameters of Model I (top row) and Model II (bottom row) as a function of the total number of GW events. Bounds are plotted here assuming $\sigma_{d_L} = 1\%$ and we plot in different colours and line styles the results for different assumptions on the source distribution. We notice that, in general, the impact of the source distribution is more prominent for Model I than for Model II. However, in both cases, there are only minor differences in the individual bounds between the choices $z_0=1.5$ and $z_0=1.0$, while the case $z_0=0.5$ differs more significantly, according to the specific parameter considered. For example, in Model I, choosing a certain source distribution can lead to a variation in the bounds on $w_0$ of the order of less than $1\%$, while bounds on $w_a$ and $\Omega_0$ deteriorate faster, with differences in the relative bounds of the order of $10\%$.

 We also notice a variation in the trend of these curves depending on the parameters considered. For example, in the case of $w_0$ and $w_a$ for both Model I and II the difference between the different curves is way smaller at lower $N_{\rm gw}$ than it is at higher $N_{\rm gw}$. This is likely due to the fact that, as we showed in Sec.~\ref{sec:results}, for a low number of GW sources cosmological bounds on the DE equation of state are still highly dominated by the galaxies, while the GW contribution becomes more and more important as the number of sources increases. However, for $\Omega_0$ and $\gamma_2^0$ we notice instead that the difference among the curve is the same approximately at any $N_{\rm gw}$, even diminishing slightly as $N_{\rm gw}$ gets bigger. This is consistent with Fig.~\ref{fig:m1_varying_specs} and~\ref{fig:m2_varying_specs}, as we notice that when $\sigma_{d_L} = 1\%$, GW contribute significantly to constraining those two parameters at all $N_{\rm gw}$. The curves in Fig.~\ref{fig:sigmas_sd} seems to hint that, for the bounds on these parameters in particular, the loss of high redshift sources has not a very severe impact, and is on the contrary partially compensated by the increased statistics at low redshifts.

 In conclusion, we find that the choice of the underlying GW source distribution can have a non-negligible effect on cosmological bounds, as could be expected. While some of the results presented in the main body of this work correspond to an optimistic observational scenario, we find that considering a more realistic source distribution in the context of future 4th-generation detectors ($z_0=1.0$) does not affect our results in a relevant way. On the contrary, considering a redshift distribution of bright GW events closer to realistic expectations for 3rd-generation GW detectors leads to a non-negligible deterioration of the cosmological constraints. This further validates our previous conclusion that GW-WL is unlikely to become a probe competitive with LSS with the next generation of GW detectors, while it remains potentially a promising cosmological probe in the farther future.

 \acknowledgments
 We thank Marco Raveri and Fabrizio Renzi for useful discussions. AB is supported by a de Sitter Fellowship of the Netherlands Organization for Scientific Research (NWO). AG acknowledges support from the NWO and the Dutch Ministry of Education. 
M.M. acknowledges funding by the Agenzia Spaziale Italiana (\textsc{asi}) under agreement no. 2018-23-HH.0 and support from INFN/Euclid Sezione di Roma. The work of SM is a part of the $\langle \texttt{data|theory}\rangle$ \texttt{Universe-Lab} which is supported by the TIFR and the Department of Atomic Energy, Government of India. AS acknowledges support from the NWO and the Dutch Ministry of Education, Culture and Science (OCW) (grant VI.Vidi.192.069).

\bibliographystyle{JHEP}
\bibliography{references.bib}

\end{document}